\documentclass[11pt,a4paper]{hiepadraft2}

\usepackage{besphysics}
\usepackage{authblk}
\usepackage{amsmath}
\usepackage{graphicx}
\usepackage{epsfig}
\usepackage{epstopdf}
\usepackage{multirow}
\usepackage{overpic}
\usepackage{colortbl}
\usepackage{tabularx}
\usepackage{enumitem}
\usepackage{url}
\usepackage{multirow}
\usepackage{setspace}
\usepackage{amssymb}
\usepackage{bm}

\usepackage{eqnarray}
\usepackage{subcaption}
\usepackage{tabularx}

\usepackage{booktabs}
\usepackage{siunitx}

\usepackage{color}
\usepackage{amsfonts,amsmath,amssymb}
\usepackage{bm}
\usepackage{titletoc}
\usepackage{makecell}
\usepackage[colorlinks,linkcolor=blue,anchorcolor=blue,citecolor=blue]{hyperref}
\usepackage{listings}
\usepackage{multirow}
\usepackage{booktabs}
\usepackage{array}

\hypersetup{
colorlinks=true,
allcolors=blue
}
\usepackage{float}
\usepackage[section]{placeins}
\usepackage{verbatim}
\usepackage{siunitx}
\usepackage[numbers,sort&compress]{natbib}
\usepackage{chapterbib} 
\usepackage{textcomp}

\usepackage{comment}
\graphicspath{{./figure_EicC_physics/}}

\usepackage{upgreek}
\usepackage{braket}
\usepackage{xcolor}

\colorlet{RED}{red}
\usepackage{authblk}
\newcommand{\bea}{\begin{eqnarray}}
\newcommand{\eea}{\end{eqnarray}}

\title{}

\author[1]{Xiaozhi Bai}
\author[2]{Xu Cao}
\author[3]{Zhe Cao}
\author[4]{Jinhui Chen}
\author[1]{Kai Chen}
\author[5]{Qibo Chen}
\author[6]{Shi Chen}
\author[7,8]{Xin Chen}
\author[2]{Yuquan Chen}
\author[9]{Zhenyu Chen}
% ...

\affil[1]{Key Laboratory of Quark and Lepton Physics (MOE) and Institute of Particle Physics, Central China Normal University, Wuhan 430079, China}
\affil[2]{Institute of Modern Physics, Chinese Academy of Sciences, Lanzhou 730000, China}
\affil[3]{Department of Modern Physics, University of Science and Technology of China, Anhui 230026, China}
\affil[4]{Key Laboratory of Nuclear Physics and Ion-beam Application (MOE), Institute of Modern Physics, Fudan University, Shanghai 200433, China}
\affil[5]{East China Normal University, Shanghai, 200241, China}
\affil[6]{University of Chinese Academy of Sciences, Beijing 101408, China}
\affil[7]{Department of Physics, Tsinghua University, Beijing 100084, China}
\affil[8]{Center for High Energy Physics, Tsinghua University, Beijing 100084, China}
\affil[9]{Key Laboratory of Particle Physics and Particle Irradiation (MOE), Institute of Frontier and Interdisciplinary Science, Shandong University, Qingdao 266237, China}
% ...

\abstracttext{
Chirality lies at the heart of low-energy QCD, governing the symmetry structure that shapes hadron masses and strong interaction dynamics. Among the most compelling open questions tied to chiral dynamics and spontaneous chiral symmetry breaking is the longstanding $\Lambda$ polarization puzzle, in which $\Lambda$ hyperons produced in unpolarized hadronic collisions exhibit a surprisingly large transverse polarization that remains theoretically unexplained.
This whitepaper presents the proposal for the Hyperon-Nucleon Spectrometer (H-NS) at the High-Intensity heavy-ion Accelerator Facility (HIAF). Leveraging the high energy and high intensity of HIAF’s proton and heavy-ion beams, the H-NS experiment will perform systematic studies of hyperon polarization phenomena and their underlying mechanisms in proton-proton ($pp$), proton-nucleus ($pA$), and nucleus-nucleus ($AA$) collisions in the fixed target mode. 
A wide-range beam energy scan, including proton beams from 3 GeV up to 9.3 GeV (HIAF) and up to 32 GeV (upgraded HIAF), will be conducted to examine the dependence of polarization on collision energy.
The spectrometer is designed with specialized detectors capable of high-precision reconstruction of final-state baryon polarizations. Among its many interesting and important measurements, H-NS will simultaneously measure hyperon and proton spin observables to explore the polarization mechanism in hadronic interactions and the spin structure of baryons---the fundamental constituents of visible matter.  Furthermore, the use of $pA$ and $AA$ collisions will enable detailed investigations of cold and hot nuclear matter effects on spin polarization. The H-NS program is set to provide unprecedented experimental constraints on the dynamics of spin in strongly interacting matter under extreme conditions, offering new insights into non-perturbative QCD. 
Its physics program and detector development will significantly benefit the future Electron-ion Collider in China.
This document outlines the key physics objectives and the conceptual design of the H-NS spectrometer.
}

\begin{document}
%%%%%%%%%%%%%%%%%%%%%%%%%%%%%%%%%%%%%%%%%%%%%%%%%%%%%%%%%%%%%%%%%%%%%%%%%%%%%%
\setlength{\baselineskip}{0.5cm}
%===========================================================================
%================== Table of contents ======================================
%=======]====================================================================
\tableofcontents
\newpage

%%%%%%%%%%%%%%%%%%%%%%%%%%%%%%%%%%%%%% EicC physics chapter %%%%%%%%%%%%%%%%%%%%%%
\begin{chapter}{Physics}

Understanding the fundamental building blocks of matter and the forces that bind them is a central pursuit of modern physics. 
The nucleon (protons and neutrons), a primary constituent of atomic nuclei, accounts for most of the visible mass in the universe. 
Yet, despite a century of study~\cite{Deur:2018roz,Boussarie:2023izj}, the complete description of its internal structure remains a profound challenge, particularly concerning the origin of its spin and mass.
The simple picture of a proton as a composite of three valence quarks (two up and one down) is insufficient. 
We now know the proton's interior is a dynamic and complex environment governed by Quantum Chromodynamics (QCD)~\cite{Gross:2022hyw}, teeming with a ``sea'' of virtual quark-antiquark pairs and the gluons that mediate the strong force.

Concerning strange quarks, this is where hyperons become invaluable probes. 
Produced in high-energy collisions, such as those at the Large Hadron Collider (LHC)~\cite{Evans:2008zzb} or the Relativistic Heavy Ion Collider (RHIC)~\cite{{PHOBOS:2004zne,BRAHMS:2004adc,STAR:2005gfr,PHENIX:2004vcz}}, hyperons 
act as messengers of the spin from the collision.
%provide a unique window into the nucleon's hidden structure. 
Crucially, hyperons are ``spin-self-analyzing'' as will be explained later 
%: their decay products' angular distribution reveals their polarization states~\cite{Lee:1957qs}
, offering a direct window into spin dynamics.
Studying hyperon polarization provides a unique and stringent test of QCD.

A general spin state for an ensemble is not described by a state vector, but by a $2 \times 2$ density matrix, $\rho$. The spin density matrix is Hermitian ($\rho=\rho^\dagger$), positive semidefinite, and normalized ($\text{Tr}\rho = 1$)~\cite{Bourrely:1980mr,Leader:2001nas}. For the spin-$1/2$ system, it can be expanded in terms of the identity matrix ($\mathbb{I}$) and the Pauli matrices $\vec{\sigma}$ as~\cite{Bourrely:1980mr,Leader:2001nas}
$$
\rho = \frac{1}{2} \left( \mathbb{I} + \vec{P} \cdot \vec{\sigma} \right).
$$
Here, $\vec{P}$ is the polarization vector, which fully characterizes the spin state of the ensemble. Its magnitude satisfies  $|\vec{P}| \le 1 $. An unpolarized ensemble corresponds to $|\vec{P}| = 0$, yielding the maximally mixed state $\rho = \tfrac{1}{2}\mathbb{I}$. A fully polarized ensemble corresponds to $|\vec{P}| = 1$, which represents a pure state. 

From this definition, the polarization vector is equivalent to the expectation value of the Pauli spin operator for the ensemble $\vec{P}  = \text{Tr}(\rho \, \vec{\sigma})$. Choosing a quantization axis (e.g., the $z$-axis), the longitudinal polarization $P_z$ describes the population imbalance between spin-up ($N_\uparrow$) and spin-down ($N_\downarrow$) states: 
$$
P_z = \frac{N_\uparrow - N_\downarrow}{N_\uparrow + N_\downarrow}.
$$

Hyperons, and the $\Lambda$ baryon in particular, serve as an uniquely effective probe for polarization because their weak decays provide a built-in spin analyzer~\cite{Lee:1957qs,Jacob:1959at,Chung:1971ri,Zhang:2023box,Zhang:2024rbl}. 
%due to their decay properties. 
%The primary advantage is that the $\Lambda$ is self-analyzing, meaning its polarization can be determined from the angular distribution of its own decay products. 
This capability stems from the fact that the $\Lambda$ decays almost exclusively ($BR \approx 64.1\%$) via the parity-violating weak interaction as $ \Lambda \to p\pi^{-}$. 
Due to the non-conservation of parity in the weak interaction, the decay is anisotropic. 
The daughter proton ($p$) is preferentially emitted along the direction of the parent $\Lambda$'s spin vector. 
By defining $\theta_p^*$ as the angle between the proton's momentum and the polarization axis of $\Lambda$ in the rest frame of $\Lambda$, the distribution can be simplified to~\cite{Lee:1957qs}:
$$
\frac{dN}{d\cos\theta_p^*} \propto 1 + \alpha_\Lambda P_\Lambda \cos\theta_p^*.
$$
Here, $P_\Lambda = |\vec{P}_\Lambda|$ is the magnitude of the polarization vector. 
$\alpha_\Lambda$ is the weak decay asymmetry parameter.
Since $\alpha_\Lambda$ is well measured and sizable~\cite{ParticleDataGroup:2024cfk}, this provides a direct, sensitive measurement of the hyperon polarization $P_\Lambda$.

The polarization of $\Lambda$ hyperons was first observed at Fermilab in an experiment using a 300 GeV proton beam on a beryllium (Be) target in the 1970s~\cite{Bunce:1976yb}, as shown in Fig.~\ref{fig:fnal76}. 
In that experiment, polarization was measured along the direction normal to the $\Lambda$ production plane. 
The results showed that the $\Lambda$ polarization increases with the hyperon transverse momentum ($p_{\rm T}$). 
At the largest measured $p_{\rm T}$, the polarization reached several tens of percent, even though both the beam and the target were unpolarized.
This surprising result--a produced particle ``self-polarizing'' even with unpolarized beams and targets---revealed a profound spin effect in hadronic reactions.

\begin{figure}[h]
\centering
\includegraphics[width=0.5\linewidth]{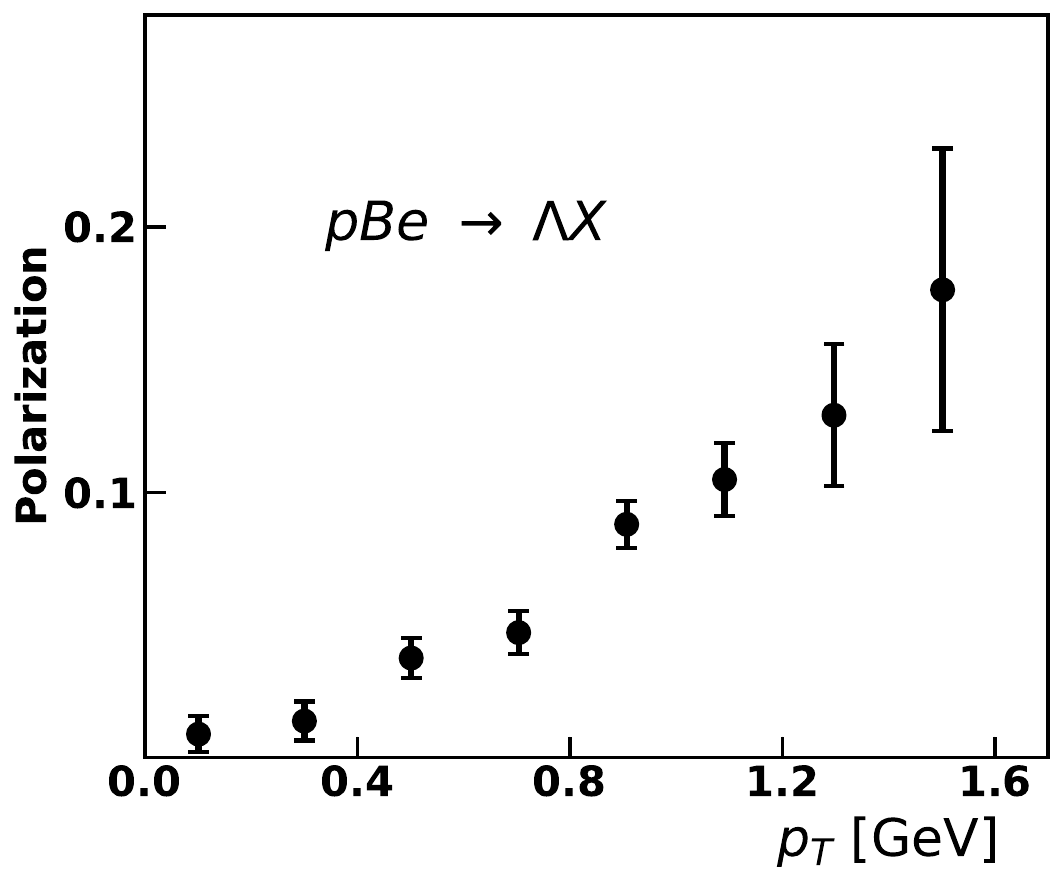}
\caption{\label{fig:fnal76}The first observable of the ``self-polarizing'' phenomenon of $\Lambda$ at Fermilab~\cite{Bunce:1976yb}. The polarization was measured along the direction normal to the $\Lambda$ production plane :$-\vec{p}_{\text{beam}} \times \vec{p}_{\Lambda}$, which is opposite to the current convention: $\vec{p}_{\text{beam}} \times \vec{p}_{\Lambda}$.}
\end{figure}

\begin{figure}[h]
\centering
\subfloat[]{
\includegraphics[width=0.45\linewidth]{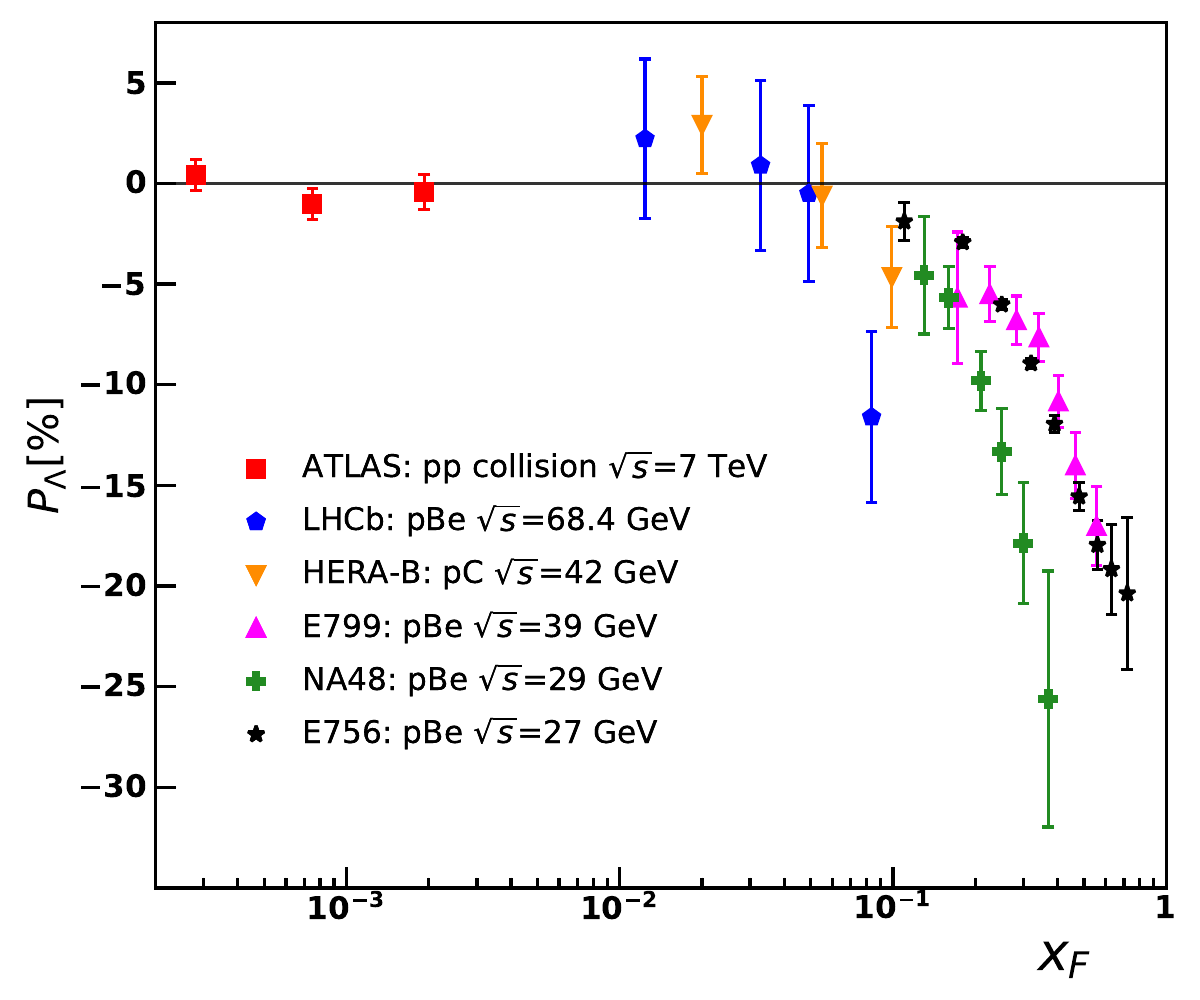} }
\centering
\subfloat[]{
\includegraphics[width=0.45\linewidth]{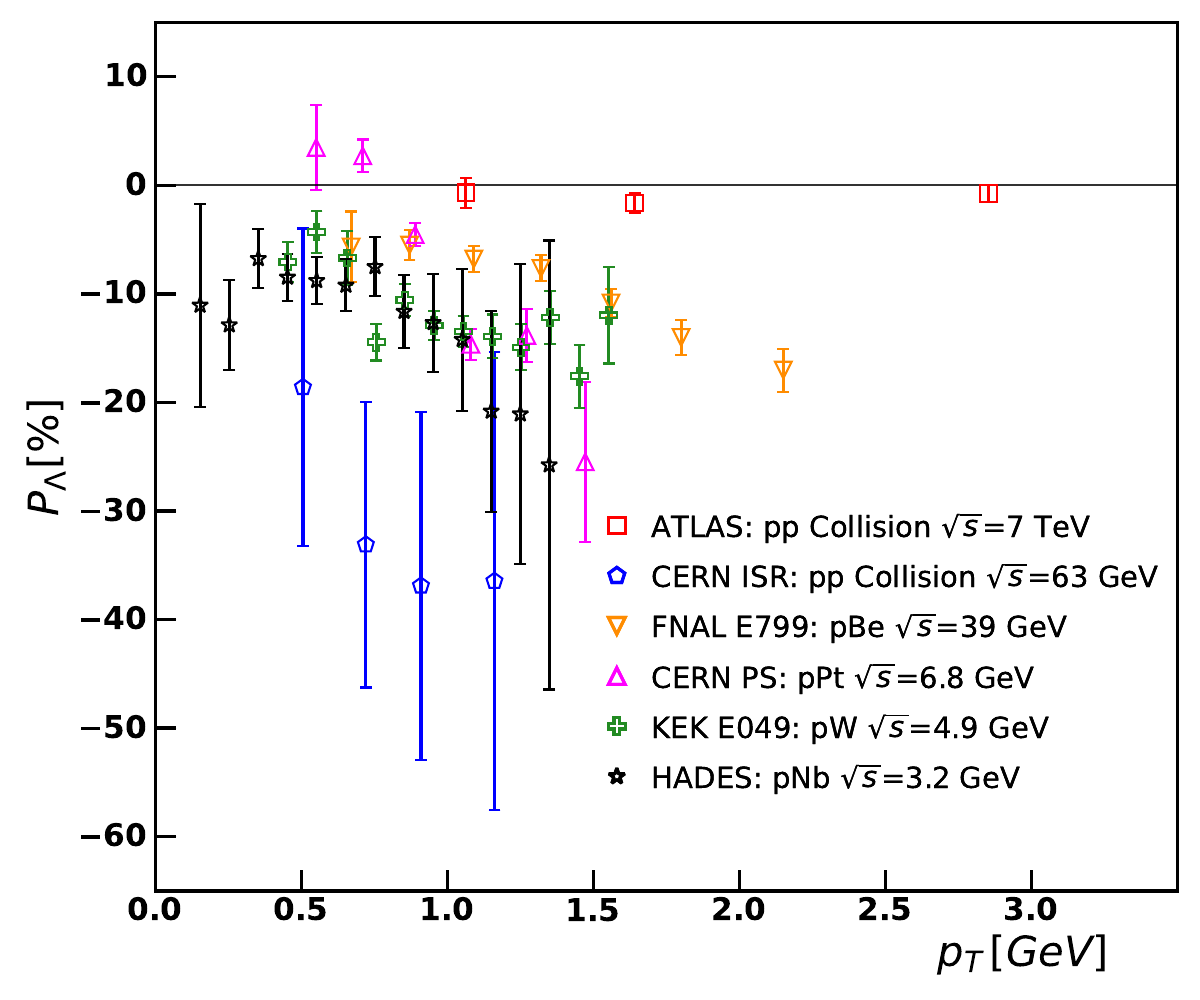}}
\caption{ \label{fig:polarization_inclusive}$\Lambda$ polarization in the $p(p/A) \rightarrow \Lambda X$ process. (a) $\Lambda$ polarization as a function of $x_{\rm F}$, with data taken from Refs.~\cite{Heller:1978ty,Ramberg:1994tk, Tkachev:1999mv,HERA-B:2006rds, ATLAS:2014ona, LHCb:2024vwi}; (b) $\Lambda$ polarization as a function of $p_{\rm T}$, with data from Refs.~\cite{Heller:1977mv,Erhan:1979xm,Abe:1983xk,Ramberg:1994tk,ATLAS:2014ona}. Overall, the $\Lambda$ polarization shows a rising trend with increasing $x_{\rm F}$ and $p_{\rm T}$.}
\end{figure}

Following the Fermilab discovery, numerous experiments at various laboratories confirmed $\Lambda$ polarization across different energies and beam/target combinations~\cite{Heller:1977mv,Heller:1978ty,Erhan:1979xm,Abe:1983xk,Ramberg:1994tk,Tkachev:1999mv,HERA-B:2006rds,ATLAS:2014ona,LHCb:2024vwi}, as illustrated in Fig.~\ref{fig:polarization_inclusive}. 
These experiments consistently found that $\Lambda$ hyperons produced in unpolarized proton-proton (and proton-nucleus) collisions have a polarization whose magnitude grows with increasing transverse momentum $p_{\rm T}$ of the $\Lambda$.
Moreover, to compare data across different collision energies, $\Lambda$ polarization was studied versus the Feynman-$x_{\rm F}$ (a variable related to the fraction of the available longitudinal momentum carried by the $\Lambda$).
These studies showed that $\Lambda$ polarization also increases with $x_{\rm F}$---hyperons carrying a larger fraction of the beam momentum tend to be more polarized. 
In particular, at moderate energies, $\Lambda$ produced very forward (high $x_{\rm F}$) can reach polarization on the order of 30–40\% in magnitude. 
Despite decades of study~\cite{Andersson:1979wj,Panagiotou:1989sv,Barni:1991qn,Soffer:1991am,Troshin:1996hd,Liang:1997rt,Felix:1999tf}, the underlying mechanism for this self-generated polarization remains an open puzzle---a long-standing question in non-perturbative QCD.
In addition to the transverse polarization in nucleon-nucleon or nucleon-nuclei events, $\Lambda$ hyperons in non-central heavy-ion collisions exhibit a distinct but related phenomenon known as global polarization relative to the reaction plane~\cite{Liang:2004ph, Liang:2004xn,Gao:2007bc}. 
Interestingly, the magnitude of $\Lambda$ global polarization is found to increase as the collision energy decreases~\cite{STAR:2007ccu,STAR:2017ckg,STAR:2018gyt,STAR:2021beb,STAR:2023nvo,ALICE:2019onw,HADES:2022enx}.
Beyond hadron–hadron collisions, $\Lambda$ polarization has also attracted broad interest in other high-energy reactions, such as lepton–nucleon scattering and electron–positron annihilation~\cite{Mulders:1995dh,Boer:1997mf,Barone:2010zz,Pitonyak:2013dsu,Wei:2014pma,Chen:2021zrr,Gamberg:2018fwy,Gamberg:2021iat,Faldt:2017kgy,Perotti:2018wxm,Salone:2022lpt,Cao:2024tvz,Zhang:2025oks}. In particular, transverse polarization of $\Lambda$ hyperons in $e^+e^-$ annihilation has been observed by the Belle Collaboration~\cite{Belle:2018ttu} and the BESIII Collaboration~\cite{BESIII:2018cnd,BESIII:2019nep,BESIII:2021cvv,BESIII:2022qax,BESIII:2023euh}. The widespread appearance of $\Lambda$ polarization across different high-energy processes makes it an important window for understanding both the production mechanism of strange hadrons and the origin of its polarization.

\begin{figure}[h]
\centering
\includegraphics[width=0.5\linewidth]{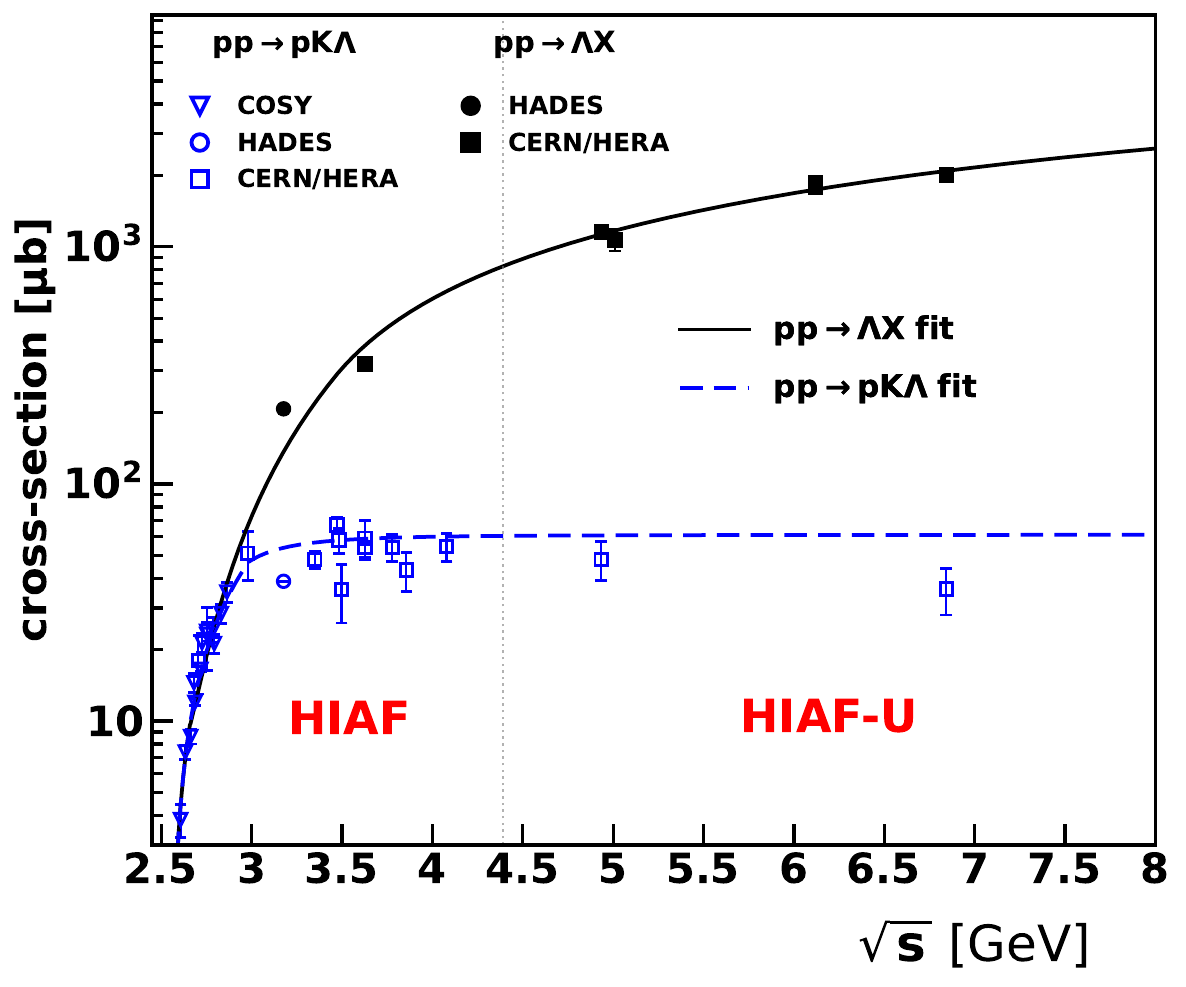}
\caption{\label{fig:cross_section}Cross sections for exclusive and inclusive $\Lambda$ production in $pp$ collisions. Blue open symbols represent the exclusive process $pp \rightarrow pK^+\Lambda$, with data taken from Refs.~\cite{Flaminio:1979ja,Balewski:1997ah,Balewski:1998pd,Sewerin:1998ky,Kowina:2004kr,Bilger:1998jf,COSY-TOF:2006tie,TOF:2010ygk,COSY-TOF:2010svd,Valdau:2007re,Valdau:2010kw,HADES:2016pau}; the blue dashed curve shows a fit to these data points. Black filled symbols represent the inclusive process $pp \rightarrow \Lambda X$, with data from Refs.~\cite{Flaminio:1979ja,HADES:2016pau}; the black solid curve corresponds to a fit to these points.}
\end{figure}

The High Intensity heavy-ion Accelerator Facility (HIAF)~\cite{Yang:2013yeb}, hosted by the Institute of Modern Physics, will provide a next-generation platform to investigate hyperon polarization. 
HIAF will deliver high-intensity beams of protons and heavy ions onto fixed targets with kinetic energies up to 9.3 GeV for proton beam, and an upgraded stage (HIAF‐U) will extend this to 32 GeV. 
This corresponds to nucleon-nucleon center-of-mass energies from just above the $\Lambda$ production threshold ($\sqrt{s}\approx 2.6\;\text{GeV}$ for $pp \to pK^+\Lambda$) up to about $\sqrt{s} \approx 8\;\text{GeV}$. 
The HIAF energy range thus bridges the low-energy threshold region to the higher ``intermediate'' energies, creating ideal conditions for systematic studies of $\Lambda$ production and polarization \cite{Zhou:2022pxl,Chen:2025eeb}.
It is important to distinguish two regimes of nucleon-nucleon interactions relevant to hyperon production:

\begin{enumerate}[leftmargin=0.5cm]
\item Near-Threshold Hadronic Regime: As the collision energy approaches the threshold for $\Lambda$ production, the reaction dynamics are dominated not by perturbative quark-gluon processes but by hadronic degrees of freedom and resonance excitations. 
In this low-energy regime, specific exclusive channels like $pp \to pK^+\Lambda$ become increasingly important, as shown in Fig.~\ref{fig:cross_section}. 
The production of a $\Lambda$ here can be described by hadronic interactions (effective field theories at the nucleon level) rather than partonic scatterings. 
For example, theoretical models based on meson exchange model~\cite{Machleidt:1987hj,Holzenkamp:1989tq} have had success in explaining the cross-section for $pp \to pK^+\Lambda$ near threshold~\cite{Laget:1991jk,Shyam:1999nm,Gasparian:1999jj,Laget:2000gq,Sibirtsev:2005mv,Liu:2005pm}.  
Recent experiments at facilities like DISTO and COSY have indeed observed significant $\Lambda$ polarization in $pp \to pK^+\Lambda$ process~\cite{Choi:1998st,COSY-TOF:2016vhv}, further validating the presence of hyperon polarization even in the threshold energy domain.

\item High-Energy QCD-Dominated Regime: At center-of-mass energies far above the $\Lambda$ threshold, hyperon production is governed by quark-gluon (partonic) interactions as described by perturbative or high-energy QCD~\cite{Gross:2022hyw}. 
In this regime, inclusive $\Lambda$ production can reveal information about the quark and gluon distributions inside nucleons~\cite{Deur:2018roz,Boussarie:2023izj}.
\end{enumerate}

Moreover, at lower collision energies, the production of additional particles is limited by kinematics, meaning event final states are relatively simple (low multiplicity). 
This makes it easier for produced $\Lambda$ hyperons to carry a large fraction of the available momentum (high $x_{\rm F}$). 
The advantage at HIAF is that high-$x_{\rm F}$ $\Lambda$ production will be abundant, and as noted, these hyperons are expected to exhibit large polarization. 
Therefore, experiments at HIAF can exploit this feature---we anticipate strong $\Lambda$ polarization signals with high statistics, which will facilitate precision measurements of the phenomenon.
The phase space coverage at HIAF in terms of $p_{\rm T}$ VS $x_{\rm F}$ is shown in Fig. \ref{fig:HIAF-phasespace}.

\begin{figure}[h]
\centering
\includegraphics[width=0.48\linewidth]{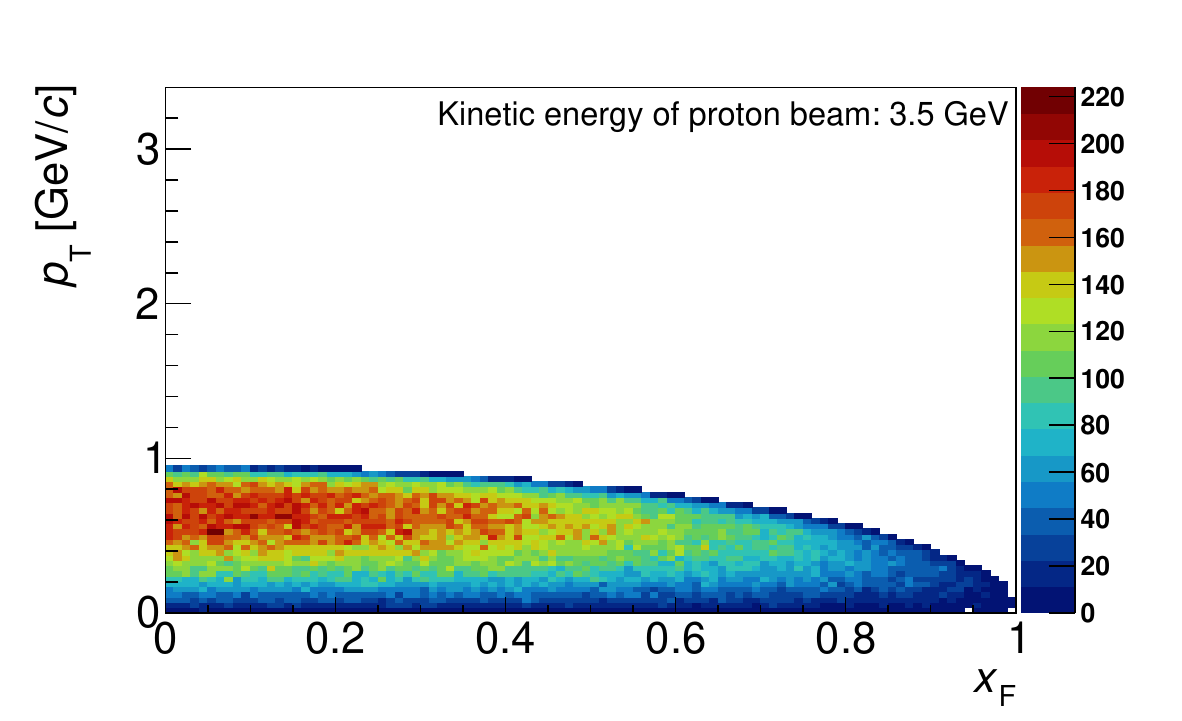} 
\includegraphics[width=0.48\linewidth]{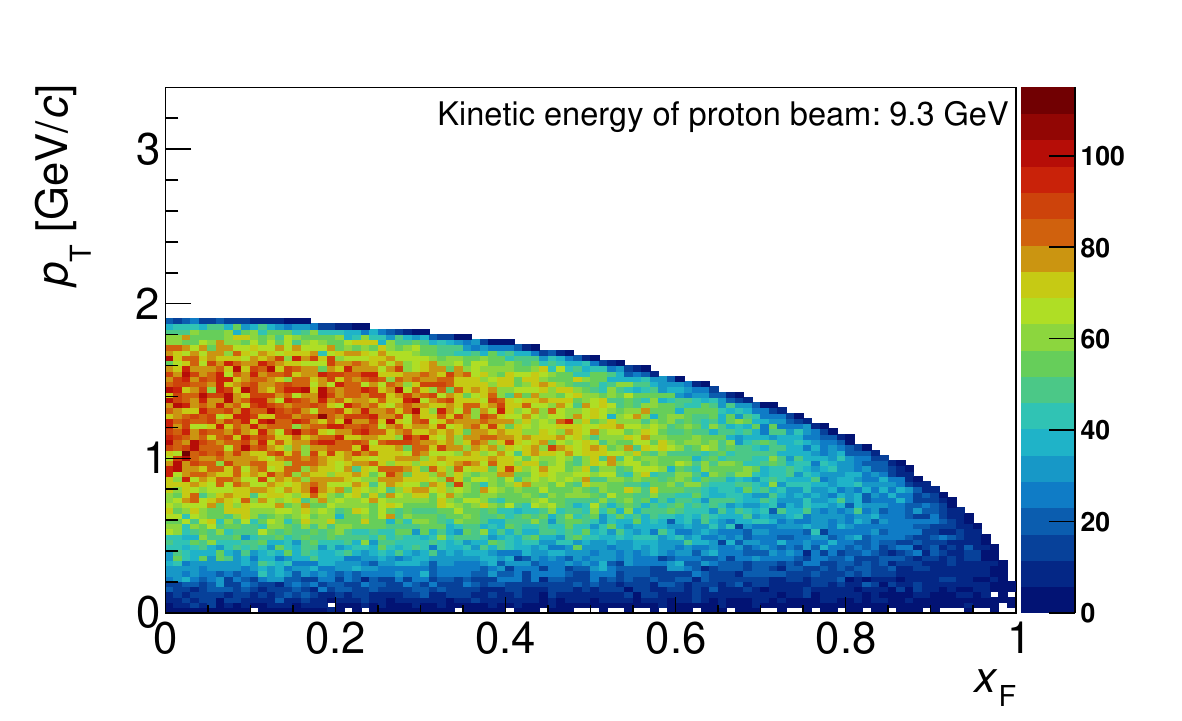}
\includegraphics[width=0.48\linewidth]{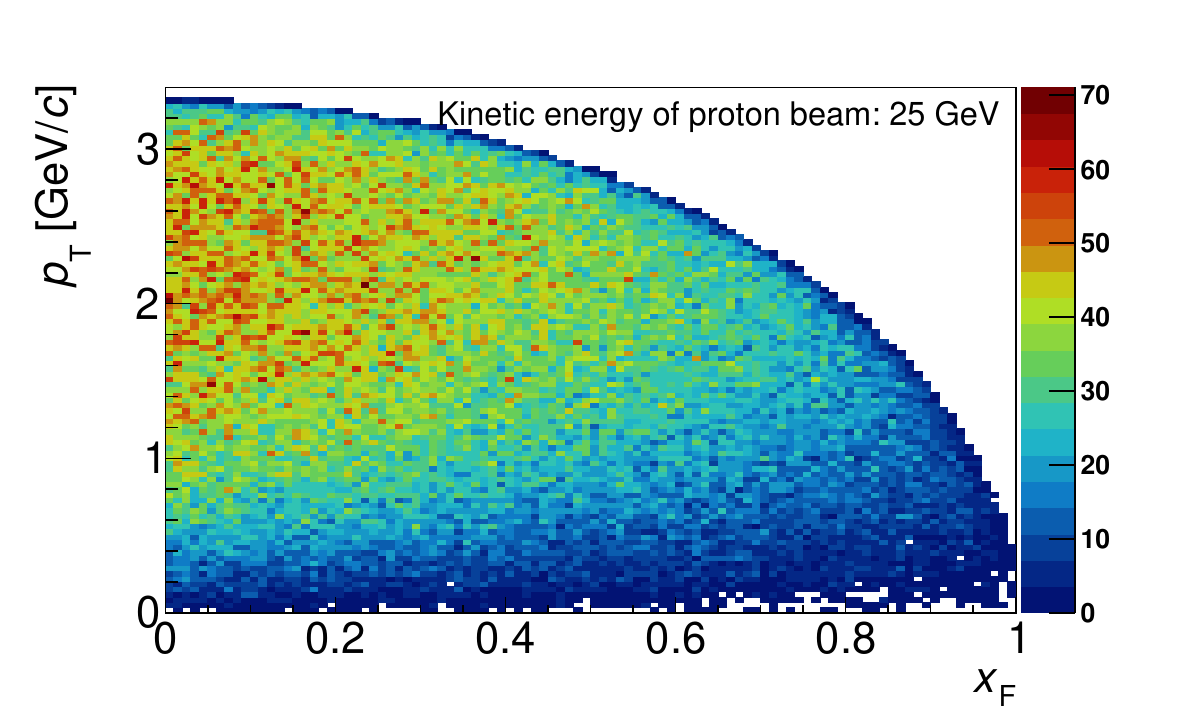}
\caption{ \label{fig:HIAF-phasespace} Phase space coverage in the center-of-mass-energy frame at HIAF for three beam energy configurations. In principle, $x_{\rm F}$ is a value from -1 to 1 in the center-of-mass-energy frame, here we show the distribution of its absolute value.  }
\end{figure}

\textbf{The proposed Hyperon-Nucleon Spectrometer (H-NS) at HIAF is designed to investigate the polarization of $\Lambda$ hyperons and protons--both spin-1/2 baryons--in $pp$, $pA$ and $AA$ collisions, with a particular emphasis on the high $x_{\rm F}$ region.} 
%The results will shed light on the origin of proton spin as well as the origin of visible matter in our universe. 
The H-NS program establishes its uniqueness through a comprehensive, multi-beam, energy-scanning approach to spin dynamics in non-perturbative QCD. 
While complementary facilities target specific regimes and physics--such as NICA at JINR focusing on moderate-energy $AA$ collisions to search for possible signs of the mixed phase and critical endpoint~\cite{KEKELIDZE2013945c,MPD:2022qhn} with potential upgrade for nucleon spin structure study, J-PARC utilizing high-intensity primary proton beams for hyperon spectroscopy and hadron structure~\cite{Aoki:2021cqa}, and the CBM experiment at FAIR aiming at discovering fundamental properties of QCD matter: the phase structure at large baryon-chemical potentials, effects of chiral symmetry, and the equation-of-state~\cite{CBM:2016kpk}. 
H-NS's defining strength is the systematic fixed-target study of polarization phenomena across $pp$, $pA$, and $AA$ collisions with a beam energy scan from 3 to 32 GeV, a range that bridges the transition from maximal polarization signals observed at lower energies to the high-density regime. 
This allows H-NS to disentangle the fundamental hyperon polarization mechanism in elementary $pp$ reactions from the subsequent modification by cold and hot nuclear matter effects in $pA$ and $AA$ collisions, all within a single experimental setup. 
By performing high-precision, simultaneous measurements of hyperon and nucleon spin observables over this broad energy landscape, H-NS will provide unprecedented, consistent data to constrain the dynamics of spin phenomena in strongly interacting systems that is inaccessible to any existing or currently approved facility.

In the following sections, we will have further brief discussions on relevant physics issues.

%\section{Baryon spectrum and structure}
\section{Associated strangeness production and hyperon polarization}
% Exclusive production of $\Lambda$

Hyperon polarization has been shown to be a very sensitive probe of the nucleon excitation spectrum in exclusive photon-nucleon, electron-nucleon and pion-nucleon collisions \cite{Cao:2013psa,Eichmann:2016yit,Ireland:2019uwn,Thiel:2022xtb,Doring:2025sgb,Burkert:2025coj}, providing insights into the internal structure of the nucleon~\cite{Zou:2005xy,An:2005cj,an:2006zf,An:2008xk,Zou:2009zz,Liu:2015ktc}.
The $pp \rightarrow pK^+\Lambda$ is a key channel for exploring the production mechanisms and polarization phenomena of strange quarks in nucleon-nucleon collisions via partial wave analysis, particularly in the energy regime of several GeV \cite{Laget:1991jk,Shyam:1999nm,Gasparian:1999jj,Laget:2000gq,Sibirtsev:2005mv,Liu:2005pm}. 
At this energy scale, it is generally understood that quark and gluon degrees of freedom inside nucleons are ``frozen'', and nucleons as a whole participate in the interaction \cite{Gross:2022hyw}. 
Phenomenological models based on meson exchange interactions~\cite{Machleidt:1987hj,Holzenkamp:1989tq} have achieved significant success in explaining the production cross-sections, deepening the understanding of the coupling of nucleon resonances to associate strangeness channels~\cite{Laget:1991jk,Shyam:1999nm,Gasparian:1999jj,Laget:2000gq,Sibirtsev:2005mv,Liu:2005pm}.
Since the final states $K^+\Lambda$ from nucleon resonances have large $s\bar{s}$ components, this reaction has the potential to investigate those nucleon resonances.
The high energy proton beam can also explore nucleon resonances with mass larger than 2.0 GeV through $pp \to p N^*$, $p \Delta^*$, $N^* N^*$ and $\Delta^* N^*$ processes \cite{Cao:2010km,Cao:2014mea,Cao:2010jj,Lenske:2018bgr}.
Such $N^*$ resonances are also a major focus of partial wave analysis in the BESIII experiment at the Beijing Electron Positron Collider II (BEPC II) \cite{BES:2004gwe,BES:2009ufh,BESIII:2012ssm,BESIII:2024vqu} and the future Super $\tau$-Charm facility (STCF) \cite{Achasov:2023gey}. The physics programs of these facilities in this field are thus mutually complementary.

\begin{figure}[h]
\centering
\subfloat[]{
\includegraphics[width=0.5\linewidth]{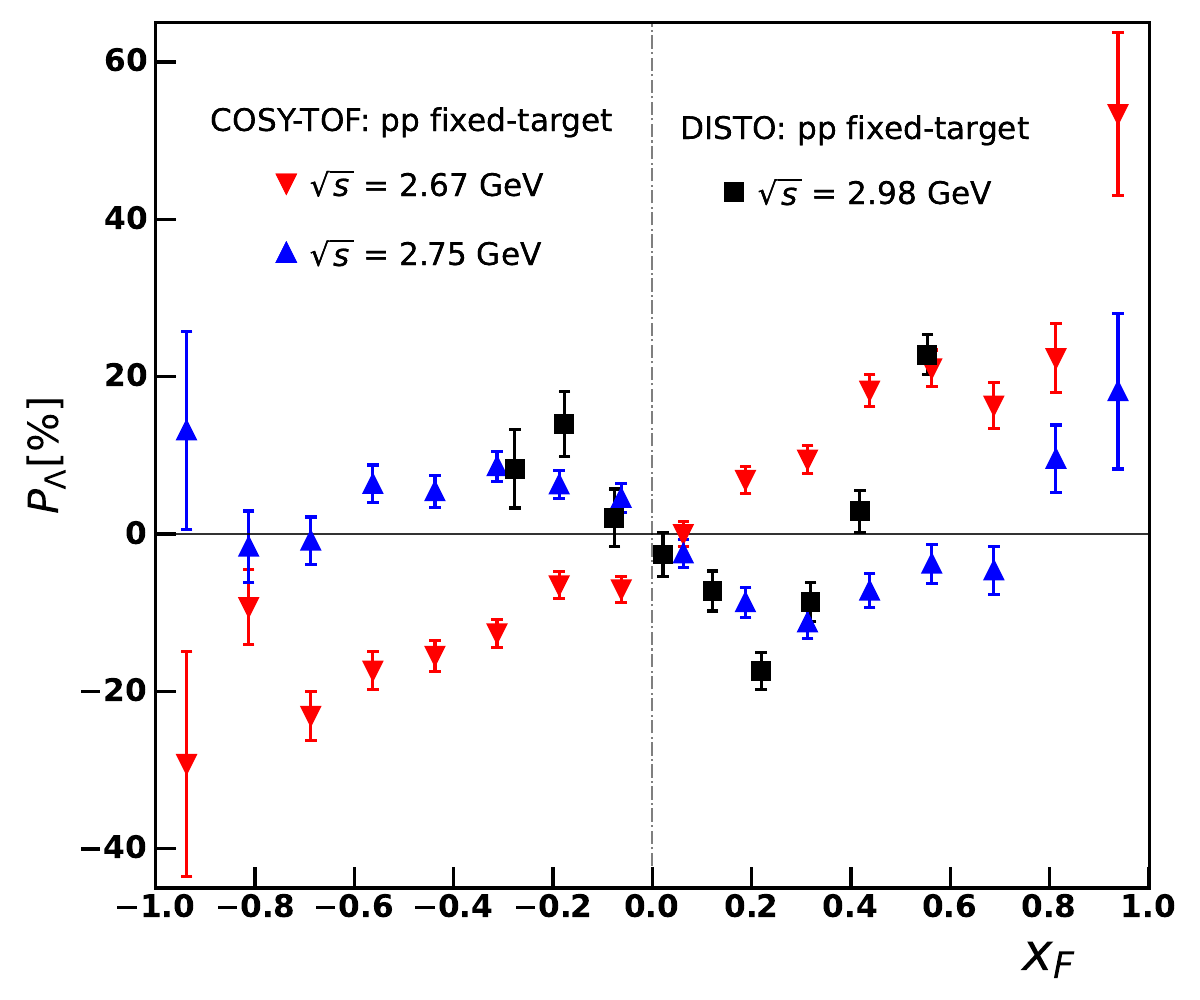} }
\centering
\subfloat[]{
\includegraphics[width=0.5\linewidth]{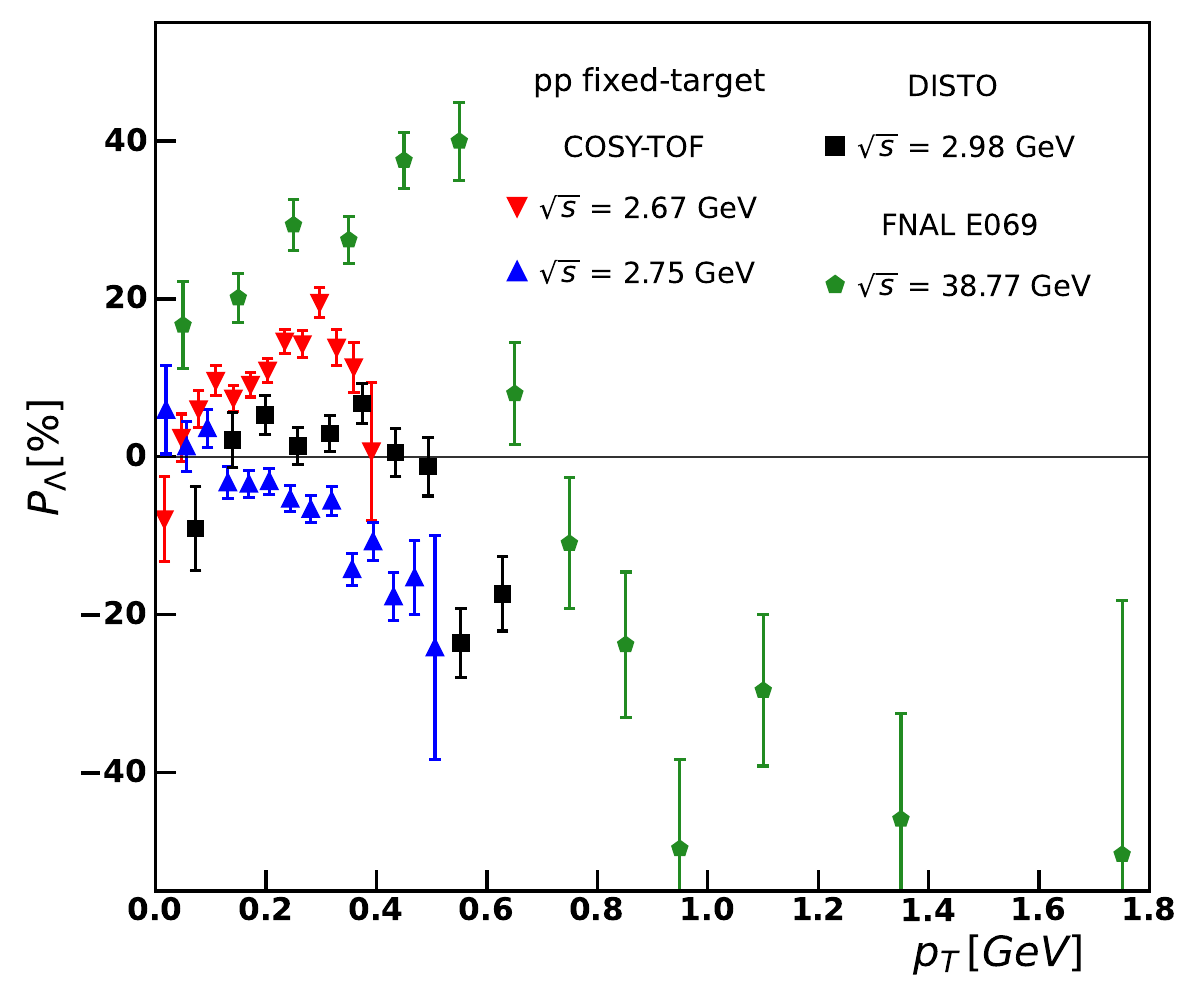}}
\caption{ \label{fig:sensitive}$\Lambda$ polarization in the $pp \rightarrow pK^{+}\Lambda$ process. (a) $\Lambda$ polarization as a function of $x_{\rm F}$, with data taken from Refs.~\cite{Choi:1998st,COSY-TOF:2016vhv}; (b) $\Lambda$ polarization as a function of $p_{\rm T}$, with data taken from Refs.~\cite{E690:2001otd,Choi:1998st,COSY-TOF:2016vhv}.  A reversal in the sign of polarization is observed with varying center-of-mass energy in (a), even across energy differences of just a few hundred MeV. In (b), the polarization not only varies with center-of-mass energy but also shows a sign flip as $p_{\rm T}$ increases.These trends highlight the complex production dynamics of $\Lambda$ hyperons in associated strangeness production.
}
\end{figure}

Recently, COSY, DISTO, and HADES collaborations have provided more detailed measurements of the differential cross-sections~\cite{Balewski:1997ah, Balewski:1998pd, Sewerin:1998ky, Kowina:2004kr, Bilger:1998jf, COSY-TOF:2006tie, TOF:2010ygk, COSY-TOF:2010svd, COSY-TOF:2015lcz, COSY-TOF:2016qxd, Valdau:2007re, Valdau:2010kw, HADES:2016pau} and polarization~\cite{Choi:1998st,COSY-TOF:2016vhv} of the produced $\Lambda$ hyperons. 
%Unlike the scaling behavior observed in inclusive Lambda production processes, 
The polarization of $\Lambda$ hyperons produced in this specific process shows a pronounced energy-dependent behavior, as shown in Fig.~\ref{fig:sensitive}. 
%Recent data indicate that in certain energy ranges, 
The $\Lambda$ hyperon polarization close-to-threshold undergoes a reversal, most noticeable in the relatively narrow energy range of a few hundred MeV. 
This phenomenon highlights the complicated production mechanisms and polarization origin involved, and would be used to extract the accurate properties of nucleon excitation in the second resonance region and hyperon-nucleon scattering through final state interaction.

On the other hand, the $\Sigma$ hyperon production is another golden channel for studying $\Delta$ resonances.
For instance, the final state $n K^+\Sigma^+$ couples exclusively to isospin-3/2 resonances~\cite{Sibirtsev:1998jf,Sibirtsev:2007sk,Xie:2007vs,Cao:2007md,Wang:2014ofa}.
New experimental data on this reaction would therefore offer detailed insights and help obtain the spectrum of $\Delta$ resonances with large $ s\bar{s} $ components.

However, the measurements of $\Lambda$ hyperon polarization in the $pp \rightarrow pK^+\Lambda$ process are currently limited to only a few discrete energy points.
The situation is even more problematic for the $pp \rightarrow n K^+\Sigma^+$ reaction, where recent experimental results of the production cross section differ by several orders of magnitude~\cite{Flaminio:1979ja,Rozek:2006ct, Valdau:2007re,COSY-HIRES:2010rot}.
Moreover, the polarization of $\Sigma$ hyperon in this reaction has never been measured.

To fully understand the underlying production mechanisms and the spontaneous polarization of hyperons in this energy range, it is essential to conduct a detailed energy scan that covers the full energy spectrum, particularly starting from the production threshold. 
Equally important are complete measurements of the four-momentum distributions of all final-state particles over the full 4$\pi$ solid angle, including Dalitz plots, various invariant mass spectra, and angular distributions. 
Such observables are crucial for disentangling the reaction mechanism, and when combined with polarization data, they can precisely reconstruct the reaction amplitude.

\section{Hyperon-nucleon interactions}

The hyperon-nucleon interaction is of wide interest and plays an important role in nuclear physics, providing essential inputs into calculations of a variety of nuclear physics phenomena, from the structure of hypernuclei to the properties of neutron stars \cite{Gal:2016boi,Tolos:2020aln,Lonardoni:2014bwa, Le:2024rkd}. 
There have been recent measurements on the $\Lambda N$ and $\Sigma N$ interactions by the CLAS experiment~\cite{CLAS:2021gur} and the E40 experiment at the Japan Proton Accelerator Research Complex (J-PARC)~\cite{J-PARCE40:2021qxa, J-PARCE40:2021bgw, J-PARCE40:2022nvq} (for a review, see Ref.~\cite{Miwa:2025adw}).
There have also been measurements at the BESIII experiment at the BEPC II~\cite{BESIII:2023trh,BESIII:2024geh,BESIII:2025eet} using the method proposed in Refs.~\cite{Yuan:2021yks,Dai:2024myk}. 
However, the experimental information is still very limited, leaving parameters in theoretical frameworks for hyperon-nucleon interactions poorly constrained~\cite{Haidenbauer:2023qhf}.
Investigations of processes with one or more $\Lambda$ hyperons in the final state provide good opportunities to constrain the hyperon-nucleon interaction through final-state interactions (FSI).

It has been proposed in Refs.~\cite{Gasparyan:2003cc, Gasparyan:2005fk} that the $S$-wave $\Lambda N$ scattering length can be extracted from the $pp \rightarrow pK^{+}\Lambda$ with a systematic uncertainty of 0.3~fm or even smaller.
If the energy of the incoming proton is high enough such that the kaon quickly flies away from the $\Lambda p$ pair,  kinematical regions can be selected such that the $\Lambda p$ distribution receives little impact from resonances that can decay into $K^+p$ or $K^+\Lambda$. 
In that case, one can focus on the $\Lambda N$ final state interaction~\cite{Sibirtsev:2005mv}. 
The method is based on the Omn\`es dispersive representation for two-body final state interaction in the elastic regime, and thus can only be applied for a single partial wave. 
Polarization information is required to separate the two possible spin (0 or 1) states of $\Lambda N$~\cite{Gasparyan:2003cc}.

One can also make use of the threshold cusp effect to extract $S$-wave two-body scattering lengths~\cite{Budini:1961bac, Cabibbo:2004gq, Meissner:1997fa} (for a review, see Ref.~\cite{Guo:2019twa}).
To have the threshold cusp visible, one needs to study a coupled-channel system such that the behavior of the line shape around the higher threshold is seen in the energy distribution of the lighter channel.
As for the $\Lambda N$, the threshold of the next coupled channel is $\Sigma N$, which is more than 70~MeV above. 
Such a separation necessitates the inclusion of pion exchanges in a full theoretical description of the coupled-channel system, which complicates the analysis.
Nevertheless, two remarkable structures have been indeed observed in the $p\Lambda$ invariant mass distribution as shown in Fig. \ref{fig:pLinv} (a).
The first one is the threshold enhancement in the $p\Lambda$ invariant masses, arising from the $p\Lambda$ FSI.
The second structure stretched in the vertical direction in Fig. \ref{fig:pLinv} is located at the $N\Sigma$ threshold.
After reasonably incorporating the contribution of nucleon resonances, the $p\Lambda$-$p\Lambda$ and $p\Lambda$-$N\Sigma$ interactions could be subsequently extracted \cite{COSY-TOF:2016qxd,Liu:2005pm,Sibirtsev:2005mv}.

\begin{figure}[h]
\centering
\subfloat[]{
\includegraphics[width=0.45\linewidth]{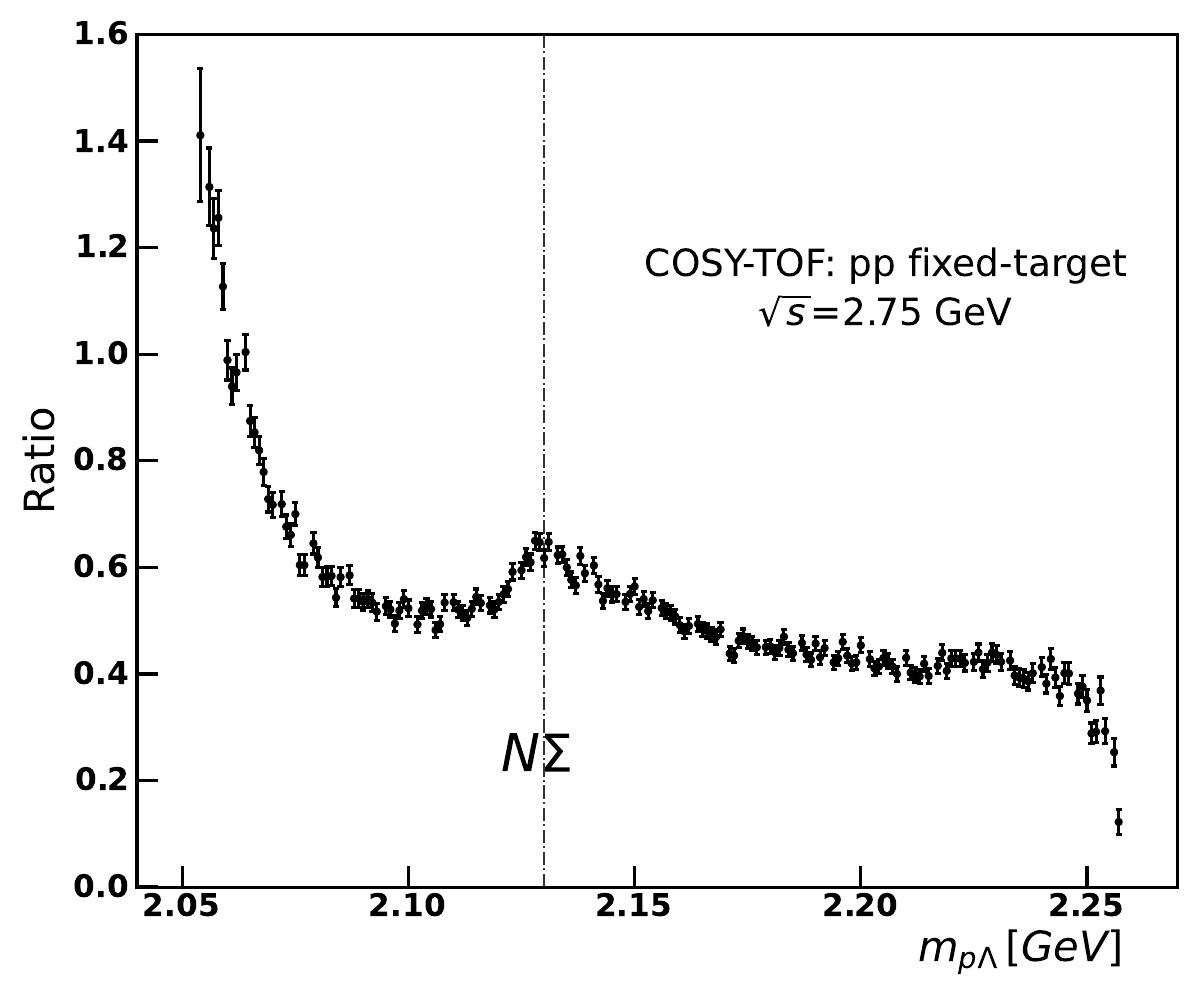}}
\centering
\subfloat[]{
\includegraphics[width=0.45\linewidth]{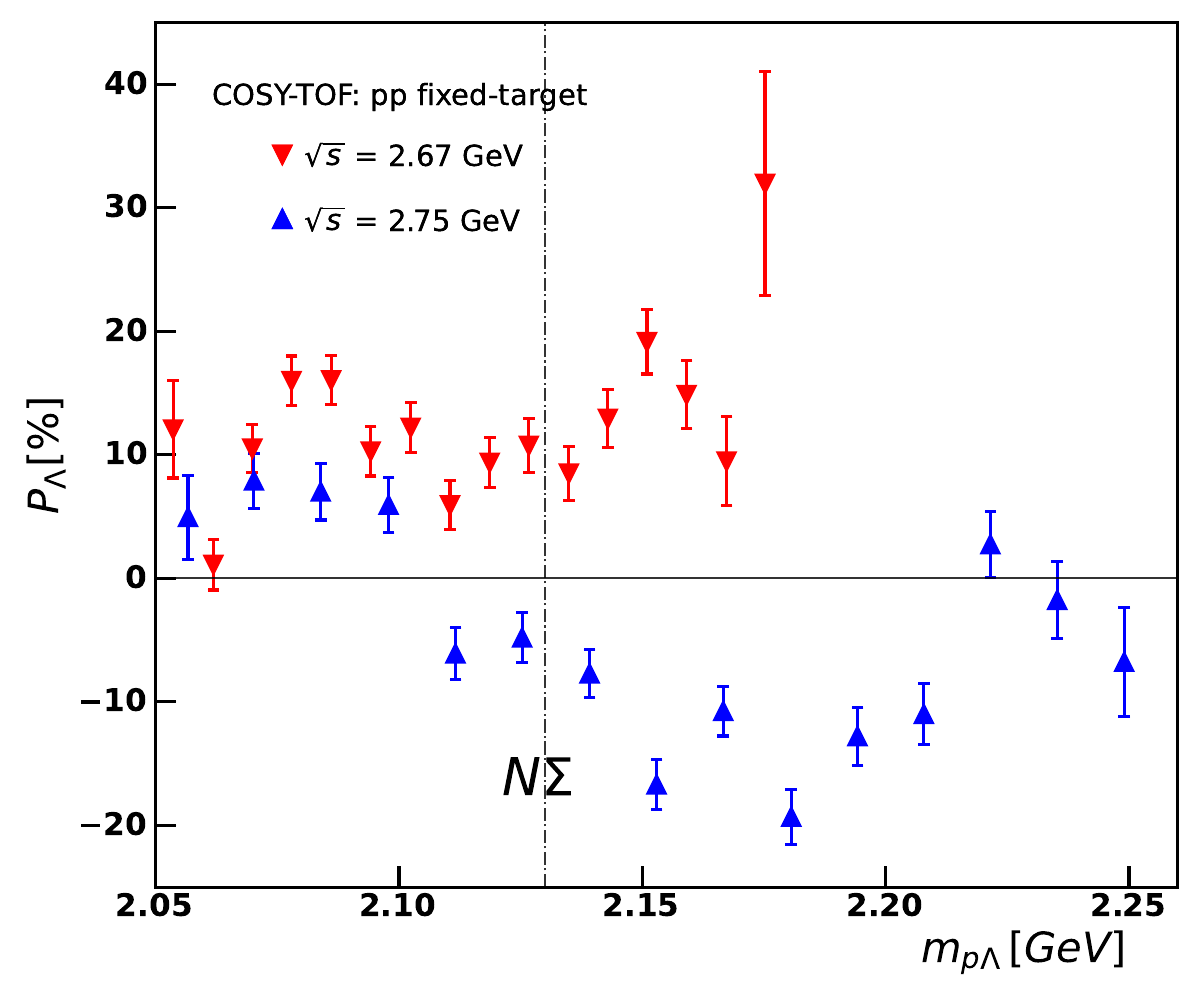}}
\caption{ \label{fig:invariant_mass}  (a) The $p\Lambda$ invariant mass distribution $m_{p\Lambda}$ in the $pp \rightarrow pK\Lambda$ reaction after dividing by the Monte Carlo phase-space simulation~\cite{COSY-TOF:2015lcz}.  (b) The $\Lambda$ polarization as a function of $m_{p\Lambda}$ at $\sqrt{s} = 2.67$ GeV (red triangles down) and $2.75$ GeV (blue triangles up)~\cite{COSY-TOF:2016vhv}. The $N\Sigma$ threshold is marked with the vertical black line.} \label{fig:pLinv}
\end{figure}

The measurement of the beam analyzing power through the asymmetry of the produced kaon using a polarized proton beam at COSY enables the extraction of the spin-triplet scattering length of $p\Lambda$~\cite{COSY-TOF:2016qxd}.
Significant $\Lambda$ polarization is observed near the $p\Lambda$ threshold, as shown in Fig. \ref{fig:pLinv} (b), but it has not yet been used to investigate $p\Lambda$ FSI.
The present sensitivity and resolution are insufficient to draw any definite conclusions about whether there is a cusp structure at the $N\Sigma$ threshold, which deserves future experimental investigation.

Moreover, this approach could yield valuable insights into strangeness $S = -2$ baryon-baryon systems, particularly regarding the coupling between $\Lambda\Lambda$ and $N\Xi$ channels through reactions such as $pp \to K^+K^+ \Lambda\Lambda$ and $pp \to K^+K^+ N\Xi$. 
The $\Lambda\Lambda$ and $N\Xi$ thresholds are located at approximately 2231\,MeV and 2260\,MeV, respectively, where the $\Xi N$ threshold further splits into $n\Xi^0$ and $p\Xi^-$ channels with a separation of roughly 5\,MeV. 
It is worth noting that theoretical investigations of the $\Lambda\Lambda$-$\Xi N$ coupled-channel systems predict a prominent cusp structure~\cite{Fujiwara:2006yh,Haidenbauer:2015zqb}.
It is important to note that a comprehensive analysis necessitates high-precision measurements of both $\Lambda\Lambda$ and $\Xi N$ final states. 
The self-analyzing polarization properties of hyperon decays or a polarized target may be used to separate the spin-triplet and spin-singlet contributions.

\section{Energy dependence of hyperon polarization}\label{Physics}

In early measurements of hyperon polarization, such as those performed by the R608 collaboration at CERN~\cite{R608:1986ltk}, the polarization for inclusive $\Lambda$ production was studied in proton-proton collisions as a function of center-of-mass energy ($\sqrt{s}$) with fixed Feynman $x_{\rm F}$ and transverse momentum ($p_{\rm T}$). 
As shown in Fig.~\ref{fig:scaling}, the data suggested that 
$\Lambda$ hyperon polarization exhibits a scaling behavior, appearing to be independent of the center-of-mass energy. However, this observation appears to contradict expectations from QCD. In the high-energy region, 
based on the QCD factorization, the transverse polarization of $\Lambda$ can not arise at twist-2 level or in leading power approximation because of the helicity or chirality conservation. It can arise at twist-3 level~\cite{Kanazawa:2000cx}. Hence, the polarized part of the cross-section is power suppressed.
%the $\Lambda$ polarization can arise from a twist-three fragmentation function, which contributes to the cross section as a power-suppressed term. 
Consequently, the 
$\Lambda$ polarization is expected to decrease with increasing center-of-mass energy. This creates a clear tension between the existing experimental data and the theoretical prediction that warrants further investigation.

The initial measurements were limited by both the precision of the data and the relatively narrow range of center-of-mass energies explored. Furthermore, the lower-energy region remains essentially uncharted in experimental studies, where non-perturbative effects are expected to play a significant role, making predictions for the energy dependence even more challenging.
It would be interesting to resolve the apparent conflict between existing measurements and QCD predictions by examining the ``scaling'' behavior across a broad energy range and elucidating the fundamental physics principles underlying this phenomenon.

\begin{figure}[h]
\centering
\includegraphics[width=0.5\linewidth]{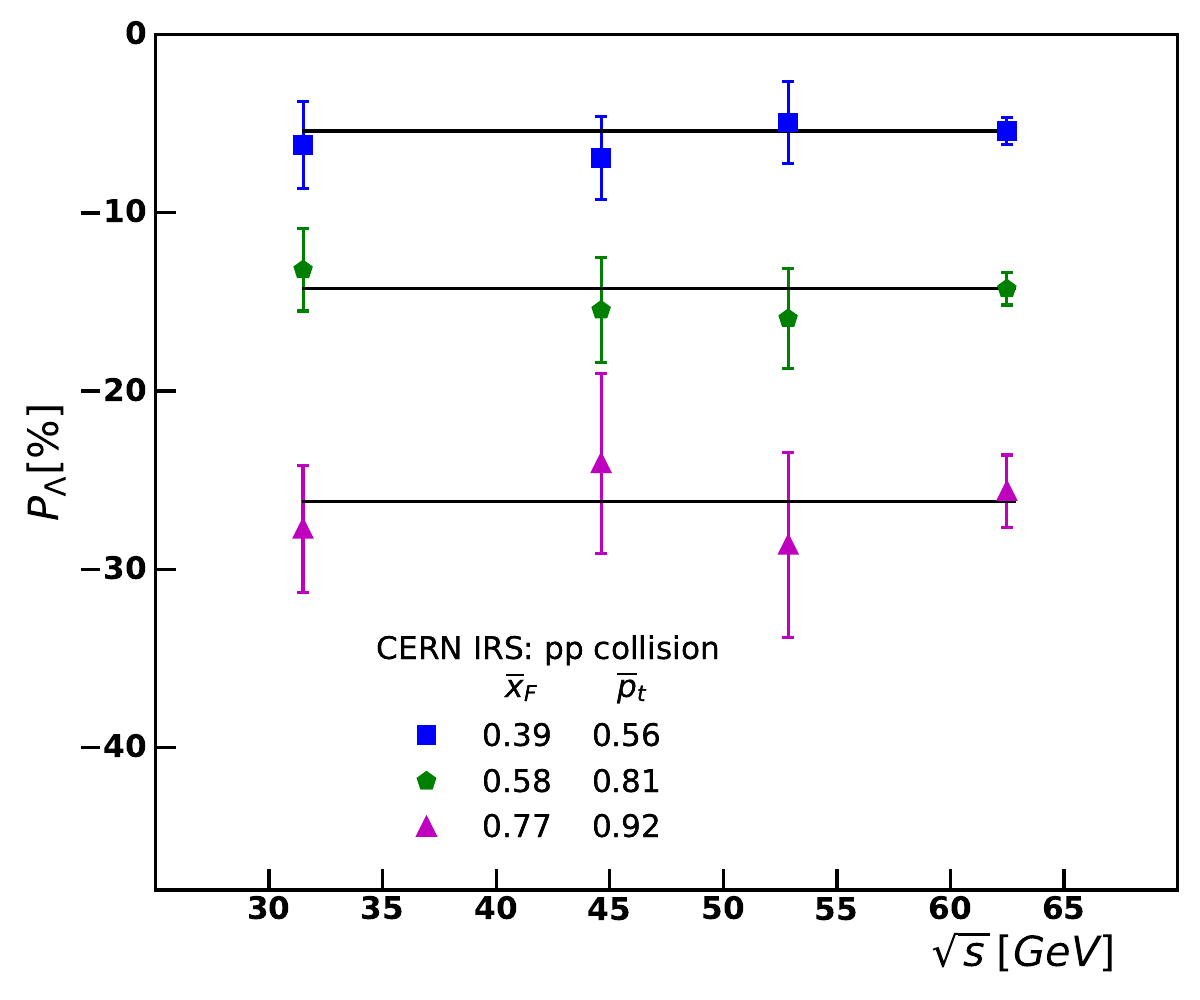}
\caption{\label{fig:scaling}Polarization of $\Lambda$ as function of $\sqrt{s}$ at fixed $p_{\rm T}$  and $x_{\rm F}$ in $pp\rightarrow \Lambda X$ process~\cite{R608:1986ltk}.}
\end{figure}

\section{Emergence of baryon spin from QCD dynamics}

A primary motivation for the study of hyperon polarization is its unique sensitivity to the non-perturbative spin dynamics of QCD hadronization. The field emerged from the seminal 1976 observation that $\Lambda$ hyperons produced in unpolarized $p+Be$ collisions exhibited significant transverse polarization (up to 20--30\%)~\cite{Bunce:1976yb}. This result presented a profound puzzle, as leading-order perturbative QCD calculations predicted negligible spin effects due to the chiral nature of the interaction~\cite{Kane:1978nd}. The observation of such large polarization implies that the spin alignment is generated not in the hard scattering, but through spin-momentum correlations during the fragmentation process.

In the framework of transverse momentum dependent (TMD) factorization, which is valid for processes sensitive to small transverse momenta ($p_{\rm T} \ll Q$), this mechanism is encoded in the polarizing fragmentation function (PFF), denoted as $D_{\rm 1T}^{\perp, h/q}(z, \bm{p}_{\rm T}^2)$. This function describes the probability for an unpolarized quark ($q$) to fragment into a transversely polarized hyperon with momentum fraction $z$ and transverse momentum $\bm{p}_{\rm T}$ relative to the quark~\cite{Anselmino:2000vs}. The PFF is a chiral-odd and naive time-reversal-odd (T-odd) function, making it a fundamental probe of the hadronization mechanism.

Complementary to the TMD framework, the collinear twist-3 factorization formalism is essential for describing spin asymmetries in single-inclusive processes where only one hard scale exists ($p_{\rm T} \sim Q$), such as inclusive proton-proton collisions ($pp \to \Lambda^\uparrow X$). In this framework, the polarization arises from quark-gluon-quark correlations encoded in the twist-3 fragmentation function, $D_{\rm T}(z)$~\cite{Koike:2017fxr}. Crucially, the two formalisms provide a unified picture: in the intermediate transverse momentum region, the twist-3 function is related to the $p_{\rm T}$-moment of the TMD polarizing fragmentation function~\cite{Gamberg:2018fwy}.

Therefore, a global analysis of $\Lambda$ polarization across complementary processes is essential. Such a program allows for the extraction of both TMD and twist-3 functions, testing the universality of spin-dependent fragmentation and mapping the 3D spin structure of hadrons in momentum space.

Beyond $\Lambda$ hyperons, the production and polarization of other hyperons, such as the $\Sigma$ and $\Xi$ hyperons, have also garnered significant experimental interest, as shown in Fig.~\ref{fig:sigma}. %For the production of hyperons in nucleon-nucleon collisions, 
Experimental studies have found that the polarization of $\Lambda$ and $\Xi$ hyperons tends to have the same sign, while the polarization of $\Sigma$ hyperons behaves oppositely. 

The naive quark model suggests that both $\Lambda$ and $\Sigma$ hyperons are composed of the same quark constituents—specifically, one strange quark ($s$) and two non-strange quarks ($u$ and $d$). However, the key difference between $\Lambda$ and $\Sigma$ hyperons lies in the quark configuration. $\Lambda$ hyperons and $\Sigma$ hyperons are both in a spin-1/2 state as well, but with a different symmetry due to their specific quark wavefunctions. This difference is reflected in the polarization directions of these particles, with $\Lambda$ and $\Sigma$ hyperons showing opposite polarization signs in reactions involving strange quark production. The study of this polarization difference between various hyperons, therefore, provides essential information about the internal quark dynamics and the spin distributions of quarks within hadrons.

\begin{figure}[h]
\centering
\subfloat[]{
\includegraphics[width=0.5\linewidth]{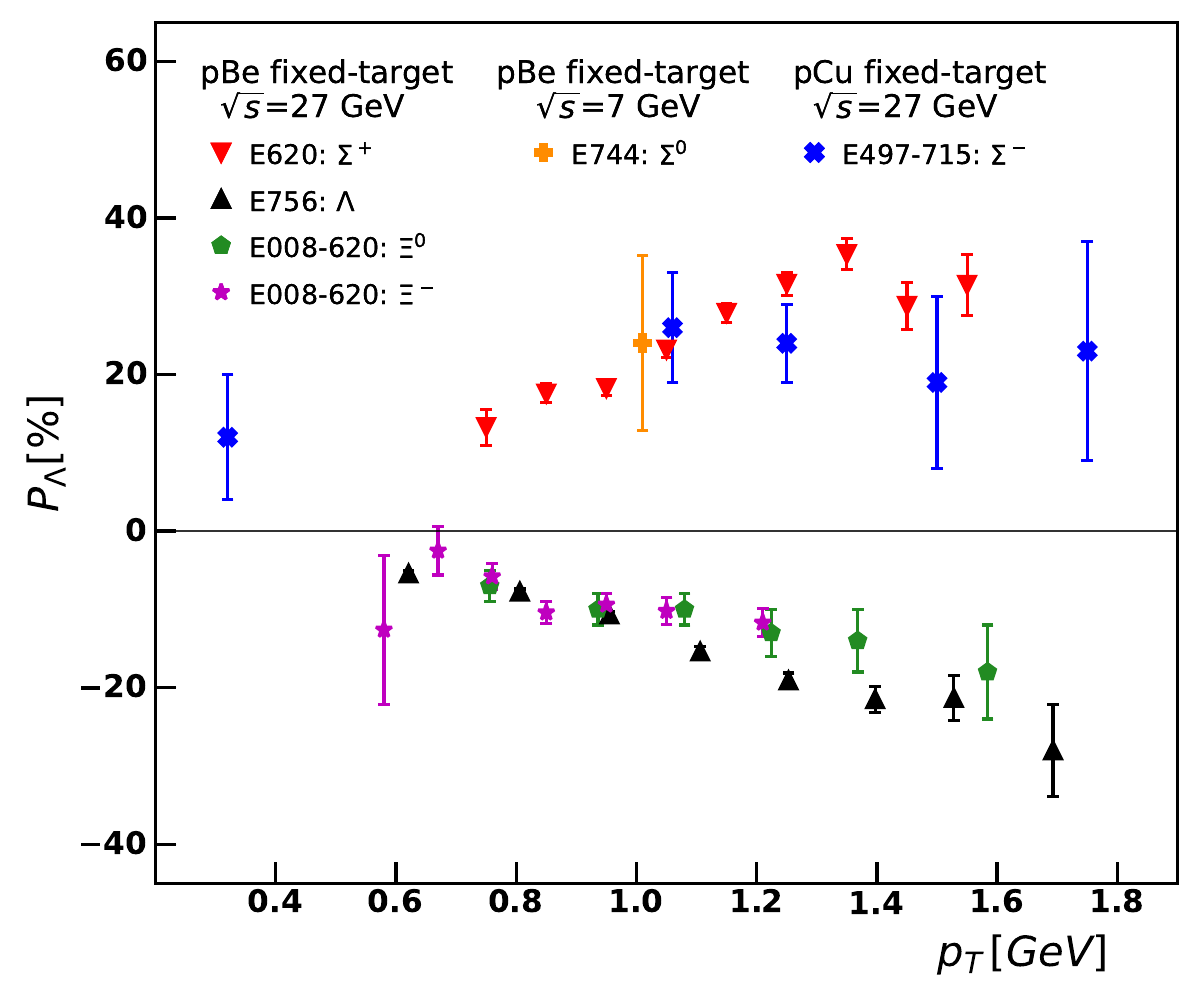} }
\centering
\subfloat[]{
\includegraphics[width=0.5\linewidth]{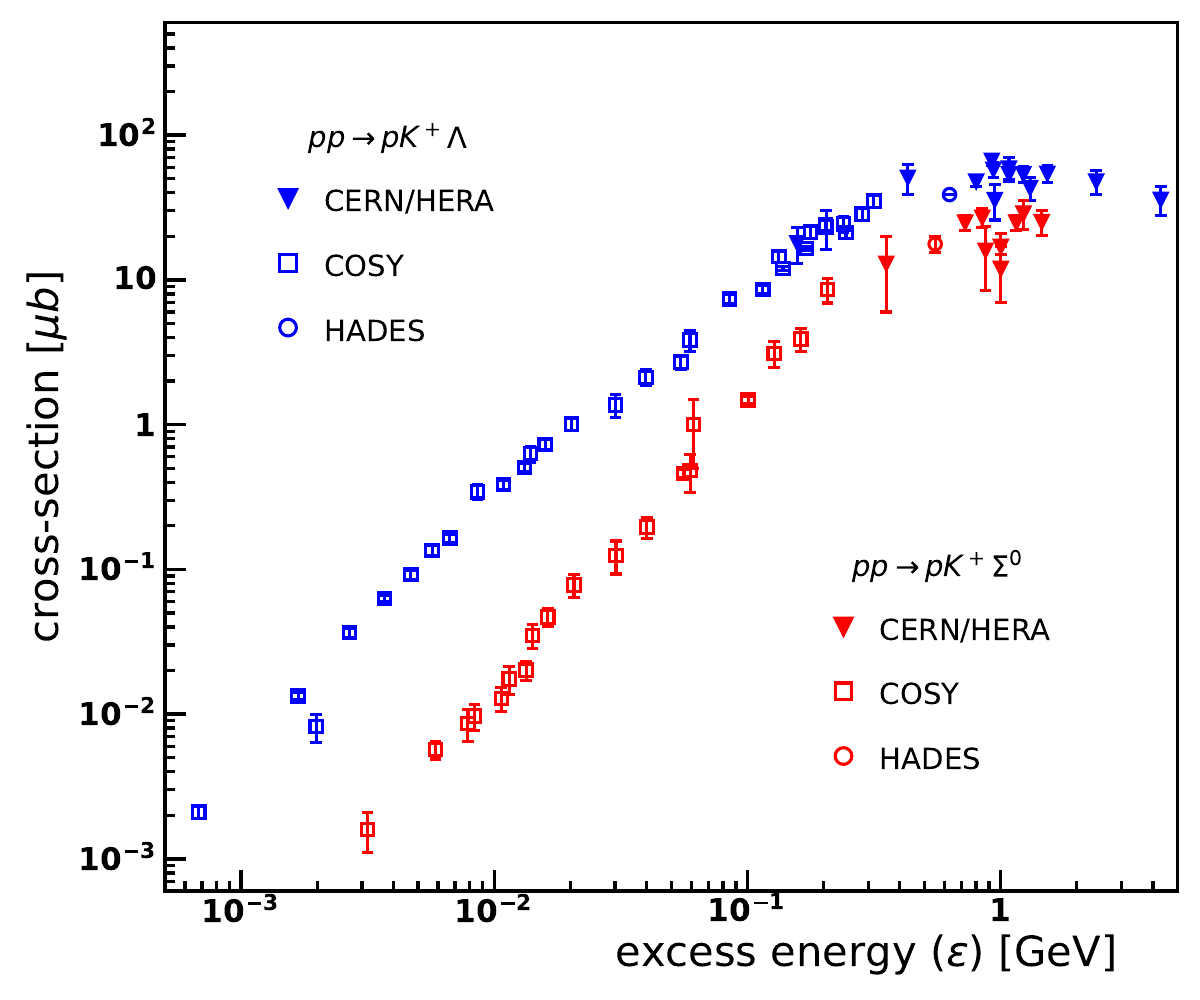}}
\caption{ \label{fig:sigma}Comparison of production cross sections for $\Lambda$ and $\Sigma$ and polarizations for different octuplet hyperons in $pp$ collisions. (a) Differences in the polarization of $\Lambda$~\cite{Lundberg:1989hw}, $\Sigma$~\cite{Wah:1984swb,Wilkinson:1987qi,Dukes:1987ys} and $\Xi$~\cite{Heller:1983ia,Rameika:1986rb} in the inclusive process $pp \rightarrow \Lambda / \Sigma/\Xi X$; (b) Differences in the production cross sections of $\Lambda$~\cite{Flaminio:1979ja,Balewski:1997ah,Balewski:1998pd,Sewerin:1998ky,Kowina:2004kr,Bilger:1998jf,COSY-TOF:2006tie,TOF:2010ygk,COSY-TOF:2010svd,Valdau:2007re,Valdau:2010kw,HADES:2016pau} and $\Sigma$~\cite{Flaminio:1979ja,Sewerin:1998ky,Kowina:2004kr,COSY-TOF:2010svd,HADES:2023dti} in the exclusive process $pp \rightarrow pK^+ \Lambda / \Sigma^0$.}
\end{figure}

\section{Parity violation, $CP$ violation and rare processes}

The study of fundamental discrete symmetries—Charge ($C$), Parity ($P$), and Time ($T$)—and their violation remains a cornerstone of high-energy physics. 

Within the Standard Model, parity violation (PV) in hadronic interactions arises from the exchange of weak gauge bosons ($W^{\pm}$, $Z^0$)~\cite{{Glashow:1961tr,Weinberg:1967tq,Salam:1968rm,Glashow:1970gm}}. The primary experimental signature is an asymmetry in the scattering cross-section for longitudinally polarized beams, defined as:
\begin{equation}
  A_{\rm LR} = \frac{\sigma^+ - \sigma^-}{\sigma^+ + \sigma^-},
\end{equation}
where $\sigma^+$ and $\sigma^-$ denote cross-sections for positive (right-handed) and negative (left-handed) beam helicities, respectively, or hyperon helicities in the final states, such as $p N \to \Lambda + X$. A non-zero $A_{\rm LR}$ would be a direct signal of PV.
In proton–proton and proton–nucleus collisions at center-of-mass energies of a few GeV, PV effects induced by $W^{\pm}$ and $Z^{0}$ exchange are strongly suppressed by the large $W/Z$ masses. As a result, PV observables in this energy range provide not only a clean probe of strangeness-conserving weak quark interactions~\cite{Haxton:2013aca,Eversheim:1991tg,Balzer:1980dn,Kistryn:1987tq,TRIUMFE497:2001hga}, enabling tests of both the traditional Desplanques, Donoghue, and Holstein (DDH) meson-exchange framework~\cite{Desplanques:1979hn} and modern effective field theory (EFT)  descriptions~\cite{deVries:2013fxa,Viviani:2014zha,deVries:2015gea,Phillips:2014kna,Schindler:2015nga,deVries:2014vqa}, but also a potentially sensitive window into physics beyond the Standard Model. These observables have therefore attracted broad theoretical and experimental interest~\cite{Bedaque:2002mn,Ginges:2003qt,Erler:2004cx,Ramsey-Musolf:2006vfz,Bernard:2006gx,Bernard:2007zu,Epelbaum:2008ga,Holstein:2010zza,Machleidt:2011zz,Schindler:2013yua,Gardner:2017xyl,Hammer:2019poc,deVries:2020iea,Davoudi:2020ngi}.

Within the Standard Model, $CP$ violation is governed by two main parameters: the complex phase in the Cabibbo-Kobayashi-Maskawa (CKM) matrix \cite{Cabibbo:1963yz,Kobayashi:1973fv}, which influences weak interactions, and the QCD angle $\bar{\theta}$~\cite{Jackiw:1976pf}, which relates to $CP$ violation in strong interactions. $CP$ violation has been established in the meson sector~\cite{{Christenson:1964fg,BaBar:2001pki,Belle:2001zzw,LHCb:2019hro}} and very recently in the baryonic decay $\Lambda_b \to pK^-\pi^+\pi^-$ \citep{LHCb:2025ray}. While experimental research on $CP$ violation has aligned with the predictions of the Standard Model, the observed degree of $CP$ violation is insufficient to fully account for the matter-antimatter asymmetry in the universe~\cite{Sakharov:1967dj,Canetti:2012zc}.

Rare production and decay processes of hadrons provide powerful and complementary probes to search for new sources of $CP$ violation and for physics beyond the Standard Model. 
Here a few rare processes are listed, including strangeness-changing processes with $|\Delta S =1|$, Lepton-Number-Violating processes ($0\nu \beta\beta$-like processes), rare and forbidden hyperon processes: 

\begin{itemize}

\item $|\Delta S =1|$ processes: $p p \to \Lambda p \pi^+$ ($p\pi^+$ could be from $\Delta^{++}$ decay), $pp \to \Sigma^+ p$, $pp \to pp K_S$ or $pp K^- \pi^+$. Processes involving parity-violating nucleon-nucleon interactions are currently the only experimental avenue for probing hadronic weak interactions that conserve strangeness~\cite{Desplanques:1979hn}.
Within the Standard Model, the contributions of flavor-changing charge currents in hadronic weak interactions involving changes in strangeness or charm are heavily suppressed and are rarely accessible experimentally. On the other hand, strangeness-changing hadronic weak processes can be severely limited by statistics and significantly obscured by purely strong-interaction backgrounds, making them challenging to isolate and study. Therefore, further experimental and theoretical investigations are critically needed. Anyway, observation of the  $|\Delta S =1|$ processes is a direct signature of hadronic weak interactions~\cite{Glashow:1961tr,Salam:1968rm};

\item  Lepton-Number-Violating processes:  $pp \to nn \mu^+\mu^+$, $N^*N^* \mu^+\mu^+$ ($N^* \to p \pi^-$), and $pp \to \Lambda \Lambda \mu^+ \mu^+$ ($0\nu \beta\beta$-like processes). Although proton-proton scattering experiments cannot reach the sensitivity of nuclear decays,  they provide access to decay modes with muons in the final states, which are forbidden in nuclear environments due to limited phase space~\cite{Rodejohann:2011mu, Deppisch:2012nb}. Furthermore, it could be theoretically easier to determine
the proton-proton scattering amplitude than the nuclear matrix element. Therefore, these neutrinoless double $\beta$-like processes constitute an ideal and complementary probe in the broader search for Lepton-Number-Violating signal~\cite{Cai:2017mow,Harz:2021psp, Babu:2022ycv};

\item Rare and forbidden hyperon processes: high-statistics strange-baryon samples at H-NS may enable searches for flavour changing neutral current (FCNC) decays such as $\Sigma^+\to p\ell^+\ell^-$, hidden-sector candidates in $Y\to BX$, $X\to e^+e^-$, and baryon- or lepton-number-violating hyperon decays such as $\Lambda\to \pi^\pm \ell^\mp$, $\Lambda\to K^\pm \ell^\mp$, $\Lambda\to\bar{p}\pi^+$, and $\Xi\to p\ell^-\ell^-$~\cite{HyperCP:2005mvo,LHCb:2025evf,McCracken:2015coa,HyperCP:2005sby}.

\end{itemize}

As a high-luminosity accelerator facility, HIAF will produce large samples of nucleons and strange hadrons in its energy regime. This provides an important experimental platform to study parity-violating and ($CP$)-violating effects, and on possible signatures of physics beyond the Standard Model.

\section{Global polarization of hyperon}

The mechanism for the polarization of $\Lambda$ hyperons in relativistic nucleus-nucleus collisions is different from the one discussed above. In nucleon-nucleon collisions the hyperon's polarization arises from a non-perturbative vacuum effect, while it is a medium effect in nucleus-nucleus collisions involving the collective motion of many particles. Another difference is that the polarization in nucleon-nucleon collisions is normally with respect to the production plane formed by the beam and the hyperon's momenta which is different for different hyperons even in one event, while in nucleus-nucleus collisions the so-called global polarization is with respect to the reaction plane formed by the beam direction and the impact parameter which is the same for all particles in one event.

\begin{figure}[t]
\centering
\includegraphics[width=0.5\linewidth]{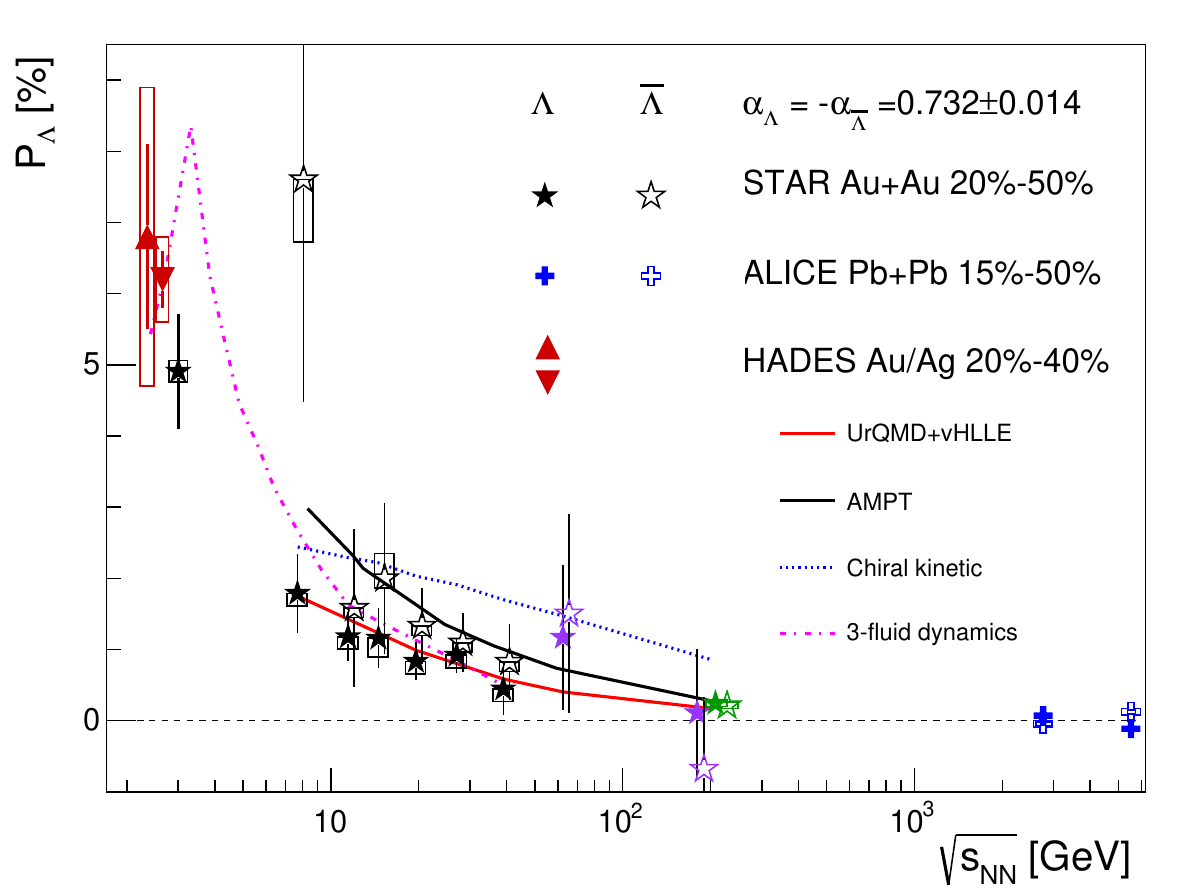} 
\caption{ Global polarization of $\Lambda$ hyperon and $\overline{\Lambda}$ hyperon in non-central heavy-ion collisions at different center-of-mass energies~\cite{Chen:2024aom}.}
\label{fig:GlobalPH_HIC}
\end{figure}

The global polarization of hyperons was first predicted by Liang and Wang two decades ago \citep{Liang:2004ph} as a spin-orbit coupling effect in the scattering of quarks by a static potential. The potential scatterings of quarks can be generalized to more realistic quark-quark \citep{Gao:2007bc} and parton-parton scatterings \citep{Zhang:2019xya}, and the spin-orbit coupling can turn into the spin-vorticity coupling \citep{Betz:2007kg,Becattini:2013fla,Fang:2016vpj} when taking ensemble average over all incident momenta \citep{Zhang:2019xya}.

The STAR collaboration measured the global polarization of $\Lambda$ and $\overline{\Lambda}$ hyperons in 2017 and global vector spin alignment in 2023 \cite{STAR:2017ckg,STAR:2022fan}. In statistical quantum field theories, the global polarization of $\Lambda$ and $\overline{\Lambda}$ hyperons, $P_{\Lambda / \overline{\Lambda}}$, can be estimated as \cite{Becattini:2016gvu}, 
\begin{equation}
   P_{\Lambda/\overline{\Lambda}} = \frac{\boldsymbol{\omega}}{2T} \pm \frac{\mu_\Lambda \boldsymbol{B}}{T},
\end{equation}
where $\boldsymbol{\omega}$ is the vorticity of the strong interaction matter produced in nucleus-nucleus collisions, and $\boldsymbol{B}$ is the magnetic field generated in collisions. From STAR's data, one can estimate that the average vorticity of the strong interaction matter is as large as $(9 \pm 1) \times 10^{21}$ per second, the largest one in any system in nature. Therefore, the measurement of the global polarization of $\Lambda$ hyperons reveals that the quark-gluon plasma produced in relativistic nucleus-nucleus collisions is the most vortical fluid that has ever been observed.

In principle, $P_\Lambda$-$P_{\overline{\Lambda}}$ could provide further information about the average magnetic fields. However, later measurements by the STAR collaboration found no significant splitting between $P_\Lambda$ and $P_{\overline{\Lambda}}$ \cite{STAR:2023nvo}. These results set upper limits on the average magnetic fields, which are $9.4 \times 10^{12}\; {\rm T}$ at $\sqrt{s_{NN}} = 19.6$ GeV and $1.4 \times 10^{13}\; {\rm T}$ at $\sqrt{s_{NN}} = 27$ GeV in Au + Au collisions at a 95\% confidence level.

The global polarization in high-energy collisions can be well described by relativistic hydrodynamics, kinetic theories, and statistical quantum field theories, see recent reviews \cite{Becattini:2020ngo, Hidaka:2022dmn, Niida:2024ntm, Becattini:2024uha, Chen:2024afy} and references therein for more details. 

Recently, the HADES \cite{HADES:2022enx} and STAR \cite{STAR:2021beb} collaborations reported measurements of the global polarization of $\Lambda$ hyperons in low-energy nucleus-nucleus collisions. 
These results are shown in Fig.~\ref{fig:GlobalPH_HIC}. One can see that the magnitude of $\Lambda$'s global polarization continues to increase as the collision energy decreases, even to the production threshold for $\Lambda$ hyperons. The elliptic flow measurements of identified particles at RHIC show that the onset of deconfinement of hadrons is around 4.5 GeV~\cite{STAR:2025owm,Chen:2024aom}, which indicates the system produced at 3 GeV is dominated by hadron gas. This raises two natural questions: \emph{How can we understand the global polarization of $\Lambda$ hyperons at such low energies? Is the physical interpretation of vorticity-induced polarization still valid in this regime?} To address these questions and better understand the spin dynamics in heavy-ion collisions across different regions of QCD phase diagram, systematic experimental and theoretical studies in low-energy nucleus-nucleus collisions are necessary. 

\section{Global  polarization  of light (hyper-)nucleus} 
Similar to  hyperons, the lightest hypernucleus, the hypertriton ($^3_\Lambda\text{H}$), can also acquire a global polarization in non-central relativistic nucleus--nucleus collisions~\cite{Sun:2025oib}. Based on the spin-dependent coalescence model~\cite{Sun:2025oib,Liu:2023nkm,Sheng:2020ghv,Lv:2024uev,Zhang:2024hyq}, the spin polarization of nucleons and $\Lambda$ hyperons can be transferred to $^3_\Lambda\text{H}$ when they recombine to form the $^3_\Lambda\text{H}$ at the late stage of the collision.  The $^3_\Lambda\text{H}$ polarization can be measured via two-body weak decay $^3_\Lambda\text{H}\rightarrow \pi^-+^3\text{He}$. The angular distributions of $^3$He ($^3\overline{\text{He}}$) from the decays of $^3_\Lambda\text{H}$~(${^3_{\bar{\Lambda}}}\overline{\rm H}$) with different spin-parity assignments are summarized in Tab.~\ref{tab:spin}. It can be seen that different spin structure of  $^3_\Lambda\text{H}$(${^3_{\bar{\Lambda}}}\overline{\rm H}$) leads to distinct angular distribution of the daughter $^3$He nucleus. \emph{Therefore, the spin polarization and decay pattern of $^3_\Lambda\text{H}$ provide a novel tool to decipher whether its spin is $1/2$ or $3/2$}~\cite{Sun:2025oib}. 

For recent experimental progress on light (anti-)(hyper-)nuclei production, see Refs.~\cite{STAR:2010gyg,STAR:2011eej, ALICE:2015rey, STAR:2019wjm,ALICE:2020zhb,STAR:2021orx, ALICE:2022zuz,ALargeIonColliderExperiment:2021puh,ALICE:2022xiu,ALICE:2022ugx,ALICE:2022sco,STAR:2022hbp,STAR:2023fbc,ALICE:2024djx,ALICE:2025byl,A1:2026sjf,Chen:2023mel,Chen:2018tnh}. For discussions on effects of hadronic re-scattering, Pauli blocking, wavefunction size, and critical fluctuations on the production of light nuclei, see Refs.~\cite{Sun:2022xjr,Wang:2025wim,Sun:2018mqq,ALICE:2026tqr,Sun:2017xrx}.
\begin{table}[!h]
     \centering
     \begin{tabular}{|c|c|c|c|}
     \hline
         $J^\pi$& structure &decay mode & $\frac{dN}{d\cos\theta^*}$   \\
        
         \hline
          $\frac{1}{2}^+$ & $\Lambda(\frac{1}{2}^+)-np(1^+)$ &$^3_\Lambda\text{H}\rightarrow\pi^-+^3$He& $\frac{1}{2}(1-\frac{1}{2.58}\alpha_\Lambda\mathcal{P}_\Lambda\cos\theta^*)$  \\
          \hline
          $\frac{1}{2}^+$& $\Lambda(\frac{1}{2}^+)-np(0^+)$ &$^3_\Lambda\text{H}\rightarrow\pi^-+^3$He&$\frac{1}{2}(1+\alpha_\Lambda\mathcal{P}_\Lambda\cos\theta^*)$    \\
          \hline
          $\frac{3}{2}^+$& $\Lambda(\frac{1}{2}^+)-np(1^+)$ &$^3_\Lambda\text{H}\rightarrow\pi^-+^3$He&$ \frac{1}{2} \big{(}1-\mathcal{P}_\Lambda^2 (3\cos^2\theta^*-1)\big{)}$    \\ 
          \hline
          \hline 
          $\frac{1}{2}^-$ & 
          $\bar{\Lambda}(\frac{1}{2}^-)-\overline{np}(1^-)$ &
          ${^3_{\bar{\Lambda}}}\overline{\rm H}\rightarrow\pi^++^3\overline{{\rm He}}$& 
          $\frac{1}{2}(1-\frac{1}{2.58}\alpha_{\bar{\Lambda}}\mathcal{P}_{\bar{\Lambda}}\cos\theta^*)$  \\
          \hline
          $\frac{1}{2}^-$& $\bar{\Lambda}(\frac{1}{2}^-)-\overline{np}(0^-)$ &${^3_{\bar{\Lambda}}}\overline{\rm H}\rightarrow\pi^++^3\overline{{\rm He}}$&$\frac{1}{2}(1+\alpha_{\bar{\Lambda}}\mathcal{P}_{\bar{\Lambda}}\cos\theta^*)$    \\
          \hline
          $\frac{3}{2}^-$& $\bar{\Lambda}(\frac{1}{2}^-)-\overline{np}(1^-)$ &${^3_{\bar{\Lambda}}}\overline{\rm H}\rightarrow\pi^++^3\overline{{\rm He}}$&$ \frac{1}{2} \big{(}1-\mathcal{P}_{\bar{\Lambda}}^2 (3\cos^2\theta^*-1)\big{)}$    \\
          \hline          
     \end{tabular}
     \caption{Spin-parity, internal structure, decay mode, and decay angular distribution of $^3_\Lambda\text{H}$ and ${^3_{\bar{\Lambda}}}\overline{\rm H}$. The parameters $\alpha_\Lambda$, $\mathcal{P}_\Lambda$, and $\theta^*$ denote the decay parameter of $\Lambda$, the spin polarization of $\Lambda$, and the angle between the momentum of $^3$He and the spin quantization axis of the $\Lambda$ hyperon in the rest frame of $^3_\Lambda\text{H}$, respectively. Table taken from Ref.~\cite{Sun:2025oib}.}
     \label{tab:spin}
 \end{table}

Although protons and neutrons are also expected to be globally polarized in non-central collisions, direct measurements require baryon polarimeter~\cite{Liang:2025owx} in the detector and are thus experimentally challenging.
In the case that the hypertriton has the spin structure listed in the first line of Tab.~\ref{tab:spin}, the proton spin polarization ($\mathcal{P}_p$) can be reliably reconstructed through a simple linear relation given by~\cite{Liu:2025kpp}
\begin{eqnarray}
\mathcal{P}_p \approx \frac{1}{4}\left(3\mathcal{P}_{^3_\Lambda\mathrm{H}} + \mathcal{P}_\Lambda\right). \label{eq:NucleonPolarization}
\end{eqnarray}
This relation directly reflects the fact that, in the spin wave function of the $^3_\Lambda\text{H}$~($J^\pi=\frac{1}{2}^+$~\cite{STAR:2017gxa}), the $\Lambda$ spin aligns with the total spin with probability $1/3$ and anti-aligns with probability $2/3$. \emph{Such a linear relation provides an  avenue for accessing the spin polarization of protons  in nuclear collisions.}

As the collision energy decreases to $\sqrt{s_{NN}}<3$ GeV, the production rates of hyperons and hypernuclei are strongly suppressed, making polarization measurements much more challenging. In this regime, instead of hyperons and hypernuclei, one may study the polarization of nuclear matter through unstable nuclei such as $^4$Li. The $^4$Li nucleus has a ground state with $J^\pi=2^-$ and three excited states. Because the excitation energies of $^4\text{Li}$ (a few MeV) are much smaller than the typical temperature of the hadronic matter produced in heavy-ion collisions, the populations of these states are approximately proportional to their statistical spin degeneracies, namely $5:3:1:3$. Consequently, the yield-averaged angular distribution of the decay products ($^4\text{Li}\rightarrow p+^3\text{He}$) is given by~\cite{Zheng:2025ngn}
\begin{eqnarray}
\label{coalescence average angular distribution}
    \frac{dN}{\sin \theta^* d\theta^*} 
    &\approx&\frac{1}{2} \left[1-\frac{1}{36}\left(5 \mathcal{P}_N^2+8 \mathcal{P}_N \mathcal{P}_L+11 \mathcal{P}_L^2\right)(3\cos ^2\theta^* -1) \right].
\end{eqnarray}
Here, $\mathcal{P}_N$ denotes the nucleon spin polarization, and $\mathcal{P}_L$ denotes the contribution from $p-$wave orbital motion. Since the values of $\mathcal{P}_N$ and $\mathcal{P}_L$ are typically small ($\sim 0.02$), the spin alignment of $^4$Li is expected to be of order $10^{-4}$. Yet, spin correlations among nucleons or phase-space anisotropies may significantly enhance the spin-alignment signal of $^4$Li. \emph{Future measurements on the spin alignment of $^4$Li  offer a promising method to probe the vortical structure and spin-dependent equation-of-state of the nuclear matter at high baryon densities.}

\end{chapter}

\begin{chapter}{Hyperon-nucleon spectrometer at HIAF}

\section{The physics requirements}
To address the physics goals of hyperon production and polarization measurements---spanning proton-proton, proton-nucleus, and nucleus-nucleus collisions across beam energies from 3 to 25 GeV---the Hyperon-Nucleon Spectrometer (H-NS) at HIAF demands stringent detector performance. High momentum resolution ($\sim$2\% @ 1 GeV/$c$) is required for precise kinematic reconstruction, especially for low-transverse-momentum tracks in weak decay topologies. Excellent vertex resolution ($<$~\SI{500}{\micro\meter}) is critical to suppress combinatorial backgrounds and reconstruct $\Lambda$ decay vertices unambiguously, with material budget kept below 0.5\% $X/X_0$ per layer to minimize multiple scattering. Particle identification (PID) must achieve 3$\sigma$ separation between pions and kaons up to 2 GeV/$c$, and between protons and kaons up to 5 GeV/$c$, enabled by LGAD sensors with 20–30 ps time resolution. The detector acceptance must cover polar angles from $5^\circ$ to $100^\circ$ to capture the full angular distribution essential for polarization analysis. To handle event rates up to 1 MHz, the readout must be fast and radiation-hard, particularly for the calorimeters and silicon trackers operating near the target. A compact, low-material silicon trackers and forward calorimeters with excellent spatial ($\sim$3 mm) and energy resolution (3\%/$\sqrt{E}$) are crucial to efficiently reconstruct multiple hyperons in the final state. Finally, the inclusion of a baryon polarimeter using a thin carbon foil enables direct measurement of proton polarization via elastic scattering, adding a unique capability to the spectrometer. Table~\ref{tab:hns_requirements} summarizes the primary requirements in H-NS.

\begin{table}[htbp]
    \centering
    \caption{Physics Requirements for the Hyperon-Nucleon Spectrometer (H-NS)}
    \label{tab:hns_requirements}
    \begin{tabular}{@{}lll@{}}
        \toprule
        \textbf{Subsystem / Category} & \textbf{Parameter} & \textbf{Design Requirement} \\ 
        \midrule
        
        % 径迹与顶点探测器部分
        \multirow{3}{*}{Tracker \& Vertex} 
        & Momentum resolution ($\Delta p/p$) & $\sim$2\% @ 1 GeV/$c$ \\
        & Single-point position resolution & $\sim$10~$\si{\micro\meter}$ \\
        & Vertex resolution ($\sigma_{\text{vtx}}$) & $< 500~\si{\micro\meter}$ \\
        & Material budget (per layer) & $< 0.5\% \, X/X_0$ \\
        \midrule
        
        % 粒子鉴别探测器部分
        \multirow{3}{*}{PID Detector} 
        & $\pi/K$ separation & $3\sigma$ up to 2 GeV/$c$ \\
        & $p/K$ separation & $3\sigma$ up to 5 GeV/$c$ \\
        %& Time resolution (LGAD sensors) & 20--30 ps \\
        \midrule
        
        % 电磁量能器部分
        \multirow{2}{*}{Calorimeter} 
        & Spatial resolution & $\sim$3 mm \\
        & Energy resolution & $3\%/\sqrt{E}$ \\
        \midrule
        
        % 剩余的全局/其他项目
        \multirow{4}{*}{Others} 
        %& Beam energy range ($E_{\text{beam}}$) & 3--25 GeV (pp, pA, and AA collisions) \\
        & Polar angle acceptance ($\theta$) & $5^\circ$ to $100^\circ$ \\
        & Event rate capability & Up to 1 MHz (fast, radiation-hard readout) \\
        & Baryon polarimeter target & Thin carbon foil (for proton elastic scattering) \\
        
        \bottomrule
    \end{tabular}
\end{table}

\section{The conceptual design of H-NS detector}

The H-NS spectrometer, as shown in Fig.~\ref{fig:det_concept}, integrates a cylindrical tracking system and axial magnetic field for precision hyperon studies. The vertex detector employs five concentric Monolithic Active Pixel Sensor (MAPS) layers spaced from \SI{5}{\cm} to 35{\cm} radius with \SI{30}{\micro\meter} pixels and \SI{0.5}{\percent} radiation length per layer, supplemented by five forward disks positioned at \SI{40}{\cm} to \SI{100}{\cm} along the beam axis. A \SI{1}{\milli\meter} carbon target between the third and fourth layers contributes 10\% of the tracking material budget for final-state proton polarization measurements. Particle identification combines a barrel-mounted LGAD layer at \SI{45}{\cm} radius with around \SI{100}{\micro\meter} spatial and \SI{30}{\pico\second} timing resolution and a forward TOF disk at \SI{115}{\cm} providing additional tracking with around \SI{100}{\micro\meter} spatial and \SI{30}{\pico\second} timing capabilities.
The superconducting solenoid (\SI{1.5}{\tesla}, \SI{160}{\cm} length) ensures homogeneous axial field coverage up to \SI{60}{\cm} radius. Tracking terminates at a downstream PbWO\textsubscript{4} calorimeter positioned at \SI{125}{\cm} with \SIrange{20}{22}{\milli\meter} crystal modules, achieving $\sigma_E/E = 3\%/\sqrt{E}$ energy resolution for neutral particles. This integrated design optimizes hyperon decay reconstruction while minimizing material in critical tracking regions. Geometry parameters are listed in Tab. \ref{tab:hns-geometry}.

\begin{figure}[h]
\centering
\includegraphics[width=0.7\linewidth]{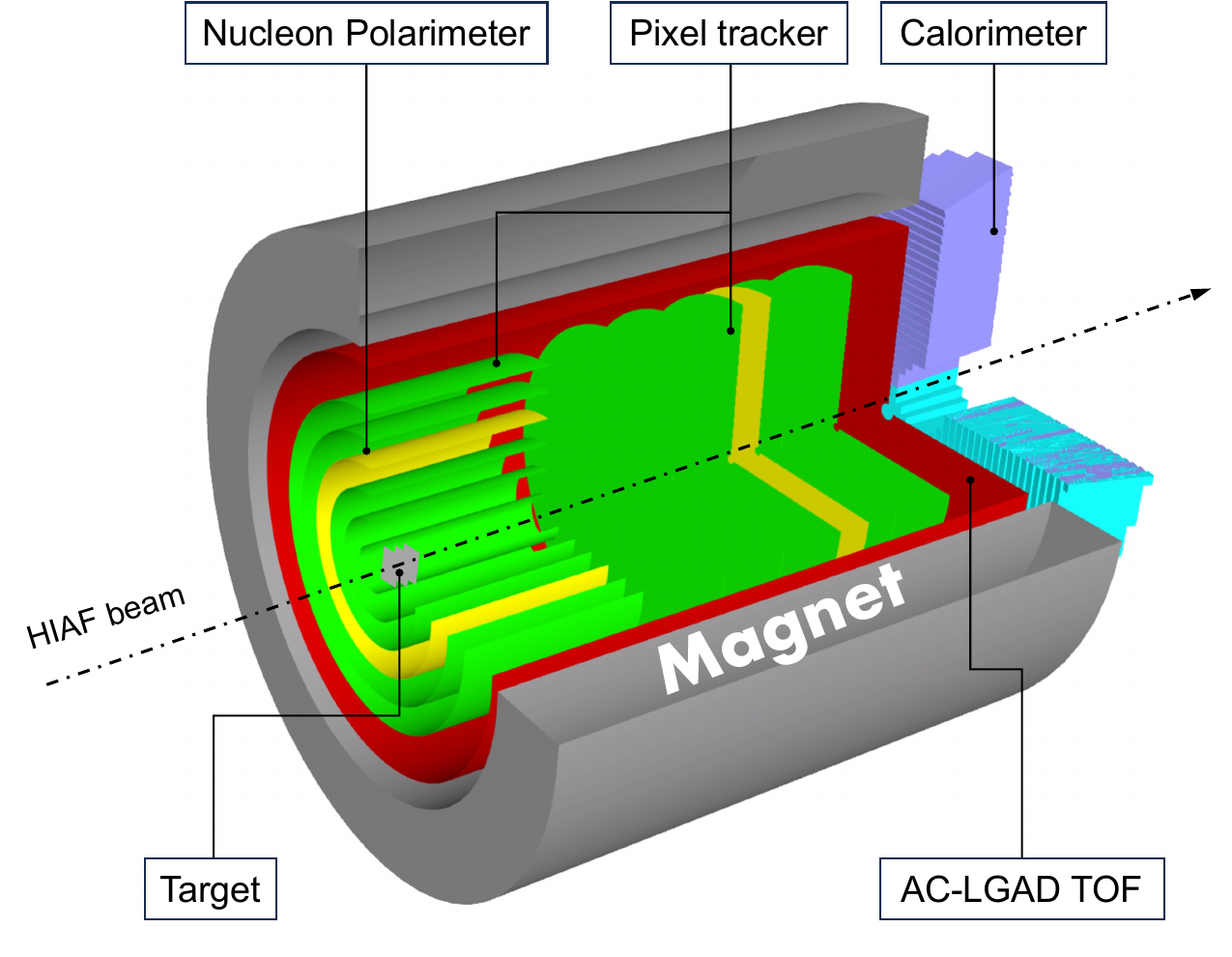}
\caption{\label{fig:det_concept}Concept of H-NS detector.}
\end{figure}

\begin{table}[htb]
\centering
\caption{H-NS detector geometry parameters.}
\label{tab:hns-geometry}
\begin{tabular}{>{\raggedright}p{4.5cm}cc}
\toprule
\textbf{Component} & \textbf{Parameter} & \textbf{Specification} \\
\midrule
\multirow{6}{*}{Vertex/Tracker (MAPS)}
& Barrel(Forward) layers & 5(5) \\
& Barrel radial coverage & \SI{5}{\cm} to 35{\cm} \\
& Forward disks position & \SI{40}{\cm} to \SI{100}{\cm} \\
& $\eta$ coverage & \SI{-0.2}{} to \SI{4.0}{} \\
& Pixel size & \SI{30}{\micro\meter} \\
& Material budget per layer & \SI{0.5}{\percent} $X/X_0$ \\
%& Target location & Layer 3-4 interface \\
\midrule
\multirow{7}{*}{PID System (LGAD)}
& Barrel(Forward) layers & 1(1) \\
& Barrel radial coverage & $\SI{45}{\cm}$ \\
& Forward disk position & \SI{115}{\cm} \\
& $\eta$ coverage & \SI{-0.2}{} to \SI{4.0}{} \\
& Hit resolution & \SI{100}{\micro\meter} \\
& Timing resolution (per track) & \SI{30}{\pico\second} \\
& Material budget per layer & $\leq$\SI{50}{\percent} $X/X_0$ \\
\midrule
\multirow{5}{*}{Calorimeter}
& Material & PbWO\textsubscript{4}/Lead-Glass \\
& Front-end position & \SI{125}{\cm} \\
& Crystal size & \SI{20}\sim \SI{22}{\milli\meter} \\
& Crystal length & \SI{20}{\cm} \\
& $\eta$ coverage & \SI{1.75}{} to \SI{4}{} \\
\midrule
\multirow{4}{*}{Magnet}
& Configuration & Solenoid \\
& Field strength & \SI{1.5}{\tesla} \\
& Radial coverage & \SI{0}{\cm}  to  \SI{60}{\cm} \\
& Z coverage & \SI{-30}{\cm}  to \SI{130}{\cm} \\
\bottomrule
\end{tabular}
\end{table}

\subsection{Magnet}

The superconducting solenoid magnet is a constituent part of the H-NS detector. It is designed to provide a 1.5 Tesla  magnetic field within an aperture that has a diameter of 1.2 m and a length of 1.6 m. The design can benefit directly from the technical solutions pioneered by large detector superconducting solenoids, such as the Collider Detector Facility (CDF) and the Compact Muon Solenoid (CMS) project. The key technology is the aluminum-stabilized superconductor which is characterized by good mechanical properties and low electrical resistivity~\cite{Mentink:2022iti}. The coil fabrication technology is also important and it has advanced along with the conductor technology, such as the inner coil winding technique, indirect cooling, quench protection scheme and so on.

The main components of the superconducting solenoid include the superconducting coil system, cryogenic system, quench protection, vacuum tank, power supply and the return yoke. The coil and yoke interact to form the desired magnetic field. Figure~\ref{Calculated_magnetic_field_distribution} shows the magnetic field distribution, simulated by the software of COMSOL Multiphysics, resulting from a single-layer winding and the return yoke. The magnetic field in the flux
return yoke is 1.28 T for the central magnetic field of 1.5 T. The coil is internally wound in an aluminum alloy cylinder to support the magnetic hoop stresses. The technology of internal winding of aluminum stabilized conductor has been chosen and successfully tested for many spectrometer solenoids from CELLO in the early 80s to ATLAS and CMS at LHC today~\cite{Baynham:2006oxc}. The solenoid cold mass is cooled by helium in tubes attached to the coil support cylinder. Indirect cooling simplifies the cryostat design and reduces the quantity of liquid helium it contains, effectively eliminating the overpressure hazard associated with the cryostat.

Aluminum stabilized conductor gives high stability against quenches due to the large electrical conductivity of aluminum at low temperature. It is able to withstand large thermal perturbations before a normal conducting zone starts to spread. The quench protection system is designed to dissipate stored energy by means of an external dump resistor. In addition, aluminum strips are glued inside the coil surface to increase the quench propagation. The hot spot temperature shall be limited to 100 K for safety reasons.  

Stainless steel is selected for the vacuum vessel of the cryostat to ensure mechanical stability. The coil is suspended from the outer vacuum vessel by the radial and axial ties. A radiation shield is positioned between the coil and the cryostat outer vessel, which is cooled by helium gas at temperatures ranging from 50 K to 80 K. To compensate the cryogenic losses of the solenoid, a refrigeration power is required. A closed loop helium satellite refrigerator supplies adequate refrigeration power at both 50 K for the intermediate temperature shields and at 4.5 K for the forced flow refrigeration of the coil cold mass.

\begin{figure}[tbp]
\centerline
{\includegraphics[width=0.6\linewidth]{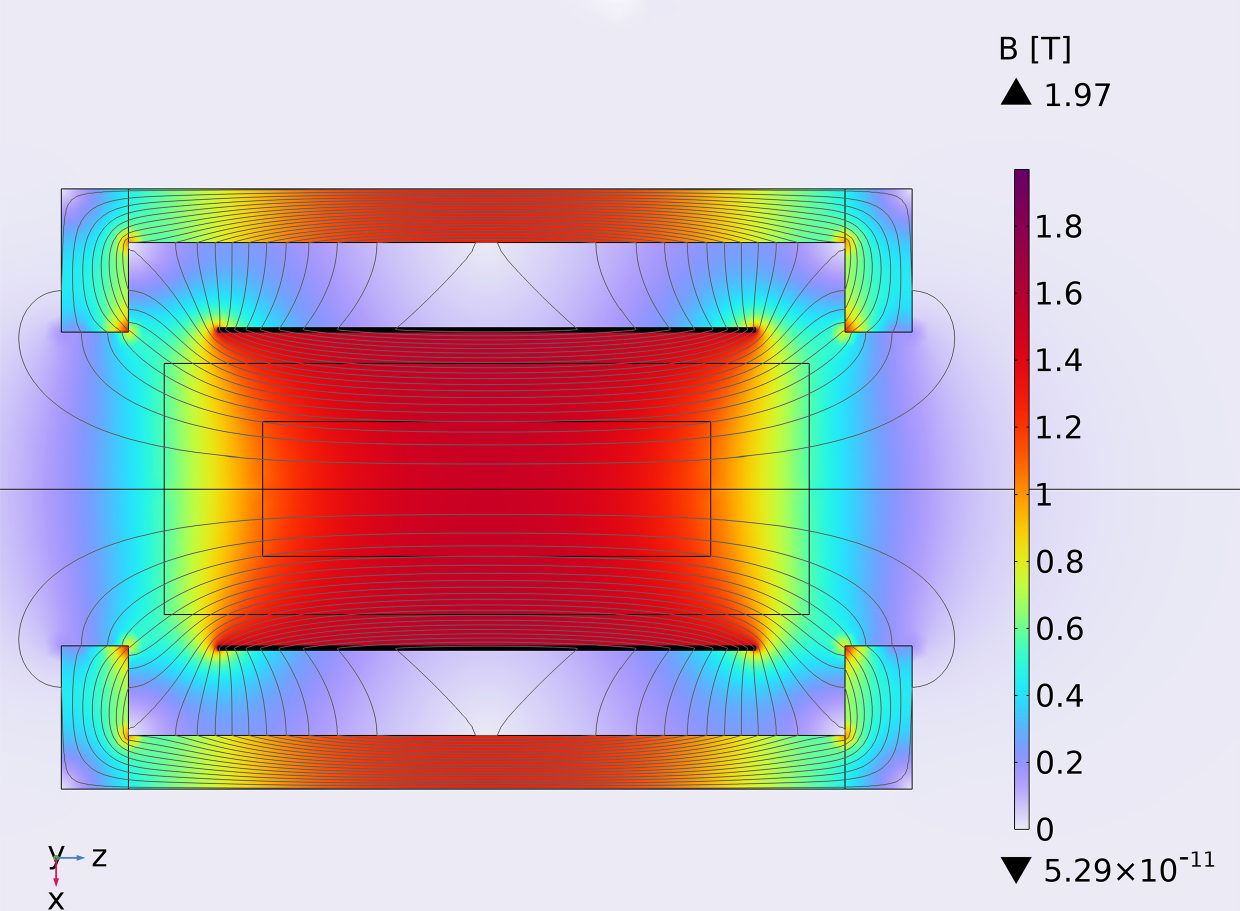}}
\caption{Calculated magnetic field distribution within and surrounding the solenoid.}
\label{Calculated_magnetic_field_distribution}
\end{figure}

\subsection{Targets}

\begin{figure}[tbp]
\centering
\subfloat[]{
\includegraphics[width=0.35\linewidth]{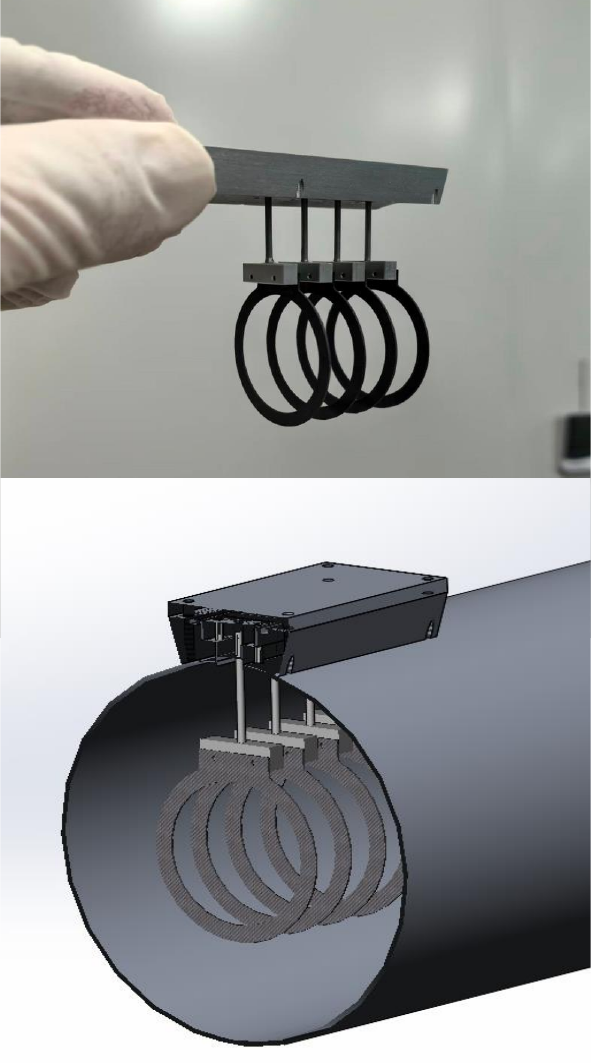}}
\centering
\subfloat[]{
\includegraphics[width=0.5\linewidth]{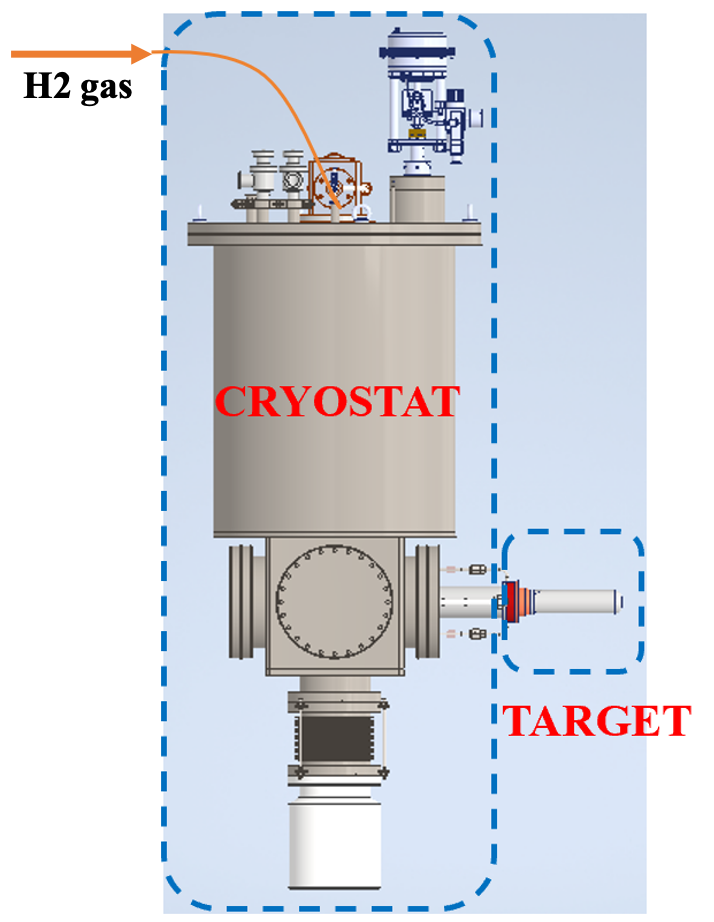}}
\caption{ \label{fig:A_LH2_tgt}Supporting frame of the heavy-ion targets (a) and the liquid hydrogen target (b).}
\end{figure} 
\textit{Unpolarized targets}---In the first stage, the H-NS will operate with unpolarized targets. As shown in Fig.~\ref{fig:A_LH2_tgt}-(a), heavy-ion targets will be made of foils of materials containing carbon, calcium or titanium, attached to a movable insert.
A liquid hydrogen (LH2) target (Fig.~\ref{fig:A_LH2_tgt}-(b)) with the required geometry to fit with the pixel silicon tracker will be developed. The typical density of the LH2 is 75 mg/cm$^{3}$ at 16 K. The LH2 target system consists of four parts: 1) the target cell, 2) the cryostat, 3) the hydrogen gas circuit, and 4) the control-command section. The circulation of the hydrogen would be driven by the thermosiphon principle, similar to the STRASSE LH2 target \cite{Liu:2023hwf}. The cryostat and the LH2 target cell located in the experimental area, as shown in Fig.~\ref{fig:A_LH2_tgt}-(b), will be operated in a vacuum. 

\vspace{0.2cm}
\noindent\textit{Polarized targets}---Following the successful implementation of the unpolarized targets, further development will focus on polarized target systems. A polarized $^3$He gas target will be realized using the metastability-exchange optical pumping (MEOP) technique~\cite{PhysRev.132.2561}. In particular, recent advances in high-field MEOP~\cite{Li:2023rcp} will enable $^3$He polarization within the 1.5 T magnetic field of the H-NS. The target will employ a double-cell design~\cite{Maxwell:2021ytu} compatible with the existing LH2 cryogenic modules.
Solid polarized proton (deuteron) targets will be developed using the dynamic nuclear polarization (DNP) technique~\cite{Crabb:1997cy}. 
The target sample is made of solid beads of proton-rich (or deuteron-rich) materials (e.g.~butanol, lithium or ammonia) doped with paramagnetic impurities. Radical electrons are first polarized in a strong polarizing magnet at very low temperature (5~T@1~K or 2.5~T@0.5~K). Spin flips induced by resonant microwaves continuously transfer the electron polarization to the proton (deuteron). To avoid particle blocking caused by the polarizing magnet, the target is envisioned to operate in the frozen spin mode~\cite{Dutz:1994mx, Bradtke:1999zg, Keith:2012ad}. That is, after the polarization build-up, the polarizing magnet is moved away from the target sample, and a holding magnetic field, produced by coils close to the target sample, is switched on for data taking. To maintain the polarization, the target sample is placed in a cryostat cooled to 20~mK using a customized dilution refrigerator.
It is envisaged that the auxiliary components of the liquid hydrogen target, such as the supporting platform and the rail system, can be extended to accommodate polarized targets in the future.
\subsection{The Monolithic Active Pixel Sensors based tracker}

\subsubsection{Introduction of MAPS}
The MAPS technology provides high granularity, low power consumption, and consequently low material budget, as well as fast readout speed in one device. Therefore, it is considered the best detector technology to satisfy the requirements of the H-NS vertex and the tracking detectors. In addition, the integration of charge collection and readout capabilities into one silicon substrate is well-suited for the required level of integration and acceptance coverage of the H-NS. 
The MAPS detector has been developed for several generations. The first generation was deployed in the STAR Heavy Flavor Tracker (HFT)~\cite{Contin:2017mck}, which is the first application of MAPS in high energy physics experiment, and the material budget is about 0.4\%  $X/X_{0}$ per layer. Then, the ALICE ITS2~\cite{ALICE:2013nwm} used the ALPIDE sensor~\cite{AglieriRinella:2017lym}, which is fabricated in a commercial 180 nm Complementary Metal Oxide Semiconductor (CMOS) imaging process provided by Tower Jazz. It provides better charge collection properties, radiation hardness and signal processing capabilities compared to traditional MAPS, and the material budget is about 0.4\%  $X/X_{0}$ per layer for both the inner and outer barrels, respectively.

\begin{figure}[h]
\centering
\includegraphics[width=0.6\linewidth]{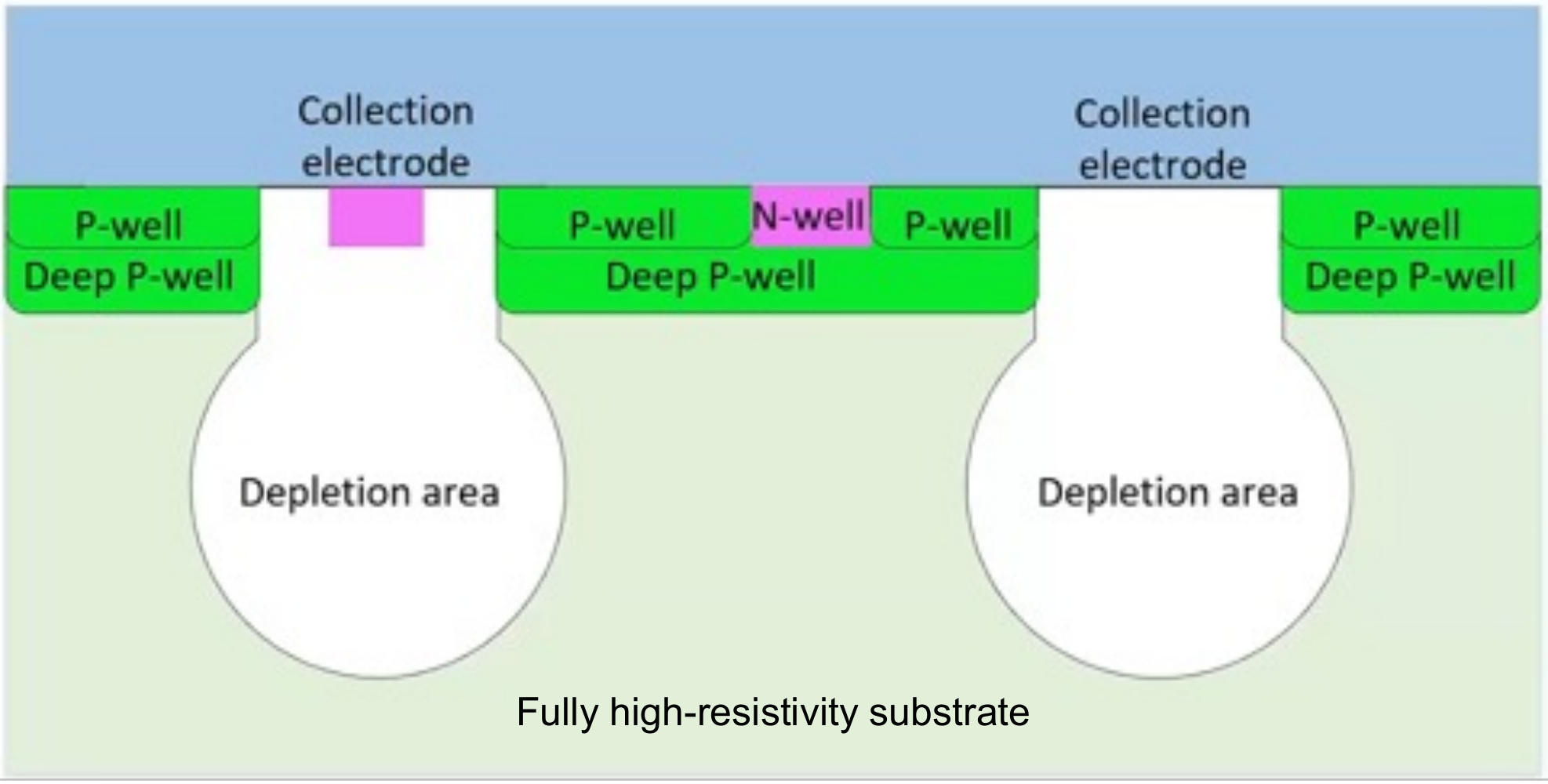}
\caption{\label{fig:maps_concept}Schematics cross-section of a MAPS pixel in the 130 nm imaging CMOS  with the deep p-well feature.}
\end{figure}

\subsubsection{MAPS development for H-NS}
The 130 nm CMOS technology by the GSMC has been selected for the implementation of the Pixel Chip (MIC6-v4, the fourth generation of MAPS chip developed by CCNU) for the H-NS tracking detectors. Figure~\ref{fig:maps_concept} shows a schematic cross section of a pixel in this technology. The pixel chip consists of a single silicon die of about 15 mm $\times$ 30 mm, which incorporates a high-resistivity silicon substrate (sensor active volume), a matrix of charge collection diodes (pixels) with a pitch of the order of \SI{30}{\micro\meter}, and the electronics that perform signal amplification, digitisation and zero-suppression. Only the information on whether or not a particle was crossing a pixel is read out, and the power consumption is expected to achieve less than 40 mW/cm$^2$.
To the design of the MIC6-v4 chip, the high resistive p-type substrate material is selected to grow on top of a low-resistivity silicon wafer used for standard CMOS manufacturing. When a charged particle passes through the silicon sensor's active volume, it liberates charge carriers (electrons and holes) in the semiconductor material. The released charge is then collected by the sensors' electrodes, which can be speeded up by allowing a reverse bias on the substrate to form a depletion area.

The geometry and requirements of the H-NS pixel tracker provide a natural grouping of five barrel layers in barrel region and five disc layers in forward region. The barrel layers are in two separate barrels (Inner Barrel and Outer Barrel), each with different specifications. The Inner Barrel (IB) of the H-NS tracker consists of the three innermost layers, also referred to as Inner Layers (Layers 0 to 2), while its Outer Barrel (OB) contains the two outermost layers, also referred as Outer Layers (Layers 3 to 4). The barrel layers are azimuthally segmented in units named Staves, which are mechanically independent. Staves are fixed to a support structure, half-wheel shaped, to form the Half-Layers. The term Stave will be used to refer to the complete detector element. The forward layers are designed in five disc layers, and each layer is azimuthally segmented in units named ladder, which consists of different number of pixel chips.  
Figure~\ref{fig:det_tracker_concept} shows the geometry layout of MAPS-based tracking detector for the H-NS. 
\begin{figure}[h]
\centering
\includegraphics[width=0.6\linewidth]{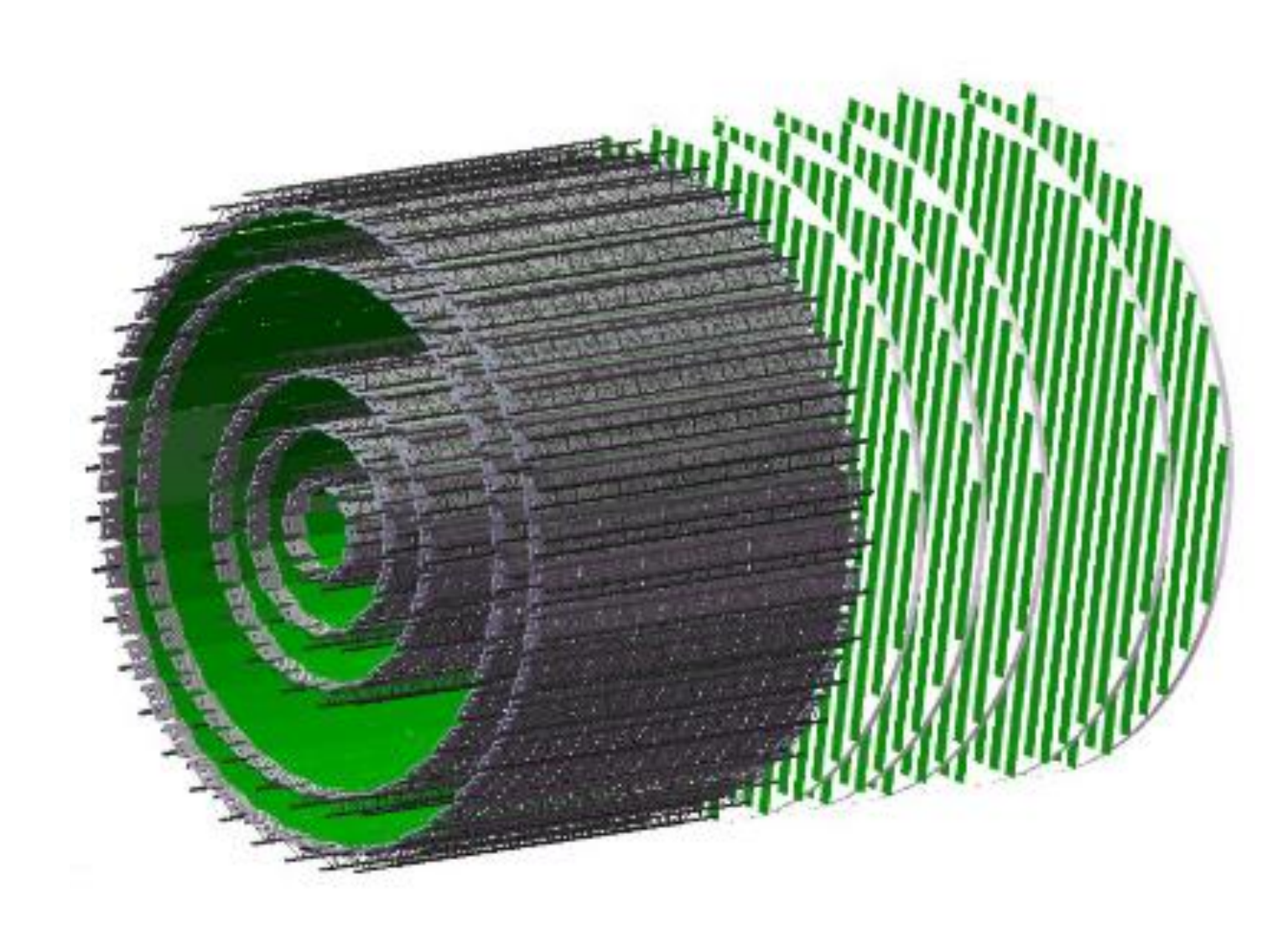}
\caption{\label{fig:det_tracker_concept}Conceptual design of MAPS-based tracking detector for the H-NS.}
\end{figure}

On the basis of the above design and considerations, broad coverage of 4.0 pseduo-rapidity unit from middle- to forward-rapidity. The total MAPS area is around 5.2 m$^2$, consisting of 2.9 m$^2$ for barrel region and 2.3 m$^2$ for forward region. The geometrical and technical parameters are listed in Tab.~\ref{tab:table_MAPS_geometry}.
\begin{table}[!htb]
\begin{center}
\caption{Geometrical parameters of the H-NS MAPS tracker.}
\label{tab:table_MAPS_geometry}
\begin{tabular}{lccccc|ccccc}
\hline\hline
\multicolumn{1}{c}{ } &
\multicolumn{5}{c}{Barrel} &
\multicolumn{5}{c}{Forward} \\
\hline
  & L0 & L1 & L2 & L3 & L4 & D0 & D1 & D2 & D3 & D4  \\
\thead{Radius / \\ Z position (cm)}  & 5 & 11 & 17 & 29 & 35 & 40 & 55 & 70 & 85 & 100\\
\thead{Active area (cm$^{2}$)} & 1310 & 3058 & 4368 & 7426 & 9173 & \multicolumn{5}{c}{4009} \\
\thead{Nr. staves /\\ ladders} & 12 & 14 & 20 & 34 & 42 & \multicolumn{5}{c}{114} \\
\thead{Nr. MAPS chips} & 336 & 784 & 1120 & 1904 & 2352 & \multicolumn{5}{c}{1028} \\
\thead{Length / inner \\outer radius (cm)}   & \multicolumn{5}{c}{42} & \multicolumn{5}{c}{2 / 38} \\
\thead{MAPS chip \\dimensions (mm$^{2}$)} & 
\multicolumn{10}{c}{15 $\times$30} \\
\thead{MAPS chip \\thickness (\SI{}{\micro\meter})} & 
\multicolumn{10}{c}{50 or 100} \\
\thead{Pixel size (\SI{}{\micro\meter}$^{2}$)} & 
\multicolumn{10}{c}{30 $\times$27} \\
\thead{Nr. Pixels per\\ chip ($r\phi \times z$)} & 
\multicolumn{10}{c}{497 $\times$980} \\
\thead{Material budget per layer ($X/X_{0}$)} & 
\multicolumn{10}{c}{$\leq$ 0.5\%} \\
\hline
\end{tabular}
\end{center}
\end{table}

\subsection{The Low Gain Avalanche Diode based time-of-flight detector}
The Low Gain Avalanche Diode (LGAD) is a novel detector technology developed by the RD50 collaboration to address the need for radiation-tolerant high precision timing detector by the HL-LHC~\cite{Pellegrini:2014lki}. After a decade of extensive research and development of the community, the technology is now considered mature and adopted by both the ATLAS and CMS experiments to construct the timing detector for their phase 2 upgrade~\cite{LGAD:ATLASHGTD,LGAD:CMSETL}. With highly resistive ($>$ 1000 $\Omega \cdot$cm) \SI{50}{\micro\meter} thick epitaxial silicon wafers, excellent time resolution of about 30 ps per hit has been achieved for minimum ionising particles. The working principle of the LGAD sensor is illustrated in Fig.~\ref{fig:DCLGADPrinciple}. A $p^+$ doped layer (commonly called gain layer) is implanted near the PN junction, which provides a high electric field when sufficient bias voltage is supplied. The performances of the gain layer can deteriorate with irradiations. The radiation hardness can be improved by co-implanting Carbon in the gain layer. The carbonated LGAD designed and fabricated by a few vendors can achieve a time resolution better than 70 ps after having received a fluence of $2.5 \times 10^{15}$ cm $^{-2}$  n$_\textrm{eq}$~\cite{Li:2022ukj,Li:2022auc}.

\begin{figure}[htbp]
    \centering
    \begin{subfigure}{0.3\textwidth}
        \includegraphics[width=\textwidth]{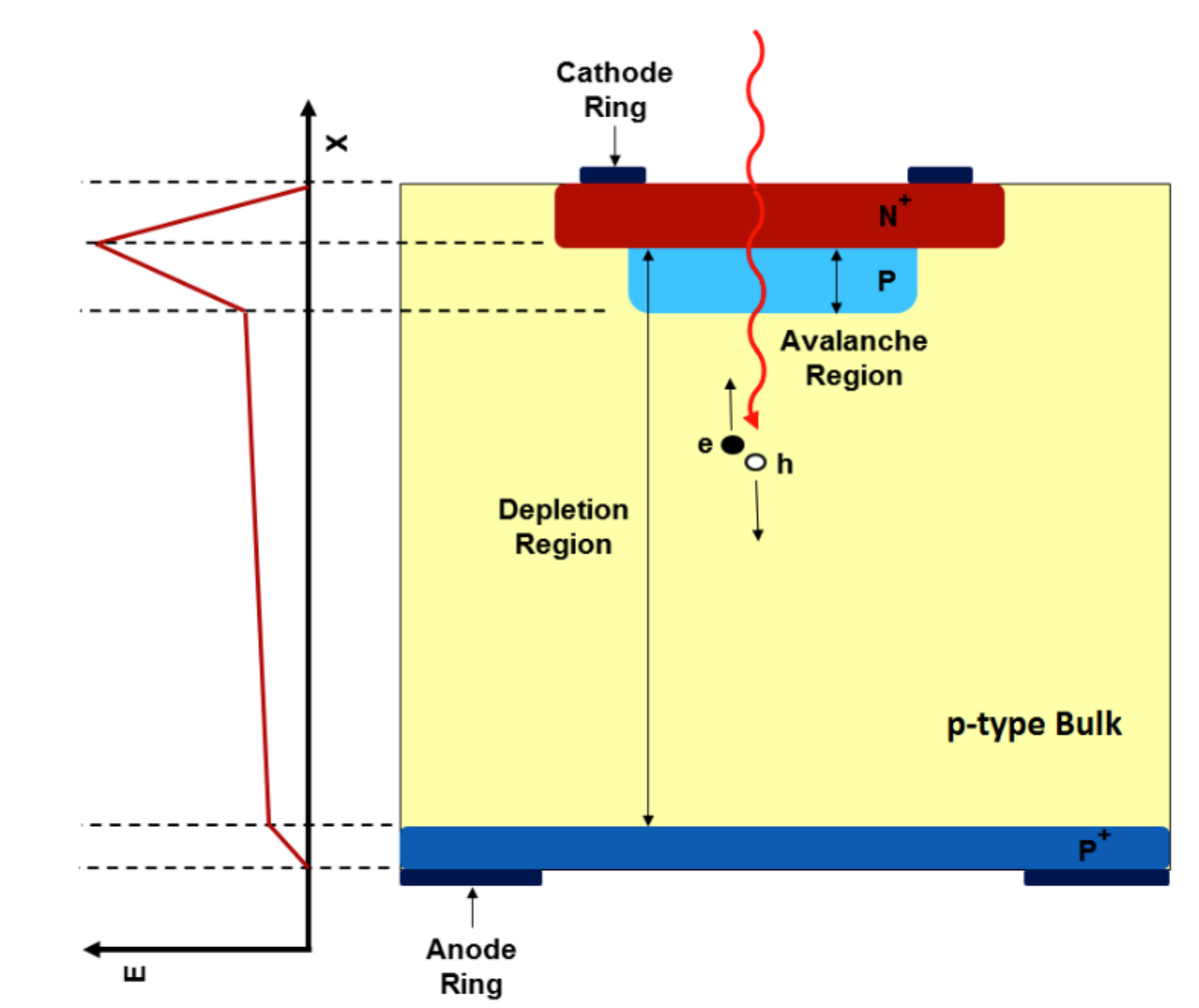}
        \caption{LGAD}
        \label{fig:DCLGADPrinciple}
    \end{subfigure}
    \hfill
    \begin{subfigure}{0.66\textwidth}
        \includegraphics[width=\textwidth]{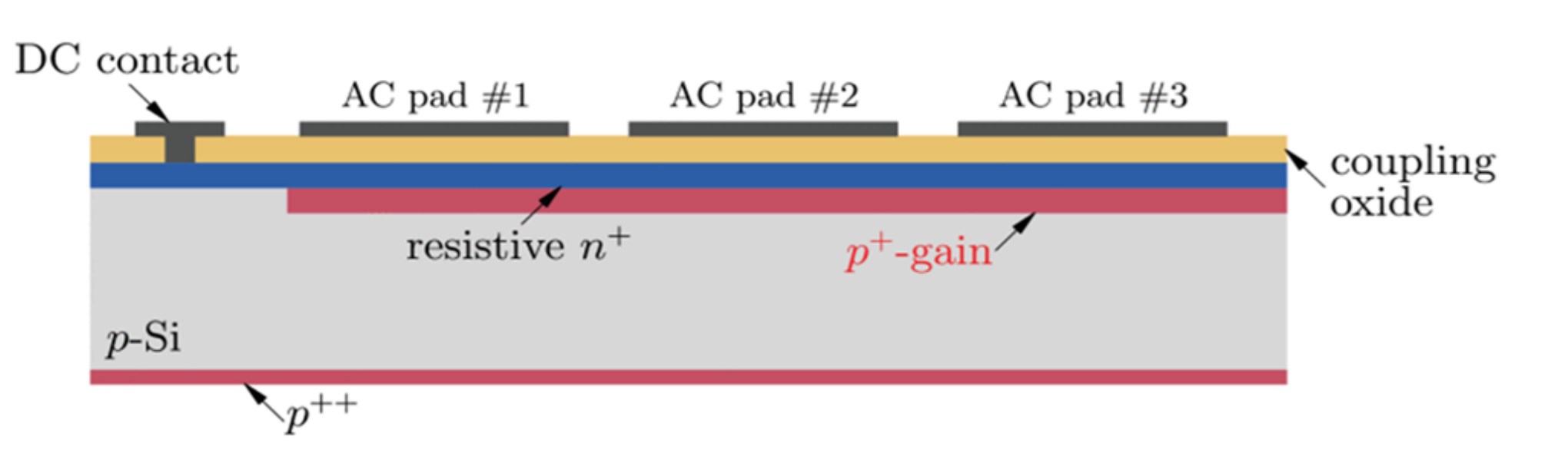}
        \caption{AC-LGAD}
        \label{fig:ACLGADPrinciple}
    \end{subfigure}
    \caption{Schematics cross-section of LGAD and AC-LGAD structures (size and position).}
    \label{Fig:LGADprinicple}
\end{figure}

The granularity of the LGAD is achieved by segmenting the gain layers, which can cause a non-negligible fraction of detector region without gain when the segmentation is too fine. For example, the fill factor is about 70\% for a granularity of 300 \SI{}{\micro\meter} $\times$ 300 \SI{}{\micro\meter}. This limitation is solved by the invention of AC-LGAD. As illustrated in Fig.~\ref{fig:ACLGADPrinciple},  in AC-LGAD design, the gain layer is not segmented and a dielectric layer is deposited between the gain layer and segmented electrodes, enabling capacitive readout. The AC-LGAD technology is adopted by the next generation of collider experiments such as the EIC to construct Time-of-Flight detector. AC-LGAD offers great flexibility in the sensor design to achieve simultaneously good spatial and temporal resolutions, and is becoming a viable solution for 4D tracking. Using pixelated electrodes and epitaxial wafers of reduced thickness, a time resolution of 20 ps and a spatial resolution of about \SI{20}{\micro\meter} has been achieved~\cite{Dutta:2024ugh}. 

The AC-LGAD is clearly an attractive option for the H-NS TOF. The TOF concept is illustrated in Fig.~\ref{fig:TOF}. Detector layers are constructed with AC-LGAD as a cylinder in the barrel region and a disk in the end-cap region. Geometrical parameters of the H-NS TOF are summarized in Tab.~\ref{tab:table_LGAD_geometry}.

\begin{figure}[tbp]
\centerline
{\includegraphics[width=0.9\textwidth]{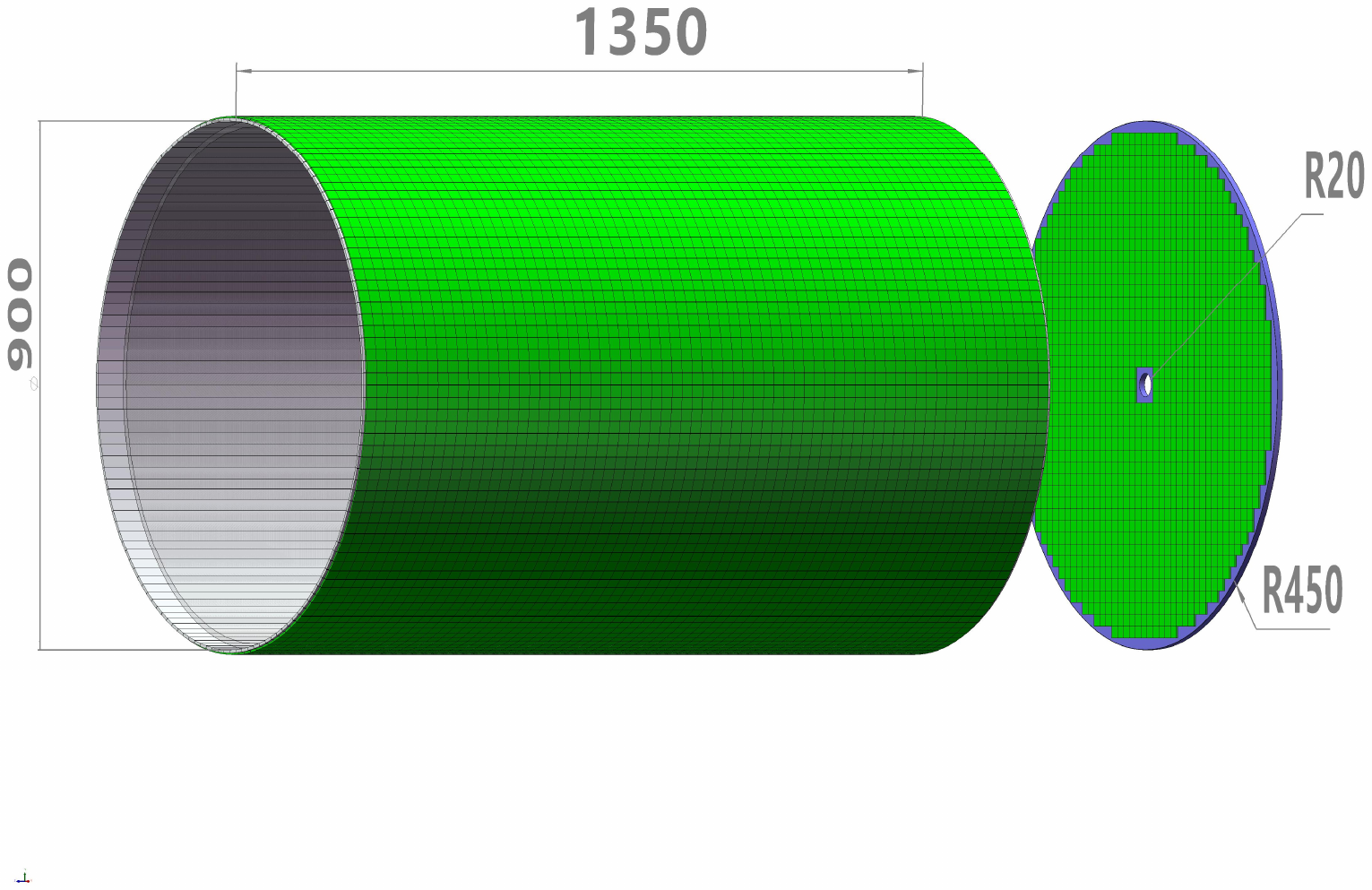}}
\caption{Illustration of the TOF concept.}
\label{fig:TOF}
\end{figure}

\begin{table}[!htb]
\begin{center}
\caption{Geometrical parameters of the H-NS TOF.}
\label{tab:table_LGAD_geometry}
\begin{tabular}{lccccc|ccccc}
\hline\hline
\multicolumn{1}{c}{ } &
\multicolumn{5}{c}{Barrel} &
\multicolumn{5}{c}{Forward} \\
\hline
  %& \multicolumn{5}{c|}{L0} & \multicolumn{5}{c}{D0} \\
\thead{Radius /  Z position (cm)}  & \multicolumn{5}{c|}{45} & \multicolumn{5}{c}{115} \\
\thead{Active area (cm$^{2}$)} & \multicolumn{5}{c|}{37520} & \multicolumn{5}{c}{5888} \\
%\thead{Nr. staves / ladders} & \multicolumn{5}{c|}{800} & \multicolumn{5}{c}{2386} \\
\thead{Nr. LGAD chips} & \multicolumn{5}{c|}{9380} & \multicolumn{5}{c}{1472} \\
\thead{Length / inner outer radius (cm)}   & \multicolumn{5}{c|}{136} & \multicolumn{5}{c}{2/45} \\
\thead{LGAD chip dimensions (mm$^{2}$)} & 
\multicolumn{10}{c}{400} \\
\thead{LGAD chip thickness (\SI{}{\micro\meter})} & 
\multicolumn{10}{c}{~300} \\
%\thead{Pixel size (\SI{}{\micro\meter}$^{2}$)} & 
%\multicolumn{10}{c}{300$\times$300} \\
%\thead{Nr. Pixels per chip} & 
%\multicolumn{10}{c}{4356} \\
\thead{Material budget ($X/X_{0}$)} & 
\multicolumn{10}{c}{$\leq$ 50\%} \\
\hline
\end{tabular}
\end{center}
\end{table}

\subsection{Nucleon polarimeter}

A measurement of the proton polarization usually utilizes the spin-dependent cross section for the proton-proton  ($pp$) or proton-carbon ($p\textrm{C}$) elastic scattering given their large analyzing powers. In the case of an unpolarized target, the differential cross section is expressed as~\cite{Bauer:2002zm, Bystricky:1976jr}: 
\begin{equation}
\frac{d\sigma}{d\phi d\!\cos\theta} =  \frac{1}{2\pi}\frac{d\sigma_0}{ d\!\cos\theta}\left[1+\mathcal{P}_yA_N\!(\theta)\cos\phi\right]
\end{equation}
Here, $\sigma_0$ represents the unpolarized cross section, $\theta$ and $\phi$ denote the polar and azimuthal angle of the scattered proton in the center-of-mass frame of the scattering. Figure~\ref{fig:principle_pol}(a) shows an illustration of the $p\textrm{C}$ scattering. $\mathcal{P}_y$ corresponds to the transverse proton polarization to be determined. $A_N(\theta)$ is the scattering angle-dependent analyzing power, which has been measured extensively~\cite{Betz:1966bon,Cheng:1967zzb, PhysRev.95.1348, Albrow:1970hn, Mcnaughton:1981ua, vonPrzewoski:1998ye, Bevington:1978zz, Besset:1980it, Onel:1989gt, Chamberlain:1957zz, Greeniaus:1979zn}.

\begin{figure}[tbp]
\centerline
{\includegraphics[width=7.0in]{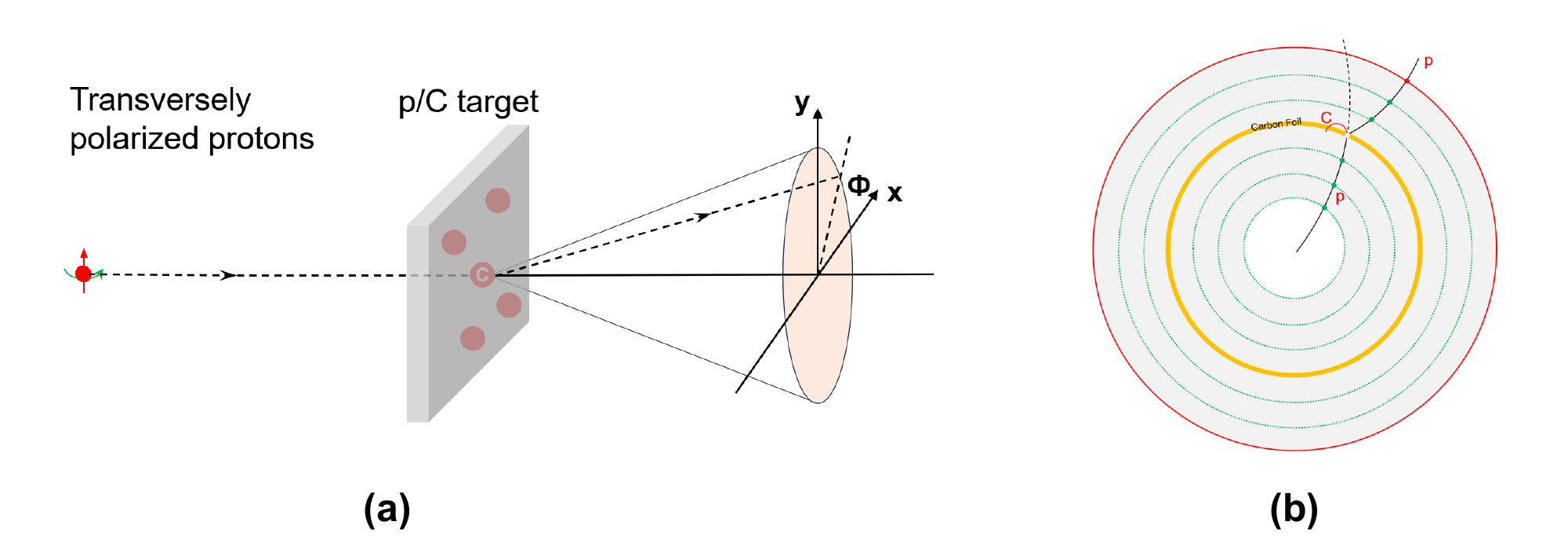}}
\caption{Principle of proton polarimeter.}
\label{fig:principle_pol}
\end{figure}

Proton polarimeters that use $pp$ or $p\textrm{C}$ scattering have been widely used, primarily as dedicated detectors to measure the polarization of the proton beam in the initial state or the polarization of the final state protons within a limited acceptance. For a large-acceptance general purpose spectrometer, a novel technique to measure the final-state proton polarization has been proposed~\cite{Liang:2025owx}. 
For H-NS, the tracking detector is composed of five layers of pixel tracker and one layer of LGAD which provides time and spatial measurements. This is ideal for integrating the nucleon polarimeter function, as illustrated in Fig.~\ref{fig:principle_pol}(b). In the design, a thin carbon (or polyethylene) layer is placed between the tracking detectors as a secondary target. The elastic scattering processes take place at the target, and the tracking detectors measure the momenta of the incident proton and scattered/recoil particles.

%\newpage
\subsection{Calorimeter}

The electromagnetic calorimeter (ECAL) system, placed at the endcap of H-NS, is designed for the detection of neutral particles including photons, neutral pions, and neutrons, and for distinguishing between neutrons and photons. The ECAL employs two different technologies to meet various detection requirements: the central region uses PbWO$_4$ crystals that provide superior energy and position resolutions, while the outer region uses lead-glass modules, offering a cost-effective solution. A dual-readout scheme combining lead glass and plastic scintillator is also under consideration as an outer region solution to further improve energy reconstruction performance. 

The ECAL is designed to provide ultra-high-precision measurements of the energetic $\gamma$-rays and to identify neutrons with moderate energy resolution. The design is similar to the design of the calorimeter~\cite{Somov:2025uiw} for $\eta$ physics in Hall-D at Jefferson Lab (JLab)~\cite{Somov:2024jiy} where PbWO$_4$ crystals are placed at the center surrounding by lead-glass blocks, as shown in Fig.~\ref{fig:ecal1}. The inner region of the ECAL consists of an array of PbWO$_4$ crystals with area of tentatively 70~$\times$~70~cm$^2$, while a hole of 4~$\times$~4~cm$^2$ at the center allows the beam passing through. Such type of hybrid ECAL have been successfully constructed and applied in several experiments at JLab, such as the HyCAL~\cite{Kubantsev:2006uf} used in the PrimEx~\cite{PrimEx-II:2020jwd} and PRAD~\cite{Xiong:2019umf} experiments which reached the energy resolution of $2\%/\sqrt{E~(\text{GeV})}$.

\begin{figure}[tbp]
\centerline
{\includegraphics[width=0.35\textwidth]{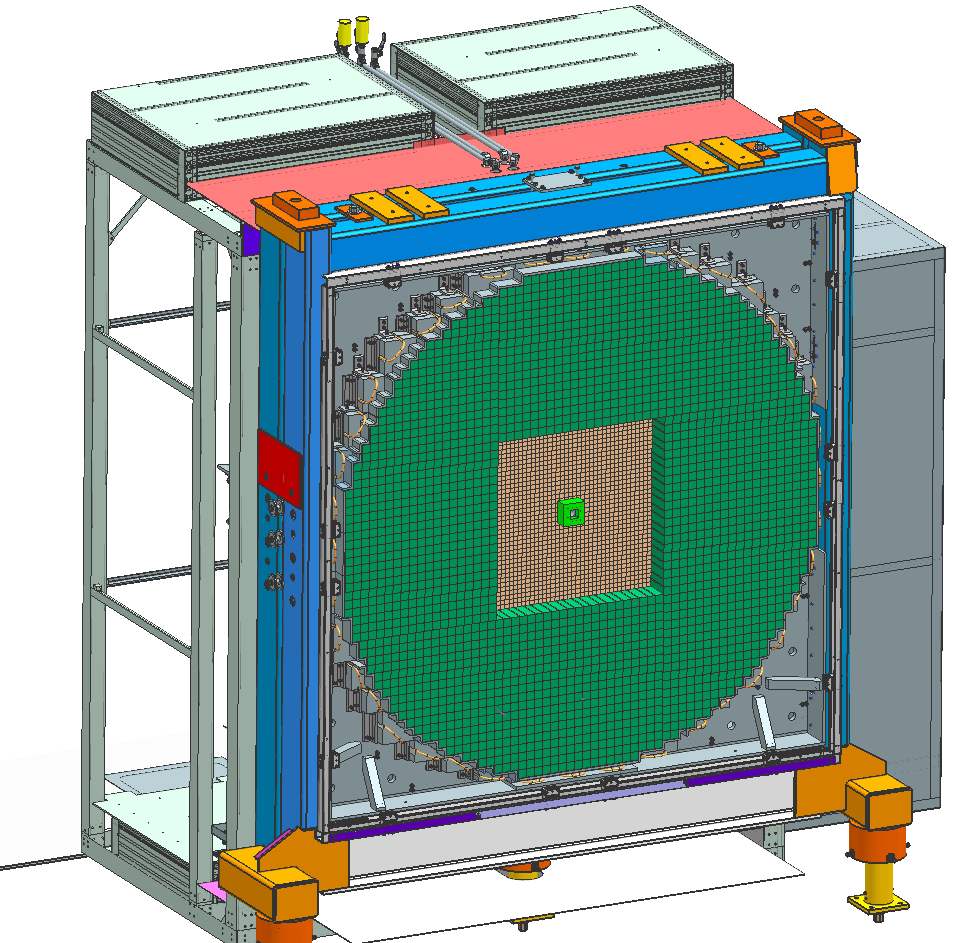}
}
\caption{The design of the lead-tungstate ECAL for $\eta$ physics at Jefferson Lab.}
\label{fig:ecal1}
\end{figure}

PbWO$_4$ crystals are widely used in modern ECALs that require both high energy resolution and fine granularity, thanks to its small Moli\`ere radius (2.19~cm) and short radiation length (0.89~cm). A typical PbWO$_4$ module has the block size of 2.05~$\times$~2.05~$\times$~20~cm$^3$ which can cover the electromagnetic showers generated by electrons and photons in a wide energy range of 0.1 to 10~GeV~\cite{Asaturyan:2021ese}. A total of 1225 PbWO$_4$ modules will be used to form a 35~$\times$~35 array, and each crystal will be directly attached to a 3/4-inch PMT. As an alternative, SiPM arrays may be used if the magnetic field of the H-NS spectrometer is too strong, which could degrade the PMT performance. The crystal array aims to achieve a high-precision energy resolution of 2–3\%/$\sqrt{E~(\text{GeV})}$, along with excellent position and timing accuracy for photon detection. In addition, the compact shower profile and superior position resolution of PbWO$_4$ are particularly beneficial for reconstructing high-momentum $\pi^0$ mesons, which are predominantly produced at small angles.

The outer region of ECAL is designed based on a lead-glass material, primarily motivated by cost considerations. Lead glass generates Cherenkov radiation, providing a fast timing response while effectively suppressing neutron-induced backgrounds. Furthermore, increasing the longitudinal granularity of the modules significantly improves photon collection efficiency by reducing light attenuation. This enhancement also enables accurate three-dimensional reconstruction of electromagnetic showers, which is crucial for particle-flow algorithms, while simultaneously increasing the effective photon collection area and improving signal quality. As a hybrid alternative, plastic scintillator layers can be inserted between lead-glass segments to enable the simultaneous readout of both Cherenkov and scintillation light. This dual-readout configuration enhances particle identification for electromagnetic showers. To compensate for the relatively low Cherenkov light yield, the photon readout via SiPMs employs a parallel connection of multiple devices within a single module, thereby increasing the effective collection area and improving signal quality.

\begin{figure}[tbp]
\centerline
{\includegraphics[width=0.6\textwidth]{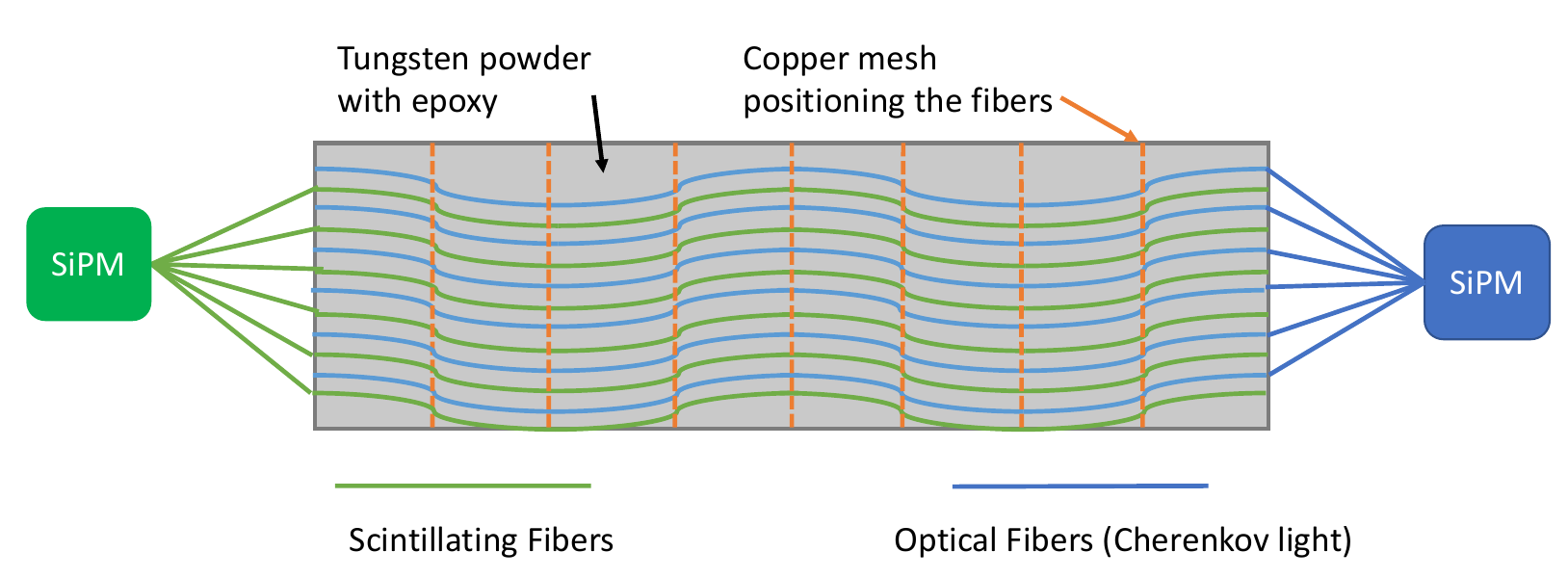} }
\caption{The schematic configuration of the SPACAL design.}
\label{fig:ecal2}
\end{figure}

The endcap of H-NS will also provide a platform for advanced calorimeter R\&D, for example, 
a Spaghetti-type calorimeter (SPACAL) design \cite{An:2022abg} as a HCAL for neutron detection.
The HCAL will employ a dual-readout design to measure Cherenkov and scintillation light simultaneously. Simulation studies show that for incident neutrons below 10~GeV, the energy resolution is highly sensitive to the longitudinal sampling ratio and its uniformity. To ensure a uniform and stable response, the baseline design will implements a SPACAL configuration, as shown in Fig.~\ref{fig:ecal2}. In this design, optical and scintillating fibers embedded in a tungsten-powder/epoxy-resin composite absorber generate the Cherenkov and scintillation light, respectively. Copper meshes provide precise fiber alignment. This structure delivers fine sampling, excellent uniformity, and superior compensation for hadronic shower fluctuations.

%\newpage
\subsection{Electronics and DAQ}

The block diagram of the readout electronics and the DAQ system for H-NS is shown in Fig.~\ref{Block_diagram_of_the_readout_electronics_system}. Each detector is equipped with dedicated front-end electronics to perform high-precision time, energy, or position measurements. After digitization, the data are preprocessed, buffered, packaged, and fed into the high-speed data interface, which are further sent to the DAQ system via optical fibers over long distance. At the same time, the readout electronics receive timing, control, and configuration information via shared optical link between the electronics and the DAQ. This timing information contains two types: one is the global system clock that all the front-end electronics are synchronized with, and the other one is the synchronization command, which is used to simultaneously reset the timestamp counters of all the front-end electronics nodes. Considering that high-precision time measurement is required for the AC-LGAD detector, this global clock features low jitter. The monitoring functionality is also essential for the readout electronics, including temperature, humidity, and current, which are gathered and transferred via the DAQ to the detector control system.

\begin{figure}[tbp]
\centerline{\includegraphics[width=6.0in]{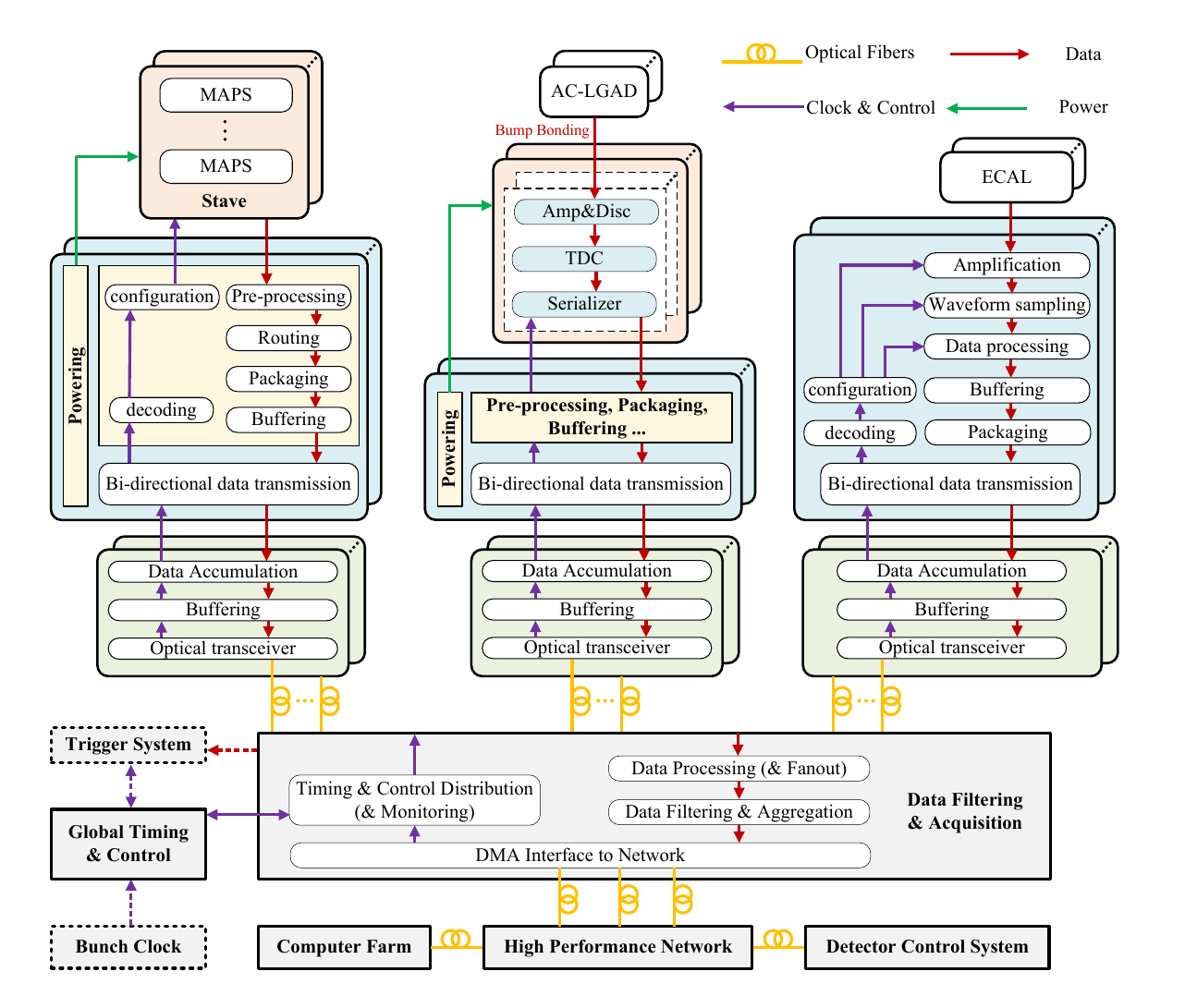}}
\caption{Block diagram of the readout electronics system.}
\label{Block_diagram_of_the_readout_electronics_system}
\end{figure}

For the MAPS-based tracker, the front-end measurement circuits are directly integrated with the sensors into one identical chip. MAPS chips are assembled into staves. The MAPS chips on these staves residing in different layers would exhibit two different data output modes: one is that each MAPS chip would output an individual serial data stream, and the other is that multiple MAPS chips on the same stave will be linked together and output one common serial data stream. Therefore, as for the readout electronics, these two modes are accommodated, to achieve which programmable flexible data routing is integrated in the front end. After this routing, the input data are further preprocessed, packaged, buffered, and finally fed into the data interface. To validate the functionality of the MAPS chips, a series of registers should be configured properly according to the configuration data received through the aforementioned downstream link. The electronics also provide the power supply to the staves, with functionality of automatic overcurrent protection and abnormal status monitoring.

For the LGAD based TOF detector, the sensor and the readout ASIC are bump bonded together, and mounted on the front-end modules. To achieve high-resolution time measurement of the weak signals generated by LGAD sensors, the first stage of the readout ASIC is a low-noise preamplifier, followed by a high-speed discriminator that generates a digital pulse -- its leading edge represents the time of arrival (TOA) and its width (time-over-threshold, TOT) corresponds to the input charge. With this TOT, time-walk correction can be performed to improve the time resolution. The time information is finally digitized by a time-to-digital converter (TDC) and output through a serial interface to the readout modules nearby the detector. In this readout ASIC, the preamplification, discrimination, and TDC stages feature low power performance. Besides, the readout modules also feature high density and are responsible for preprocessing, packaging, and buffering the data streams from multiple ASICs before forwarding them to the DAQ. To ensure overall time measurement performance, a low-jitter reference clock is recovered from the downstream coming from the DAQ.

For the calorimeter detector, the detector signals from $\mathrm{PbWO_4}$ crystals are captured by PMTs and then fed to the readout electronics. Waveform digitization is required on the electronics. Preamplifiers are used to adjust the input signal amplitudes to match the input dynamic range of the subsequent digitization circuits. To guarantee the measurement performance, high-resolution analog-to-digital conversion circuits with enough sampling speed are employed. As for the output of the readout electronics, two types of data are considered: one is the full digitized waveform of the signal and the other is the real-time charge and time extracted by on-board algorithms. The final choice would depend on the demand of the experiment.
To reduce the system complexity and to enhance the system stability, the common functional circuits in the readout electronics, such as preprocessing, routing, buffering, controlling, and high speed data transfer interface, are considered to be integrated in customized ASICs.

As for the DAQ, Fig.~\ref{Block_diagram_of_the_readout_electronics_system} illustrates the implementation of bidirectional communication with the readout electronics, facilitated by fiber optic links through the Data Filtering \& Acquisition (DFA) system. The DFA serves as both a hardware-based data filter and a custom hardware interface for the DAQ software. Incoming upstream data are first aligned and submitted to processing, which may include operations like data summation to reduce granularity for potential hardware-based trigger functionality, as indicated by the dashed line. In a trigger-based readout configuration, the Global Timing \& Control (GTC) system issues a trigger signal for trigger match. The selected data are then aggregated and transferred over a high-performance network to the computer farm, utilizing direct memory access (DMA) via PCIe or remote DMA (RDMA) over Ethernet. The DFA hardware may be implemented in PCIe or MicroTCA platform~\cite{Anderson:2016lfn,Chen:2019owc}. The final choice of the system architecture depends on factors such as the presence of a trigger system.

The system clock and the synchronization signals come from the GTC and are fed to the DFA, and further fanned out to all the front ends. Additionally, the DFA also functions as a bridge between part of the DCS and the readout electronics, enabling electronics control and monitoring. The DFA firmware therefore supports dedicated communication streams for these control functions.

\subsection{Detector control system}

The Detector Control System (DCS) is a critical component that ensures the safe and stable operation of large-scale experimental detectors. It provides real-time monitoring and precise control of essential physical parameters that influence detector performance and status, such as high-voltage and low-voltage power supplies, gas system flow rates, and environmental temperatures \cite{LI2025103345}.
The DCS is designed to guarantee operational safety and equipment protection through automated control and safety interlock mechanisms, thereby preventing potential damage resulting from parameter anomalies. High scalability is essential when designing this system to accommodate possible phased detector upgrades, a distributed architecture to eliminate single points of failure, and an intuitive user interface that promptly notifies control room operators of any issues \cite{MejiaCamacho:2023zam,Adam:2021bvk}.

To achieve these objectives, the DCS will use the Experimental Physics and Industrial Control System (EPICS 7) software framework, a widely used open-source platform for the distributed control of large scientific instruments. EPICS was chosen as the foundation of the system, valued for its scalability, a strong and active community, and a commitment to continuous maintenance \footnote{https://epics-controls.org/}.
Building on EPICS, the system employs a hybrid data storage architecture that combines a relational database (such as MySQL or PostgreSQL) for structured data, including system configuration and calibration parameters, with a time-series database for real-time equipment monitoring data \footnote{https://www.mysql.com/, https://www.postgresql.org/} \cite{You_2025}.
For user interaction, the system offers client applications based on Phoebus (a graphical platform for control system users) and lightweight web interfaces, providing flexible, convenient, and powerful access for operators and equipment experts[1]. Moreover, combining JSROOT with the web-based platform, this system enables real-time online analysis of scientific data. The architectural framework for the H-NS detector control system is illustrated in Fig.~\ref{fig:DCS-framework}.

\begin{figure}[h]
\centering
\includegraphics[width=0.7\linewidth]{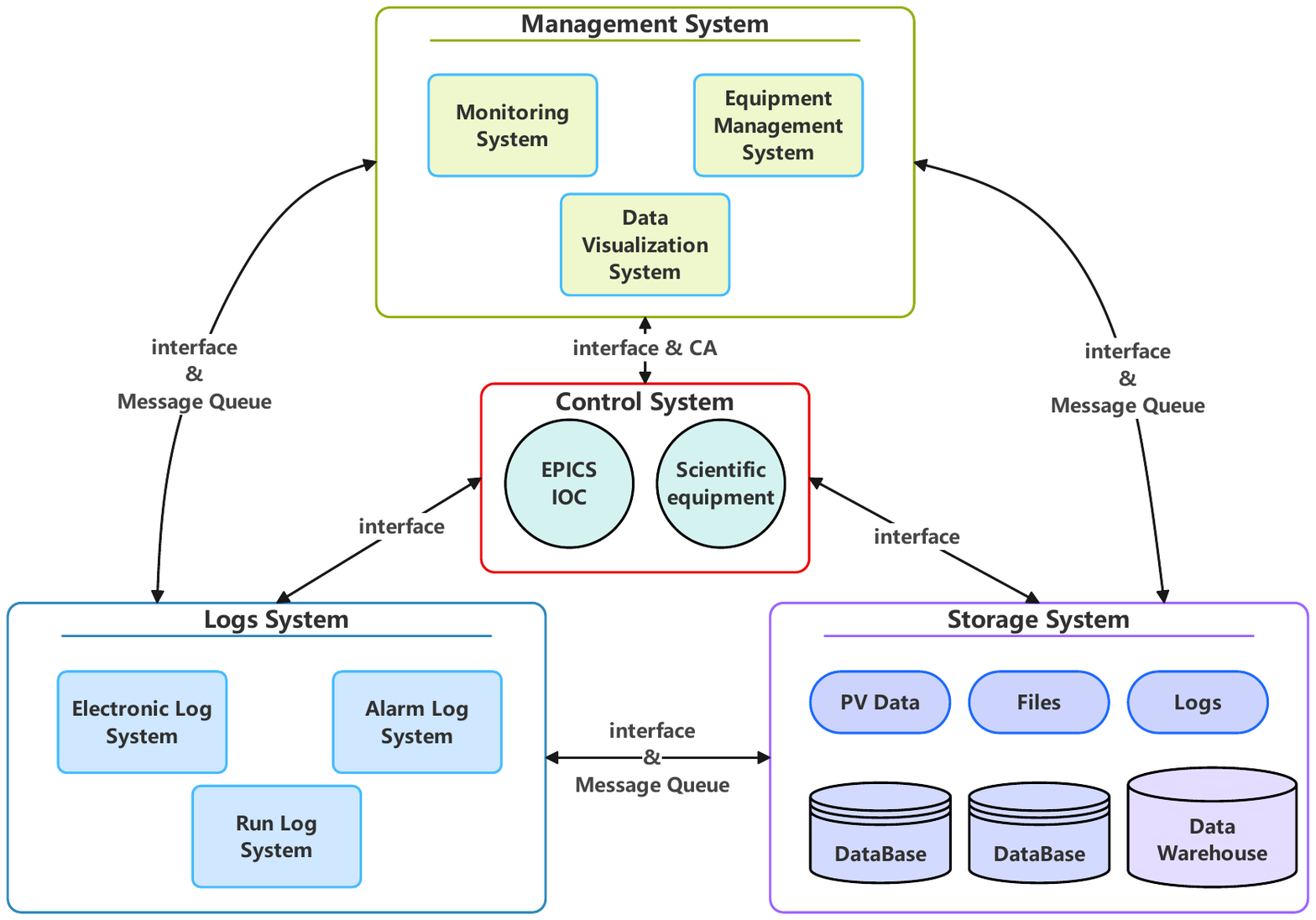}
\caption{\label{fig:DCS-framework}Architectural framework for the H-NS detector control system.}
\end{figure}

\section{HIAF complex}

The HIAF represents a next-generation platform well suited to unravel the $\Lambda$ polarization mystery. Its unique capability to deliver high-intensity proton beams with kinetic energies up to 9.3 GeV (and up to 32 GeV with the HIAF-Upgrade) covers a crucial and relatively unexplored region of nucleon-nucleon center-of-mass energy (from near-threshold, $\sqrt{s}\simeq$ 2.6 GeV, up to $\sqrt{s}\simeq$ 8 GeV). This energy range bridges two fundamental regimes of hyperon production: Near-threshold hadronic regime where dynamics are dominated by hadronic degrees of freedom and resonance excitations, and intermediate-energy QCD regime where quark-gluon interactions become increasingly important.
A key advantage of the HIAF energy domain is the prolific production of $\Lambda$ hyperons with high $x_{\rm F}$, a kinematic region where the polarization effect is known to be most pronounced, often reaching 30--40\%.

HIAF project is managed by Institute of Modern Physics, Chinese Academy of Sciences, and the construction was started on December, 2018 in Huizhou City of Guangdong Province. The main feature of this facility is to provide high intensity heavy ion beam pulse for various experiments. It is designed to provide intense primary heavy ion beams for nuclear and atomic physics, as well as other application fields. As shown in Fig.~\ref{fig:HIAF_layout}, the accelerator mainly consists of a superconducting electron-cyclotron-resonance (SECR) ion source, a continuous wave (CW) superconducting ion linac (iLinac), a booster synchrotron (BRing) and a high precision spectrometer ring (SRing). A fragment separator (HFRS) is employed as the beamline connecting the BRing and SRing, enabling the efficient separation, purification, and identification of radioactive isotopes. Typical beam parameters of HIAF are presented in Tab.~\ref{tab:HIAF_beam_parameters}. To mitigate space-charge and dynamic vacuum effects, stored ions are rapidly accelerated to high energies using fast-ramping operation at a repetition rate of 3 Hz. With both fast and slow extraction modes available in the BRing, beams can be delivered to various experimental terminals to meet different requirements, such as H-NS in the High Energy Multi-disciplinary Terminal. 
In August 2025, construction and installation of the accelerator facility were fully completed. On 28 October 2025, the beam was successfully delivered to the experimental terminals, marking the first complete and successful beam commissioning of the facility.

\begin{figure}[h]
\centering
\includegraphics[width=0.9\linewidth]{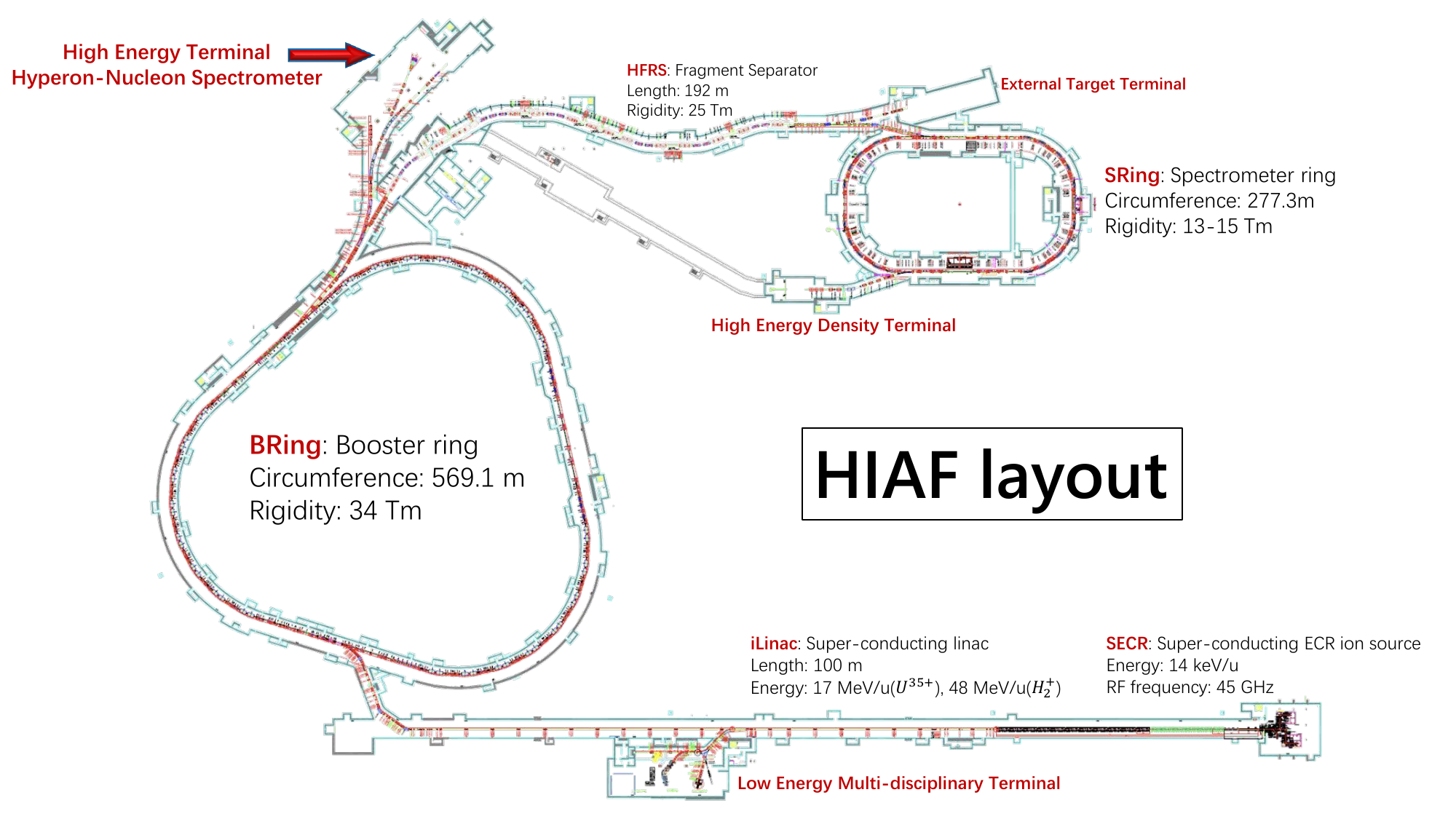}
\caption{\label{fig:HIAF_layout}Layout of HIAF complex.}
\end{figure}

\begin{table}[htb]
     \begin{center}
     \begin{tabular}{c|c|c}
     \hline\hline
       Ion species  & Kinetic Energy (GeV/u)   &  Beam intensity (ppp)    \\
     \hline
        p & 9.3  & $6.0\times 10^{12}$  \\
        $^{16}$O$^{8+} $ & 4.25  & $6.0\times 10^{11}$  \\
        $^{18}$O$^{6+} $ & 2.6  & $6.0\times 10^{11}$  \\
        $^{78}$Kr$^{19+} $ & 1.7  & $3.0\times 10^{11}$  \\
        $^{209}$Bi$^{31+} $ & 0.85  & $1.2\times 10^{11}$  \\
        $^{238}$U$^{35+} $ & 0.835  & $1.0\times 10^{11}$  \\
        $^{238}$U$^{78+} $ & 2.6  & $1.0\times 10^{10}$  \\
     \hline
     \end{tabular}
     \end{center}
     \caption{Typical beam parameters from BRing.}
     \label{tab:HIAF_beam_parameters}
\end{table}

\end{chapter}

\begin{chapter}{Simulation}

\section{The H-NS software framework: HnsRoot}
\label{sec:software_framework}

HnsRoot is an object-oriented software framework developed specifically for the simulation, reconstruction, and data analysis of the H-NS detector. Built upon the FairRoot framework~\cite{Al-Turany:2012zfk}---which provides a general-purpose foundation for detector simulations---HnsRoot incorporates specialized components tailored to the H-NS experimental program. At its core, the Run Manager orchestrates simulation initialization, event processing loops, and I/O operations via the IO Manager, using a runtime database for efficient parameter handling. Base classes are provided for key functionalities including detector geometry, magnetic field configuration, event generation, and analysis tasks, enabling modularity and extensibility for physics-specific implementations. The framework supports comprehensive detector simulation through two complementary approaches: a full GEANT4-based~\cite{GEANT4:2002zbu} simulation modeling detailed particle interactions with materials and magnetic fields, where detector geometry is defined in \textsc{ROOT} or GDML~\cite{Chytracek:2006be} formats; and a fast simulation package based on parameterizations derived from full simulations, offering significant speed-up for large-scale studies while preserving reasonable accuracy. 

For physics performance evaluation, HnsRoot integrates multiple event generators: the Pluto generator~\cite{Frohlich:2007bi} simulates $pp \to pK^+\Lambda$ process targeting $\Lambda$ production and polarization studies; GiBUU~\cite{Buss:2011mx} and \textsc{JAM2}~\cite{Nara:2019crj} handle general hadronic production in $pp \to X$ reactions; and \textsc{JAM2} also supports heavy-ion collisions ($AA \to X$) to investigate $\Lambda$ global polarization and vector meson spin alignment.
To investigate the polarization sensitivity of $\Lambda$ and $p$, we simulated 20 million events each of $pp \to pK^+\Lambda$, $pp \to X$, and $AA \to X$. The results indicate that these simulated samples are adequate for estimating the polarization uncertainty.

\section{The detector geometry and optimization}

The H-NS spectrometer employs a cylindrical tracking system with five concentric MAPS layers, complemented by five forward disks along the beam axis for comprehensive vertex detection. Particle identification is achieved through a barrel LGAD layer and a forward TOF disk, both providing \SI{100}{\micro\meter} spatial and 30~ps timing resolution.  This integrated design optimizes hyperon decay reconstruction with minimal material budget in critical tracking regions. The detector geometry of the H-NS has undergone systematic optimization to maximize physics performance while maintaining technical feasibility and cost-effectiveness. 

\label{sec:geometry_optimization}

\subsection{The strategy of optimization}
This optimization process focused on critical parameters that directly impact tracking efficiency, particle identification (PID) capabilities, and vertex resolution—all essential for achieving the primary physics objectives of measuring nucleon polarization through nucleon-nucleon elastic scattering and hyperon polarization via weak decay angular distribution analysis. The comprehensive study balanced competing requirements across the entire beam energy range from 3.5 GeV to 9.3 GeV to identify the optimal configuration.

\subsection{Fixed geometry parameters and layer specification}
Several detector parameters were predetermined based on fundamental physics requirements and technical constraints. 
The tracking system comprises five concentric layers of MAPS for precise vertexing and tracking, followed by one layer of LGAD for precise timing and PID. 
% The key parameters for each layer are summarized in Tab.~\ref{tab:detector_layers}. 
All detection layers were uniformly distributed to ensure optimal accuracy for proton polarization measurements. The innermost layer radius was established at 5 cm from the beam axis to maximize vertex resolution capabilities, providing a solid foundation for the subsequent optimization of variable parameters.

\subsection{Geometric optimization}
The profile of H-NS detector is shown in the Fig.~\ref{fig:det_opti}. The optimization study investigated four primary geometric parameters: the outer radius of the barrel LGAD detector ($R_{\rm o}$) varied from 30 cm to 70 cm, the total detector length ($L$) calculated as three times $R_{\rm o}$ (ranging from 90 cm to 210 cm), the ratio of barrel length to total detector length ($R_{\rm b}$) tested at values of 0.2, 0.33, 0.4, and 0.5, and magnet configuration comparing solenoid ($B_z = 1.5$ T) versus dipole ($B_y = 1.5$ T) options. A comprehensive simulation campaign encompassed 20 distinct detector configurations for both magnet types, generating 40 simulated samples with 2 million single tracks each. Performance evaluation utilized Pluto-generated samples of $pp \rightarrow pK^+\Lambda(\rightarrow p\pi^-)$ events at beam energies of 3.5 GeV and 9.3 GeV to ensure representative physics coverage.

\begin{figure}[htb]
\centering
\includegraphics[width=0.8\linewidth]{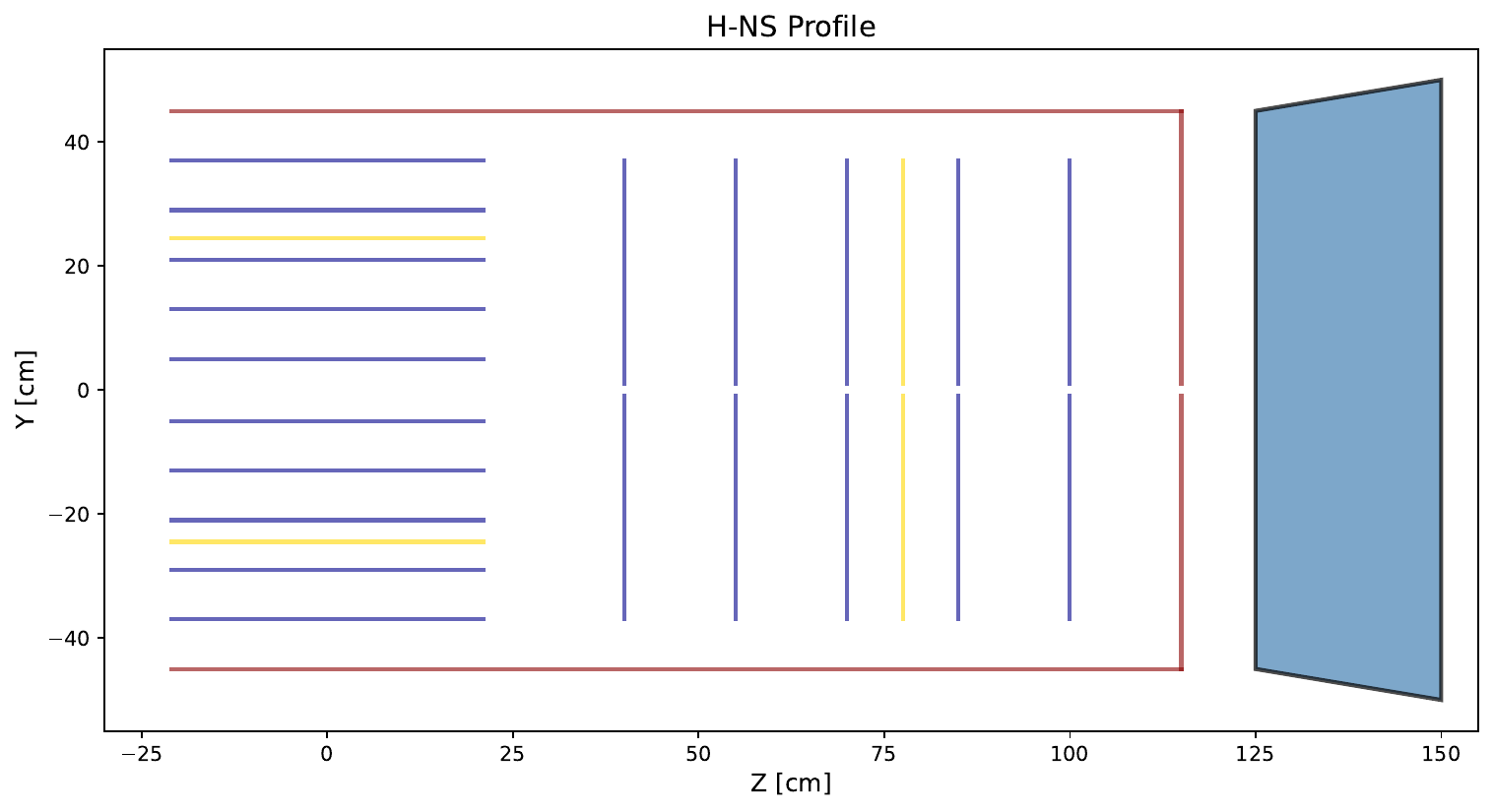}
\caption{The profile of H-NS detector.}
\label{fig:det_opti}
\end{figure}

\subsection{Performance evaluation across geometric parameters}
Tracking efficiency demonstrated a significant dependence on both beam energy and geometrical configuration as shown in~Fig.~\ref{fig:det_opti_eff}. At 3.5 GeV, efficiency increased with larger $R_{\rm b}$ values due to improved geometric acceptance but decreased with larger $R_{\rm o}$ values as low-momentum tracks failed to penetrate sufficient detection layers, reaching maximum efficiency of approximately 60\% for optimal configurations. At 9.3 GeV, efficiency increased with both larger $R_{\rm b}$ and $R_{\rm o}$ values, with larger $R_{\rm o}$ providing better coverage for tracks exhibiting longer $\Lambda$ decay lengths, showing distinctly different behavior from the 3.5 GeV case due to altered kinematic conditions. PID efficiency exhibited contrasting trends, showing negative correlation with $R_{\rm b}$ due to increased material budget degrading momentum resolution, while the correlation with $R_{\rm o}$ was energy-dependent—decreasing at 3.5 GeV as low-momentum tracks failed to reach timing detectors but increasing at 9.3 GeV due to improved momentum resolution and longer flight paths. 

\begin{figure}[htb]
\centering
\includegraphics[width=0.45\linewidth]{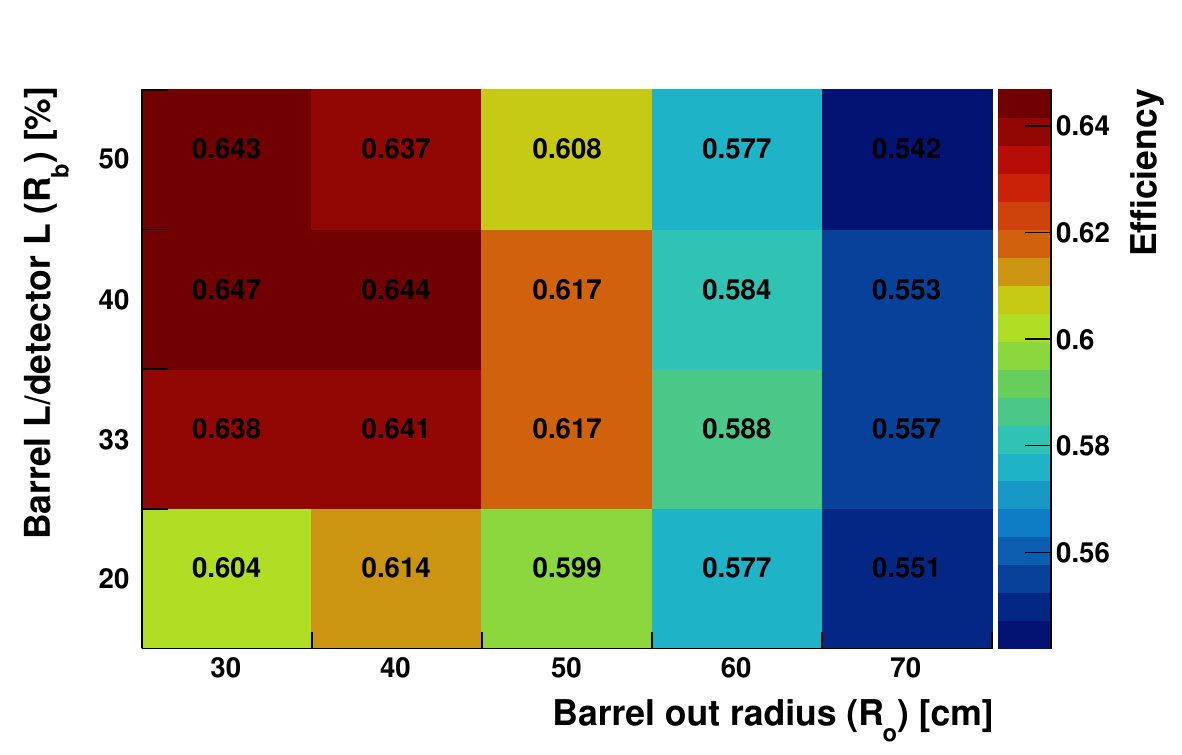}
\includegraphics[width=0.45\linewidth]{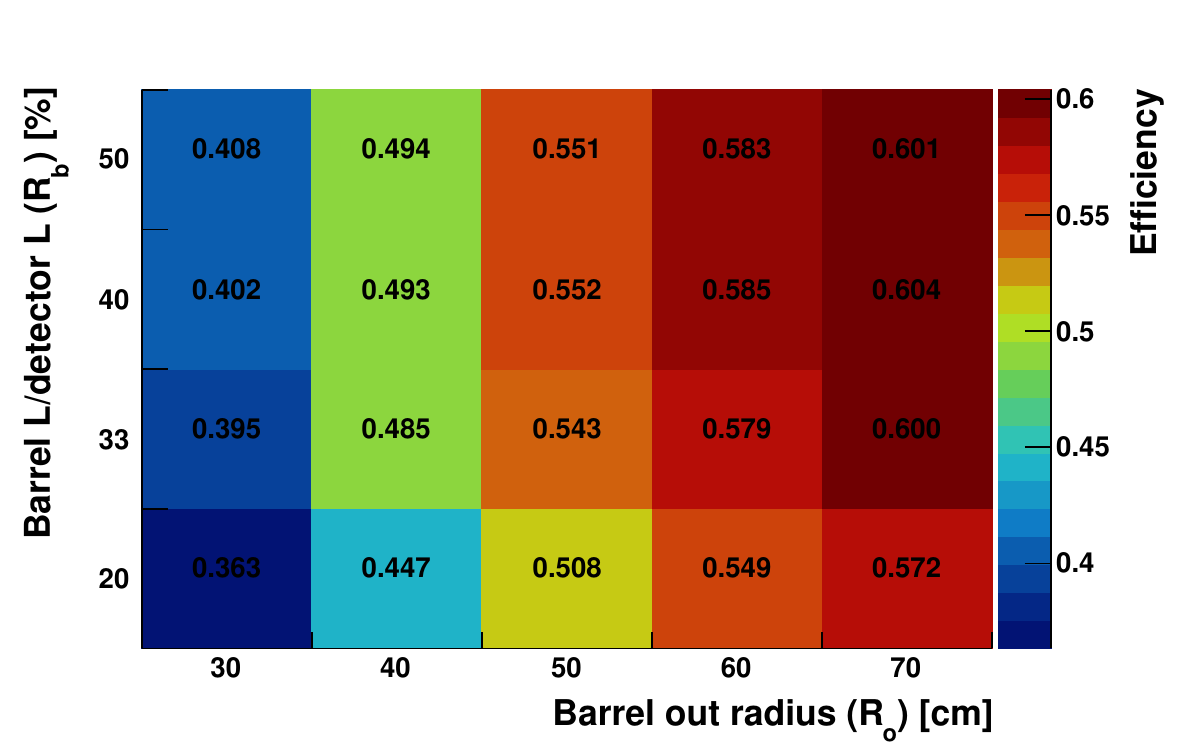}
\caption{Detection efficiency as functions of geometrical parameters. Left: Efficiencies from beam energy of 3.5 GeV. Right: Efficiencies from beam energy of 9.3 GeV.}
\label{fig:det_opti_eff}
\end{figure}

As demonstrated in the Fig.~\ref{fig:det_opti_reso}, the $\Lambda$ mass reconstruction quality consistently worsened with larger $R_{\rm b}$ values owing to increased multiple scattering from additional material, while improving with larger $R_{\rm o}$ values benefiting from the longer lever arm for enhanced track reconstruction precision, typically achieving resolutions between 1--4 MeV across configurations and beam energies.

\begin{figure}[htb]
\centering
\includegraphics[width=0.45\linewidth]{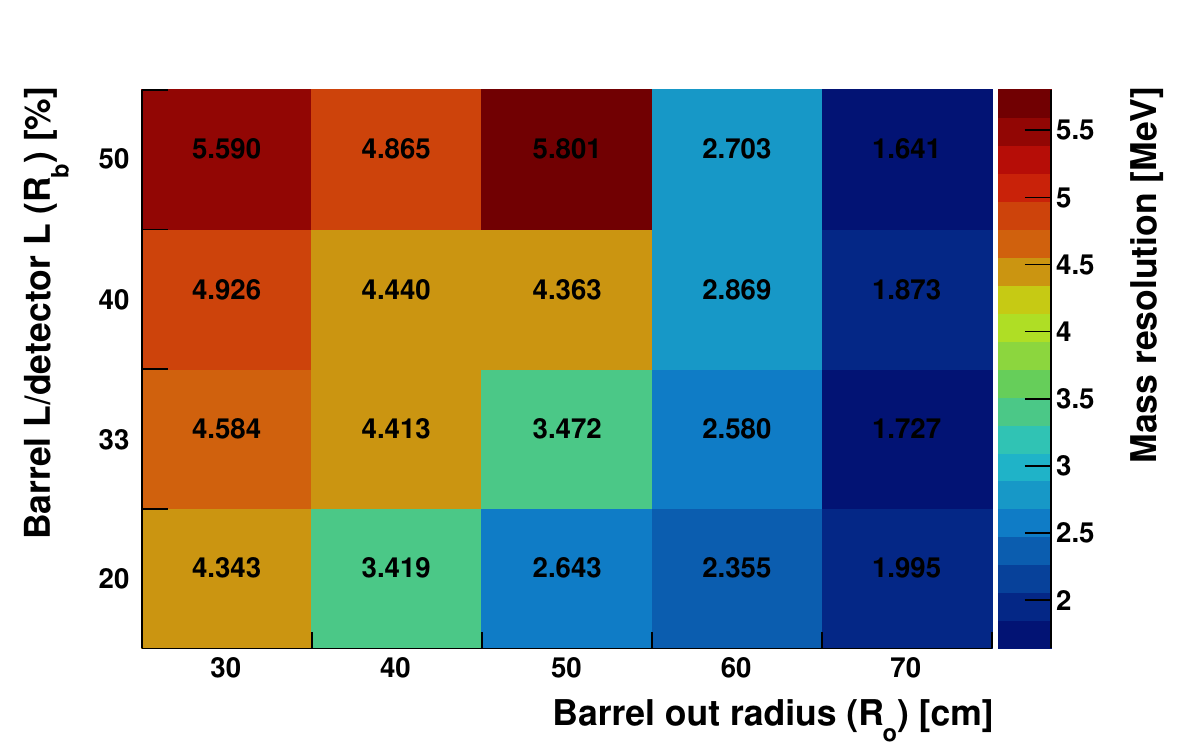}
\includegraphics[width=0.45\linewidth]{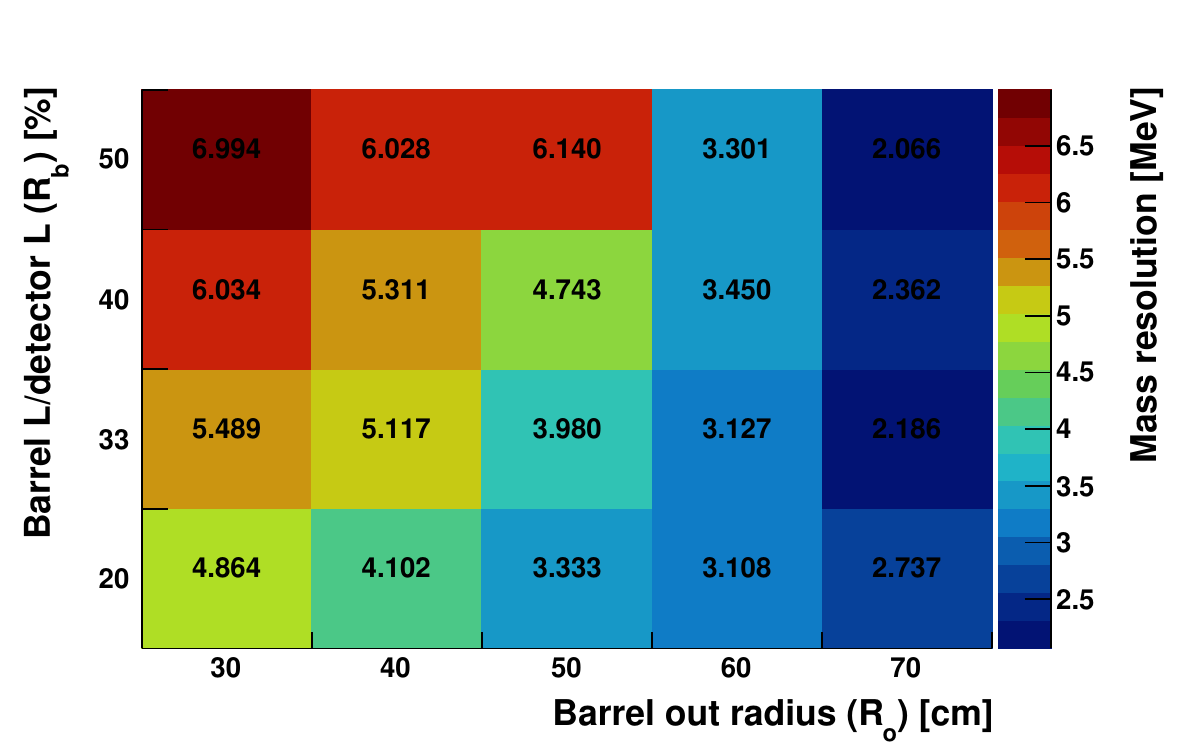}
\caption{$\Lambda$ mass resolution as functions of geometrical parameters. Left: resolutions from beam energy of 3.5 GeV. Right: resolutions from beam energy of 9.3 GeV.}
\label{fig:det_opti_reso}
\end{figure}

\subsection{Magnet configuration analysis}
The comparison between solenoid and dipole magnet configurations revealed distinct trade-offs, as detailed in Tab.~\ref{tab:magnet_comparison}. Solenoid magnets offer axial symmetry ($\phi$-symmetric), moderately higher tracking efficiency, and better technical feasibility, while also featuring simplified mechanical design and fabrication. However, they demonstrate inferior $\Lambda$ mass resolution for small-angle tracks compared to dipoles. Dipole magnets provide superior $\Lambda$ mass resolution for small-angle tracks by a factor of two improvement but suffer from $\phi$-asymmetry, reduced efficiency for low-$P_z$ tracks, and more complex engineering design challenges. The symmetry properties of the solenoid configuration better align with the physics processes under investigation, while the dipole alternative presents specific performance advantages offset by significant technical complications.

\newcolumntype{L}[1]{>{\raggedright\let\newline\\\arraybackslash\hspace{0pt}}m{#1}}

\begin{table}[htbp]
    \centering
    \caption{Comparison of solenoid and dipole magnet configurations}
    \label{tab:magnet_comparison}
    \small % 
    \begin{tabular}{c L{0.4\linewidth} L{0.4\linewidth}}
        \toprule
        & \textbf{Solenoid Configuration} & \textbf{Dipole Configuration} \\
        \midrule

        \multirow{4}{*}{\rotatebox[origin=c]{0}{\textbf{Adv.}}}
        & \textbf{Axial symmetry} ($\phi$-symmetric) 
        & \textbf{Superior mass resolution} for small-angle tracks (a factor of two improvement ) \\
        \cmidrule(r){2-2} \cmidrule(l){3-3}
        & Moderately higher tracking efficiency
        & -- \\
        \cmidrule(r){2-2} \cmidrule(l){3-3}
        & Simplified mechanical design and fabrication 
        & -- \\
        %\cmidrule(r){2-2} \cmidrule(l){3-3}
        %& Better technical feasibility 
        %& -- \\
        
        \midrule % 

        \multirow{3}{*}{\rotatebox[origin=c]{0}{\textbf{Disadv.}}}
        & Inferior mass resolution for small-angle tracks (vs. dipole) 
        & Asymmetric in $\phi$ \\
        \cmidrule(r){2-2} \cmidrule(l){3-3}
        & -- 
        & Reduced efficiency for low-$P_z$ tracks \\
        \cmidrule(r){2-2} \cmidrule(l){3-3}
        & -- 
        & Complex engineering design challenges \\
        \bottomrule
    \end{tabular}
\end{table}

\subsection{Optimized geometry selection}

Based on a comprehensive optimization, the selected geometry---inner radius $R_{\rm i} = \SI{5}{cm}$, outer radius $R_{\rm o} = \SI{50}{cm}$, length $L = \SI{150}{cm}$, and barrel ratio $R_{\rm b} = 0.33$---optimally balances tracking efficiency over the beam energy range, PID performance, mass resolution, and technical feasibility. $R_{\rm o} = \SI{50}{cm}$ balances transverse momentum acceptance for low-energy beams against $\Lambda$ decay length constraints at high energies; $R_{\rm b} = 33\%$ trades off tracking efficiency (favors larger $R_{\rm b}$) against momentum resolution (favors smaller $R_{\rm b}$). Performance strongly depends on beam energy, and the chosen solenoid configuration---despite slightly worse small-angle $\Lambda$ mass resolution than a dipole---is preferred for its axial symmetry and engineering simplicity, matching the symmetry of target physics processes. These results form a robust foundation for H-NS detector design with well-characterized performance.

\section{The performance study for the H-NS detector}
The performance studies were carried out within the HnsRoot software framework. The geometry was implemented in {\sc Geant4}~\cite{GEANT4:2002zbu} and the charged particles (e.g. pions, kaons and protons) are generated from nominal IP and cover the entire detector acceptance by multiple generators provided by the HnsRoot. The magnetic field is produced by a superconducting solenoid. The interaction between the generated particles and the detector is handled by {\sc Geant4}, in which the multiple-scattering effect and energy loss of the track are taken into account. A hit is defined as the position where the particle enters the active area of the detector. Then, these hit positions are smeared according to the detector resolution. All the hits belonging to one track are selected according to the truth-track information and used for track fitting algorithm with the Kalman filtering method~\cite{Rauch:2014wta}.

\subsection{The momentum resolution}
After the tracking fitting, the standard deviation of the $dp/p = (|\overrightarrow{p}_{\rm truth}| - |\overrightarrow{p}_{\rm reco}|) / |\overrightarrow{p}_{\rm truth}|$ can be measured and defined as the momentum resolution, where $|\overrightarrow{p}_{\rm truth}|$ and $|\overrightarrow{p}_{\rm reco}|$ are the generated and reconstructed absolute values of the particle momentum, respectively. Figure~\ref{fig:reso_p} (left) shows the momentum resolution as a function of momentum for charged pions, kaons, and protons in the pseudorapidity range $0.0 < \eta < 0.5$.  The multiple-scattering effect is more pronounced for protons and kaons below 2 GeV/$c$, therefore, worsening the resolution significantly. With the increase of the momentum, the $dp/p$ rises almost linearly. This phenomenon is understandable because the measured sagitta will be decreased for the stiffer tracks (due to the higher momentum). Figure~\ref{fig:reso_p} (right) shows the momentum-resolution results as a function of pseudorapidity in the momentum range $p = 1$ GeV/$c$. The momentum resolution is approximately constant when $\eta < 2$, and then it rises quickly. 
%In addition, the pion momentum resolution is systematically worse than other particles in most of the studied range because of its smaller mass.  
Overall, the performance is very similar for all kinds of particles. 

\begin{figure}[htb]
\centering
\includegraphics[width=0.45\linewidth]{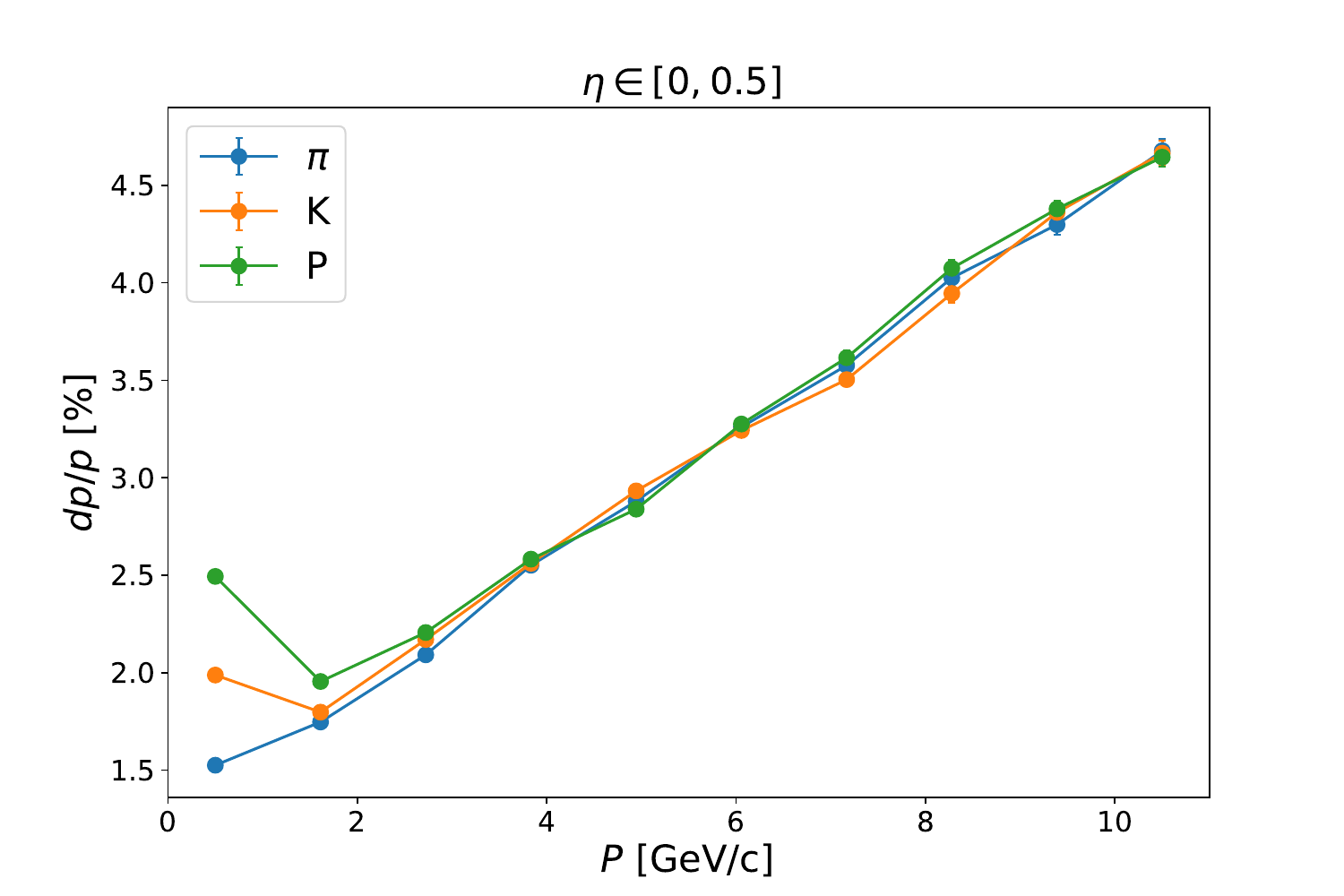}
\includegraphics[width=0.45\linewidth]{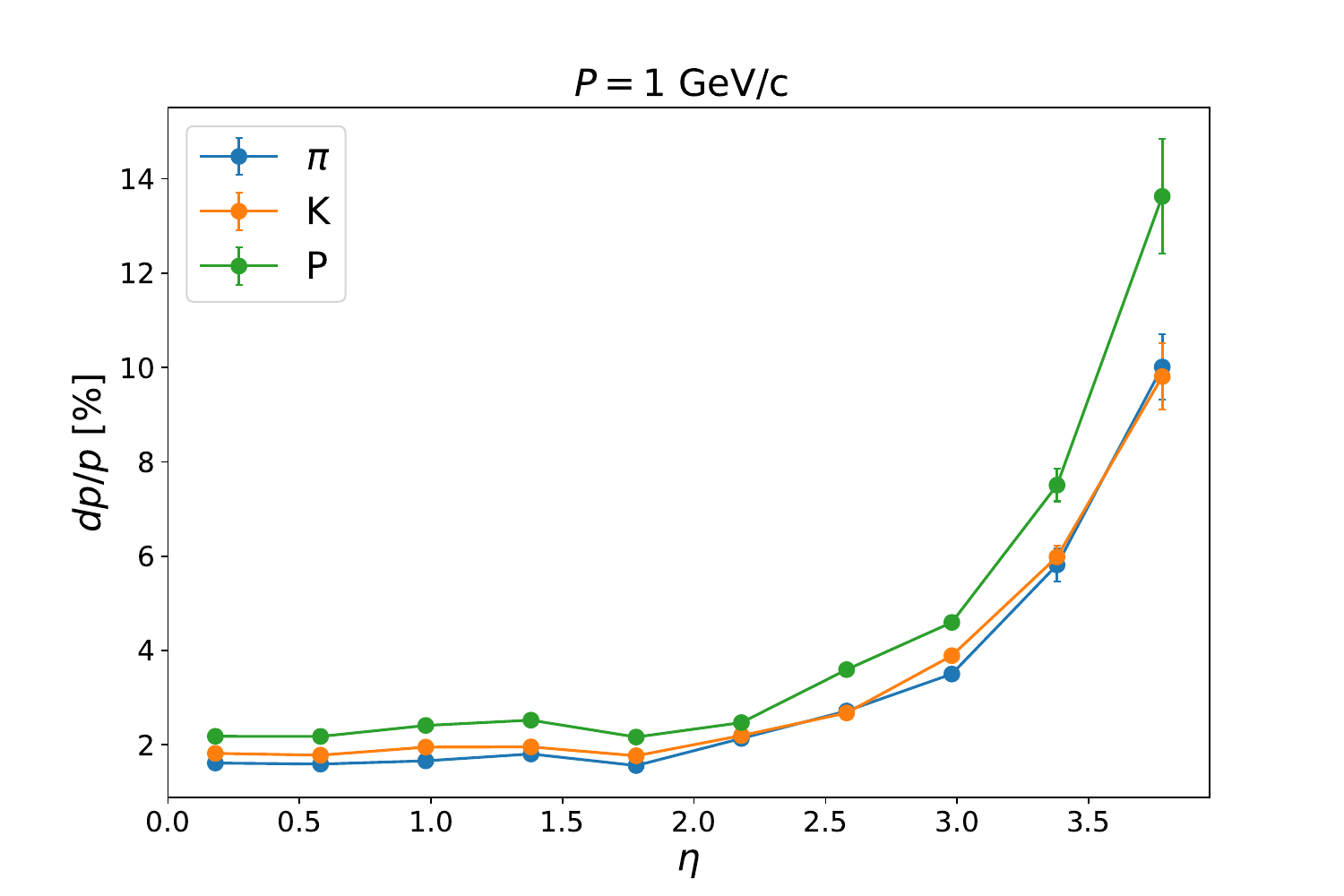}
\caption{Momentum resolution for different particles in the 1.5 T magnetic field. Left: $dp/p$ as a function of momentum in the $0 < \eta < 0.5$ range. Right: $dp/p$ as a function of pseudorapidity in the $p = 1~{\rm GeV}/ c$ range.}
\label{fig:reso_p}
\end{figure}

\subsection{The vertex and angular resolution}
The tracking system is designed not only to measure the momentum of the charged particles but also to determine the primary vertex for an event as well as the secondary vertex for the long-life particles decay. This capability is crucial for background suppression and therefore improves the sensitivity significantly for the process with long-life particles. The spatial resolution can be studied by measuring the Distance of the Closest Approach (DCA), which is defined as the spatial separation between the primary vertex and the reconstructed track projected to the $z$-axis ($DCA_z$) or to the $xy$-plane ($DCA_{r\phi}$). The DCA resolutions are shown in Fig.~\ref{fig:reso_dca}.

\begin{figure}[htb]
\centering
\includegraphics[width=0.45\linewidth]{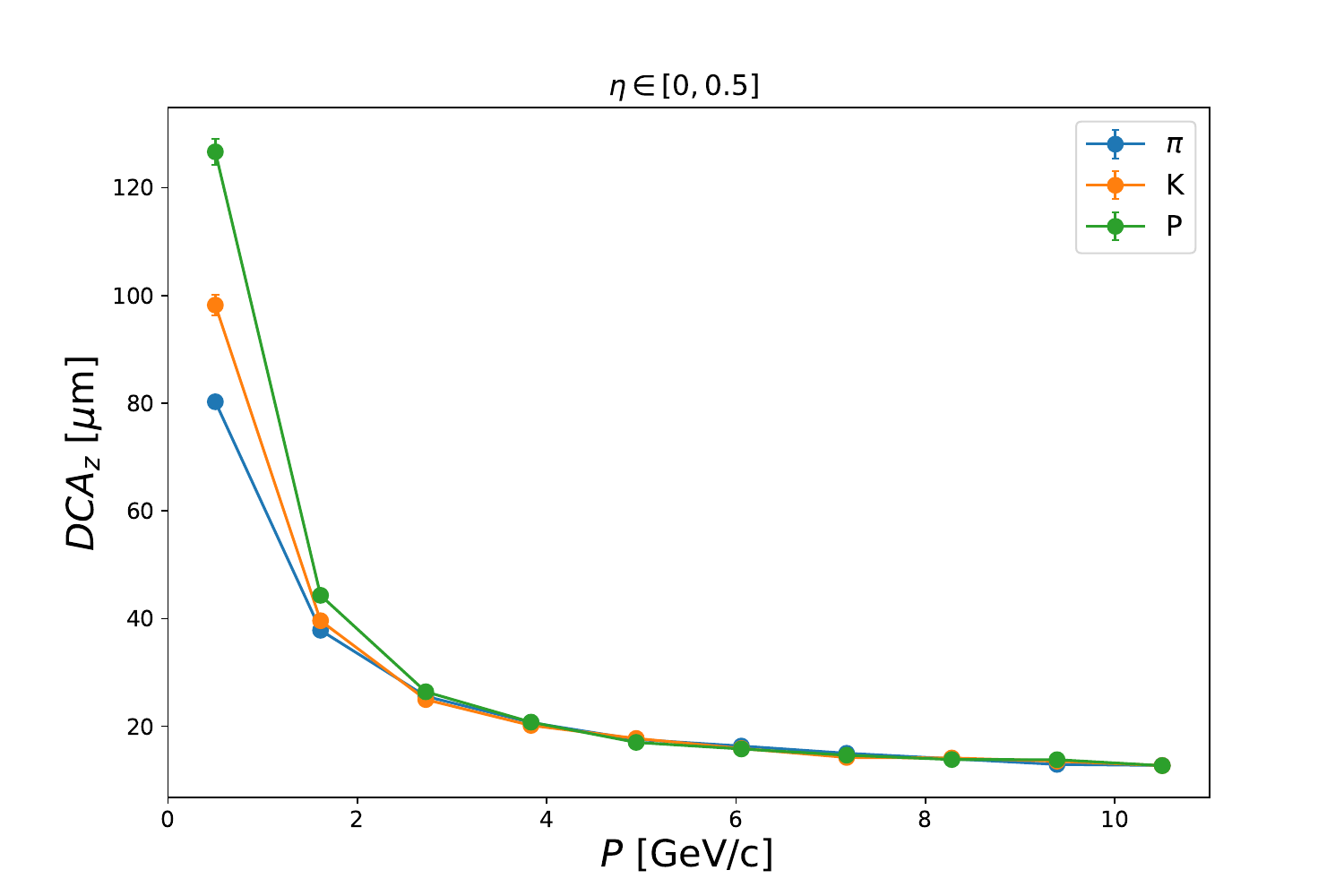}
\includegraphics[width=0.45\linewidth]{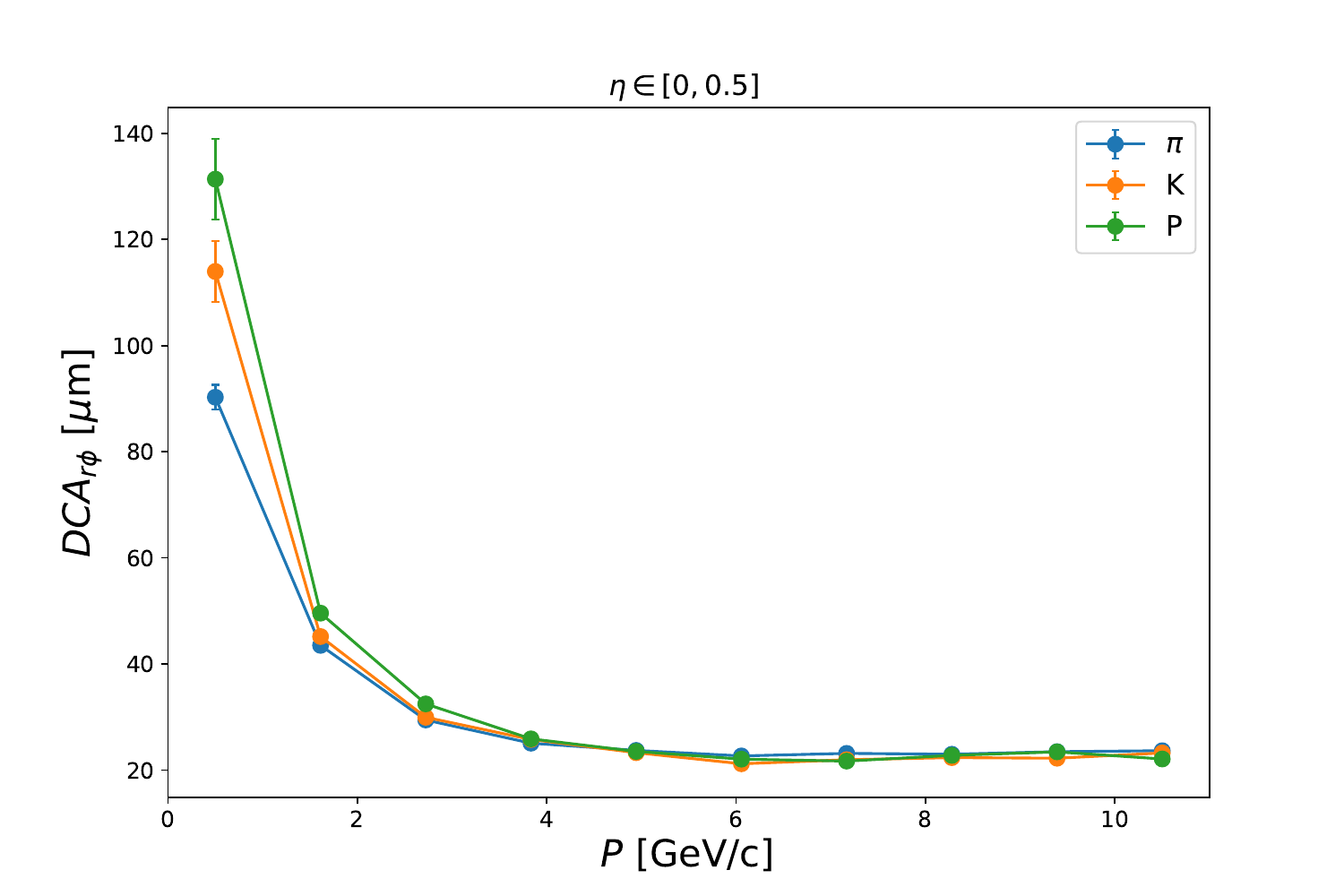}
\caption{Vertex resolution for different particles. Left: $DCA_z$ as a function of momentum in the $0 < \eta < 0.5$ range. Right: $DCA_{r\phi}$ as a function of momentum.}
\label{fig:reso_dca}
\end{figure}

The polar and azimuthal angular resolutions are quantified by the standard deviations derived from normal distributions fitted to the differences \(\Delta \theta\) (defined as \(\theta_{\text{truth}} - \theta_{\text{reco}}\)) and \(\Delta \phi\) (defined as \(\phi_{\text{truth}} - \phi_{\text{reco}}\)), respectively. These differences represent the deviations between the true angles, as generated in the simulation, and the reconstructed angles, as determined by the tracking system.
The angular resolutions are critical for accurately determining particle trajectories and interaction vertices. The polar angle \(\theta\) is particularly important for understanding the event geometry in the context of the detector's longitudinal axis, while the azimuthal angle \(\phi\) provides information about the particle's trajectory in the transverse plane.

\begin{figure}[htb]
\centering
\includegraphics[width=0.45\linewidth]{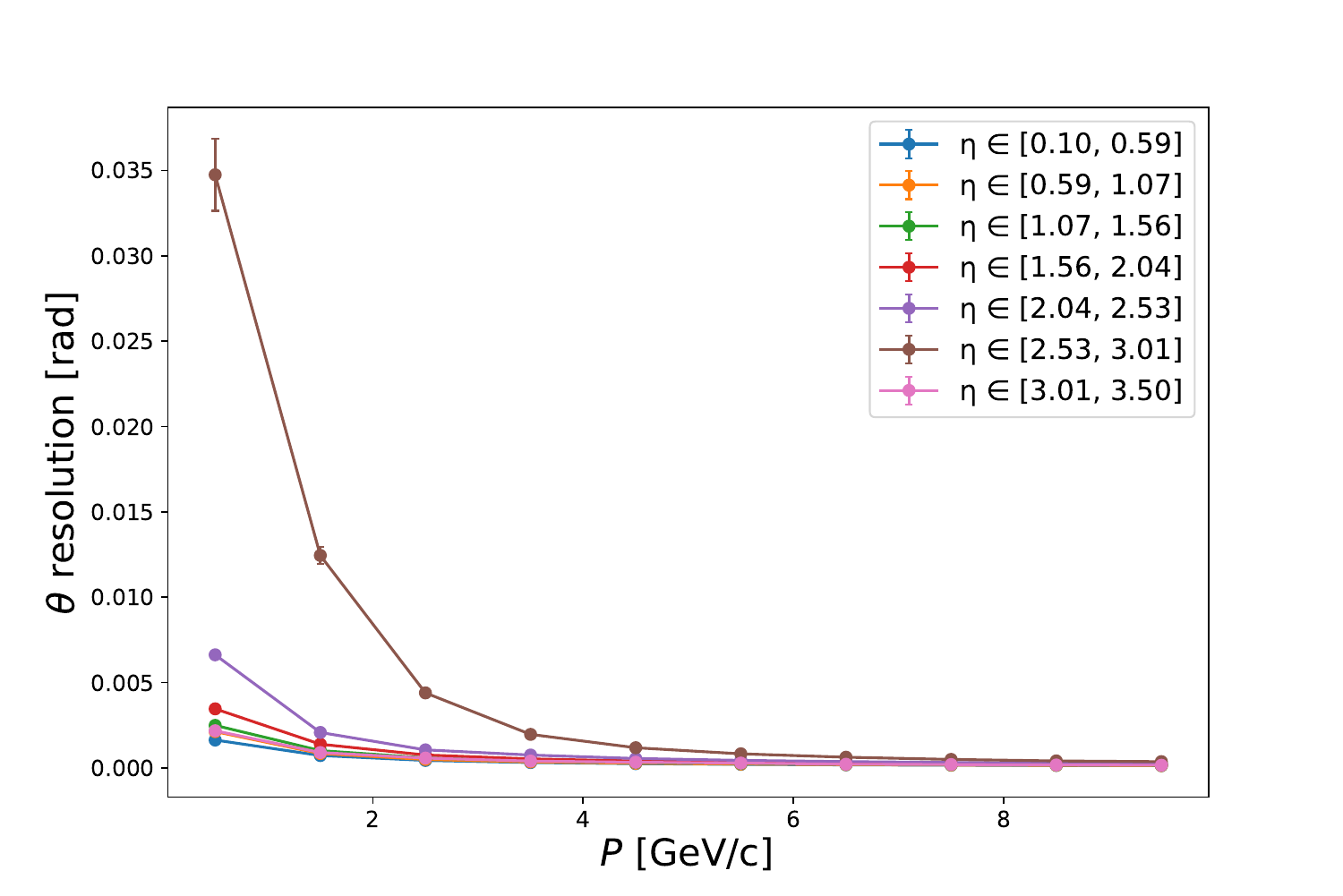}
\includegraphics[width=0.45\linewidth]{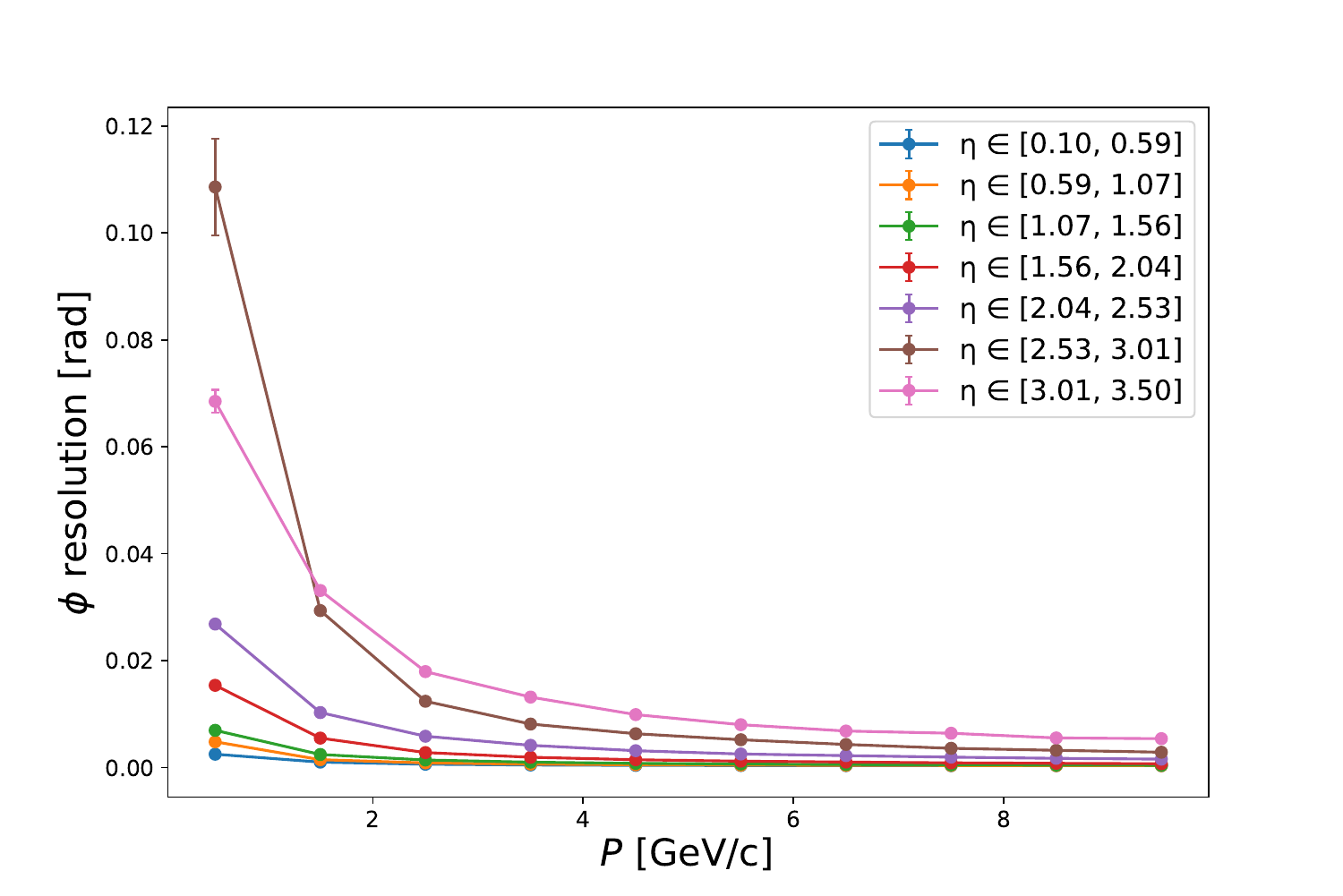}
\caption{Angular resolution as a function of momentum in different $\eta$ ranges. Left: $\theta$ resolution as a function of momentum. Right: $\phi$ resolution as a function of momentum.}
\label{fig:reso_ang}
\end{figure}

%\subsection{The angular resolution}
\subsection{The tracking and PID efficiency}

Tracking efficiency is the ratio of correctly reconstructed particle tracks to the total number of tracks that should have been found. As a key indicator of a detector's precision and reliability, achieving high tracking efficiency is vital for ensuring the quality of experimental data derived from particle interactions.

\begin{figure}[htb]
\centering
\includegraphics[width=0.65\linewidth]{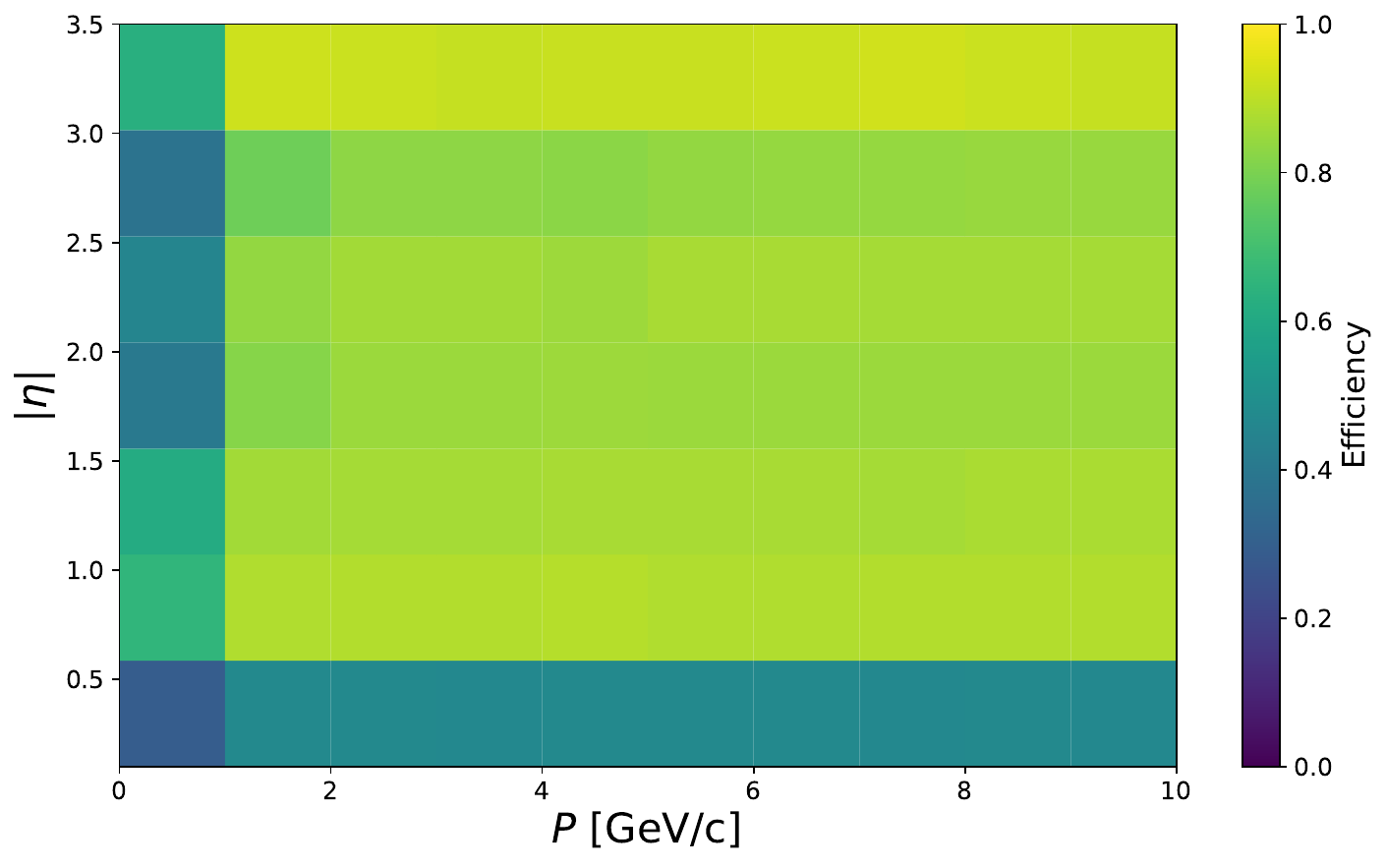}
\caption{The 2D tracking efficiency.}
\label{fig:tracking_eff}
\end{figure}

Figure~\ref{fig:tracking_eff} shows the charged pion tracking efficiency at different \(\eta\) and momentum regions. It is important to note that pattern recognition efficiency is not included in this study. The efficiency loss observed is primarily due to the detector acceptance and the quality of track fitting. To successfully reconstruct a track, we require that at least three hits be found. Consequently, in the central part of the detector (\(|\eta| < 1\)), the efficiency at low \(p_{\rm T}\) is worse than in the forward/backward regions. This is related to the minimum \(p_{\rm T}\) threshold needed for a track to reach the outer layer of the detector.

For the \(|\eta| > 1\) region, the loss of efficiency at higher \(p_{\rm T}\) is attributed to the lower acceptance at the edge of the barrel-to-endcap transition region. These regions present geometric challenges that can impede the successful reconstruction of tracks, particularly those with higher transverse momentum.

In the H-NS, PID efficiency is evaluated through a joint calculation of the measured mass and its uncertainty, incorporating both the time-of-flight (TOF) resolution $\sigma(t)$ and momentum resolution $\sigma(p)$. The squared mass is given by $m^2 = p^2 \left( \frac{1}{\beta^2} - 1 \right)$, and its uncertainty $\sigma_{m^2}$ is derived via error propagation.
%as $\sigma_{m^2} = m^2 \cdot \sqrt{ \left(2 \frac{\sigma_p}{p}\right)^2 + \left(2 \gamma^2 \frac{\sigma_\beta}{\beta}\right)^2 }$. 
Figure~\ref{fig:pid_mass2_vs_qp} shows the distribution of the squared mass of particles versus the product of particle charge $q$ and momentum $p$, based on a inclusive MC sample $pp\to X$ with beam kinetic energy of 9.3 GeV. The distributions for each type of particle are clearly visible and well distinguished.
For each particle hypothesis (e.g., $\pi$, $K$, $p$), the likelihood is computed under the assumption of a Gaussian probability density function $G(m, \sigma_m)$. The particle is assigned to the species yielding the highest likelihood. Notably, PID efficiency is not solely determined by the TOF system ($\sigma(t)$) but is equally and critically influenced by the tracker system ($\sigma(p)$), with both systems jointly dictating the final identification performance.

\begin{figure}[htb]
\centering
\includegraphics[width=0.65\linewidth]{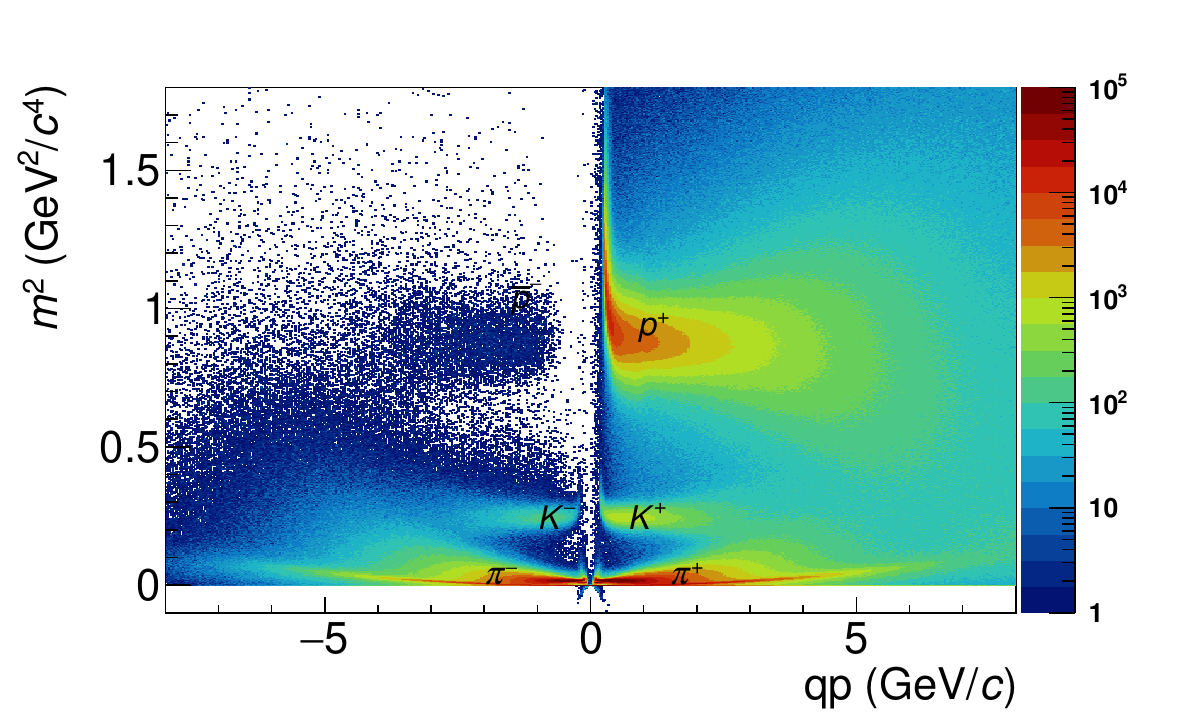}
\caption{The distribution of particle's squared mass versus the product of charge and momentum.}
\label{fig:pid_mass2_vs_qp}
\end{figure}

%\subsection{The PID efficiency}
\section{Physics projection}
\subsection{$\Lambda$ polarization in $pp(A)$}
In order to evaluate and optimize the performance of the $\Lambda$ polarization measurement at the H-NS detector, a full-process simulation, including event generation, {\sc Geant4}~\cite{GEANT4:2002zbu} simulation, tracking reconstruction, PID, signal events selection, and polarization extraction, is performed. 
The $\Lambda$ hyperon is reconstructed from its decay products, proton ($p$) and pion ($\pi^{-}$).
For the exclusive channel, $pp\to pK^{+}\Lambda$, each particle among the four final-state particles is required to register more than three hits in the detector. Except for the bachelor proton, all other particles are required to pass the PID selection criterion.
\begin{figure}[htb]
\centering
\includegraphics[width=0.45\linewidth]{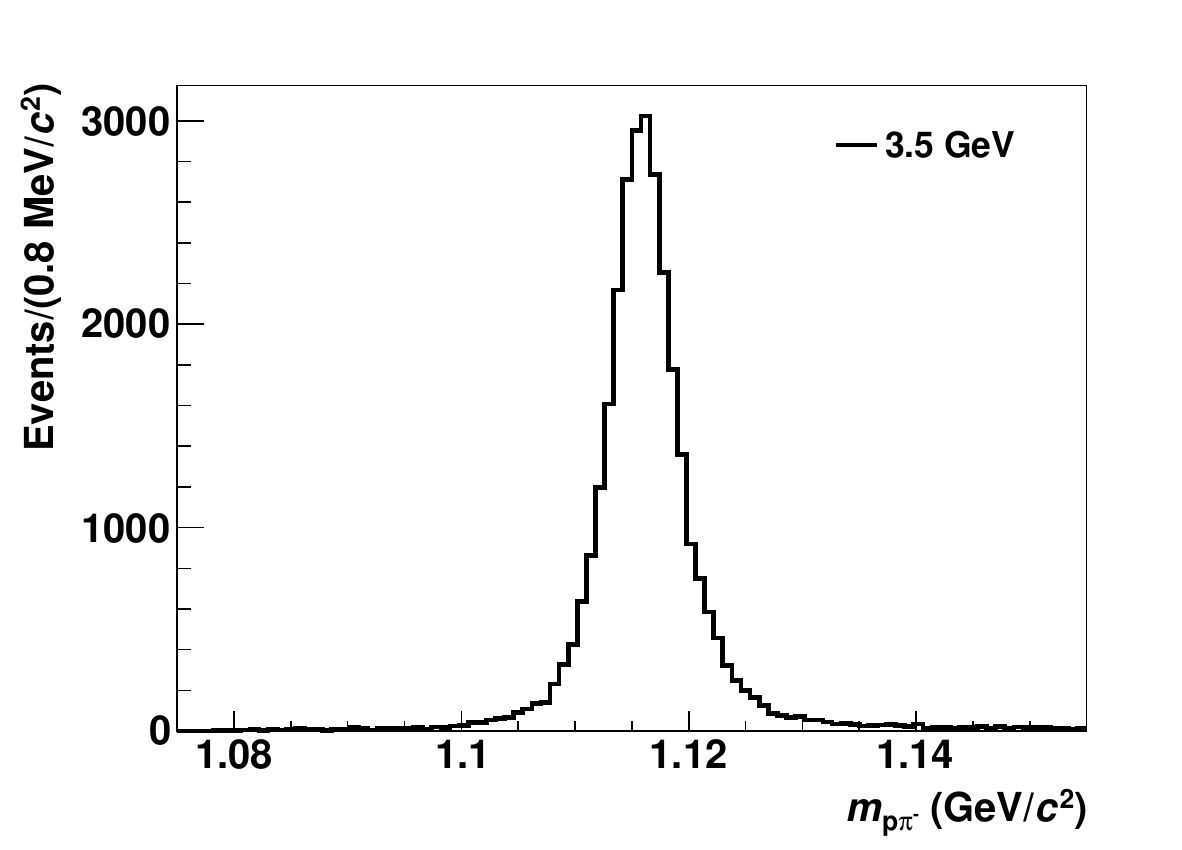}
\includegraphics[width=0.45\linewidth]{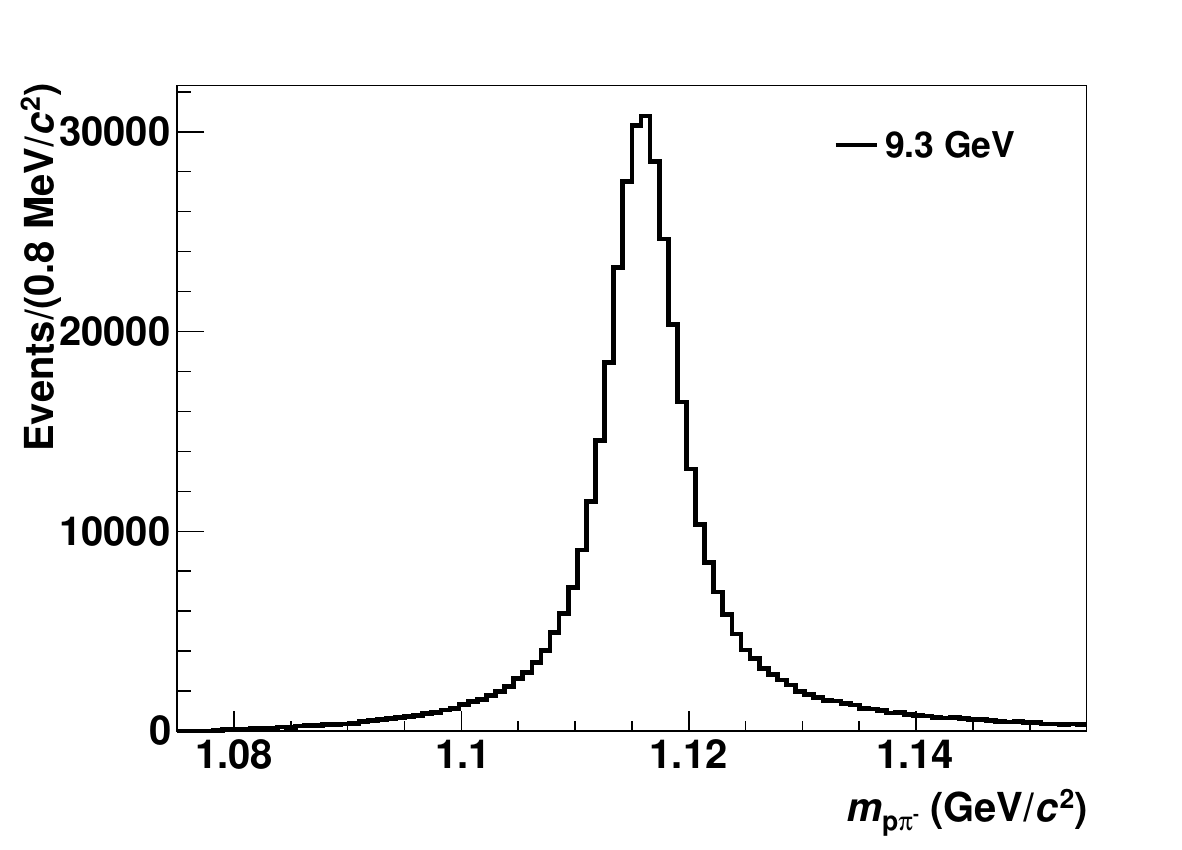}
\caption{The $\Lambda\to p\pi^{-}$ invariant mass distributions from $pp\to X\Lambda$ channel after applying all selection criteria. Left: beam kinetic energy of 3.5~GeV. Right: beam kinetic energy of 9.3~GeV.}
\label{fig:pp2XLambda_Lambda_mass}
\end{figure}
For the inclusive channel, $pp\to X\Lambda$, it is sufficient to find $p$ and $\pi^{-}$ in the final state that satisfy the hits and PID criteria, and then combine them to form the $\Lambda$ hyperon.
Furthermore, by using kinematic variables related to the $\Lambda$ decay, the background can be effectively suppressed and a clean signal can be selected, including the minimum distance between the trajectories of $p$ and $\pi^{-}$ ($DCA(p,\pi^{-})<0.4$~cm), the angle between the direction of the combined momentum and the vertex direction ($\Delta\theta<0.1$~rad), and the difference of the combined $\Lambda$ mass and world average $\Lambda$ mass from PDG ($|m_{p\pi^{-}}-m^{\rm pdg}_{\Lambda}|<0.04~{\rm GeV}$).

Using 20 million $pp$ events generated at the beam kinetic energy of 3.5~GeV and 9.3~GeV, respectively, and after applying the selection criteria described above, the $\Lambda$ mass distributions from the $pp\to X\Lambda$ channel are shown in the Fig.~\ref{fig:pp2XLambda_Lambda_mass}. The total number of reconstructed $\Lambda$ events has increased dramatically and the $\Lambda$ mass resolution gets worse with higher beam kinetic energy, as expected.
To estimate and extract the uncertainty of the $\Lambda$ polarization, the two‑dimensional distribution of $p_{\rm T}$ versus $x_{\rm F}$ is divided into $20\times20$ bins, where $p_{\rm T}$ is the transverse momentum of the $\Lambda$ and $x_{\rm F}$ is the Feynman scaling variable. 
The polarization extraction leverages polar asymmetry ($A_\theta = \mathcal{P}_y \alpha_{\Lambda}$), with $\mathcal{P}_y$ determined from fits to $N(\theta) = \frac{\mathcal{N}_{\rm obs}}{\pi}(1 + \mathcal{P}_y \alpha_{\Lambda}\cos\theta)$, where $\mathcal{N}_{\rm obs}$ is the reconstructed $\Lambda$ events, $\mathcal{P}_y$ is the $\Lambda$ polarization, and $\alpha_{\Lambda}$ is the asymmetric parameter of $\Lambda\to p\pi^{-}$ decay.
For H-NS's 1~MHz luminosity, statistical uncertainties are negligible ($\Delta\mathcal{P}_y/\mathcal{P}_y \sim \sqrt{3/\mathcal{N}_{\rm obs}}/(\mathcal{P}_y \alpha_{\Lambda})$), making systematic control of $\alpha_{\Lambda}$ the dominant precision factor.

Figure~\ref{fig:pp2XLambda_pol_xf_pt} shows, for beam kinetic energies of 3.5~GeV and 9.3~GeV, the uncertainty of $\Lambda$ polarization as a function of the two‑dimensional $p_{\rm T}$ versus $x_{\rm F}$ distribution after scaling the simulated 20 million $pp$ events up to a total of $10^{13}$ events (i.e., $\mathcal{N}_{\rm obs}$ is enlarged proportionally), running and collecting data at H-NS for a few months. It can be seen that, when H‑NS detector collects on the order of $10^{13}$ $pp$ events, the estimated uncertainty of the $\Lambda$ polarization measurement can reach the level of $0.002\%$. 
Figure~\ref{fig:pp2pKLambda_pol_xf_pt} shows the uncertainty of $\Lambda$ polarization using $10^{9}$ $pp\to pK^{+}\Lambda$ events, running and collecting data at H-NS for approximately half a month, the estimated uncertainty of the $\Lambda$ polarization measurement can reach the level of $0.06\%$.

\begin{figure}[htb]
\centering
\includegraphics[width=0.45\linewidth]{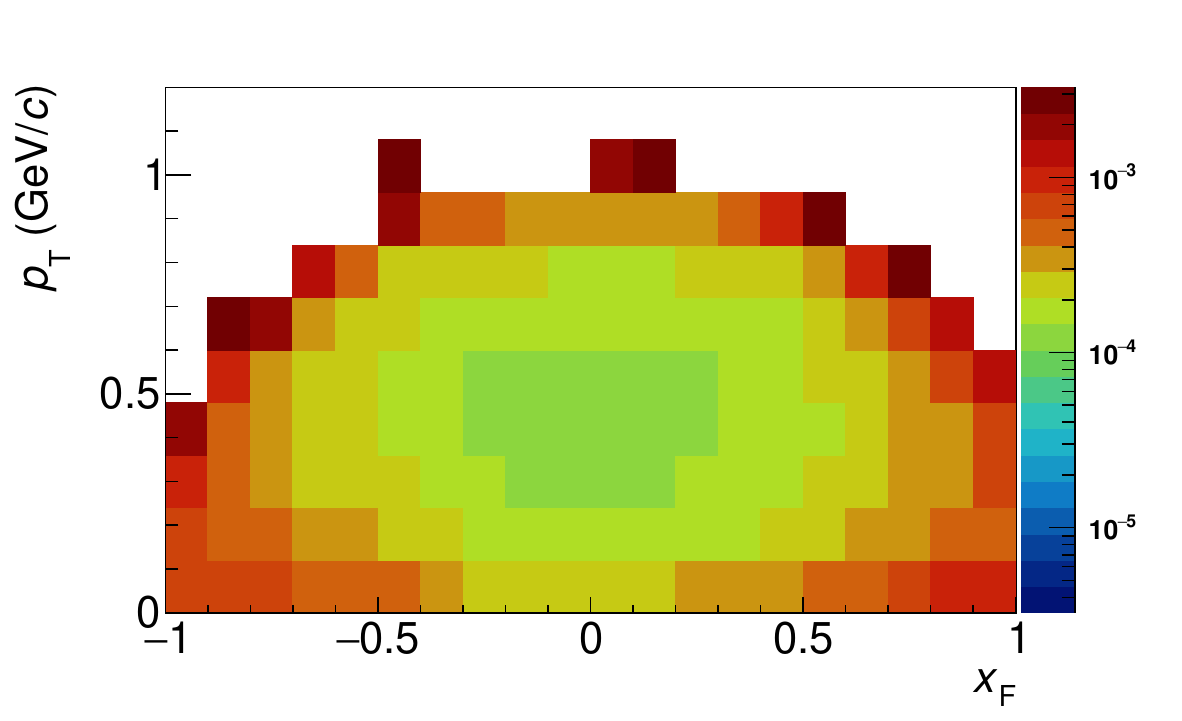}
\includegraphics[width=0.45\linewidth]{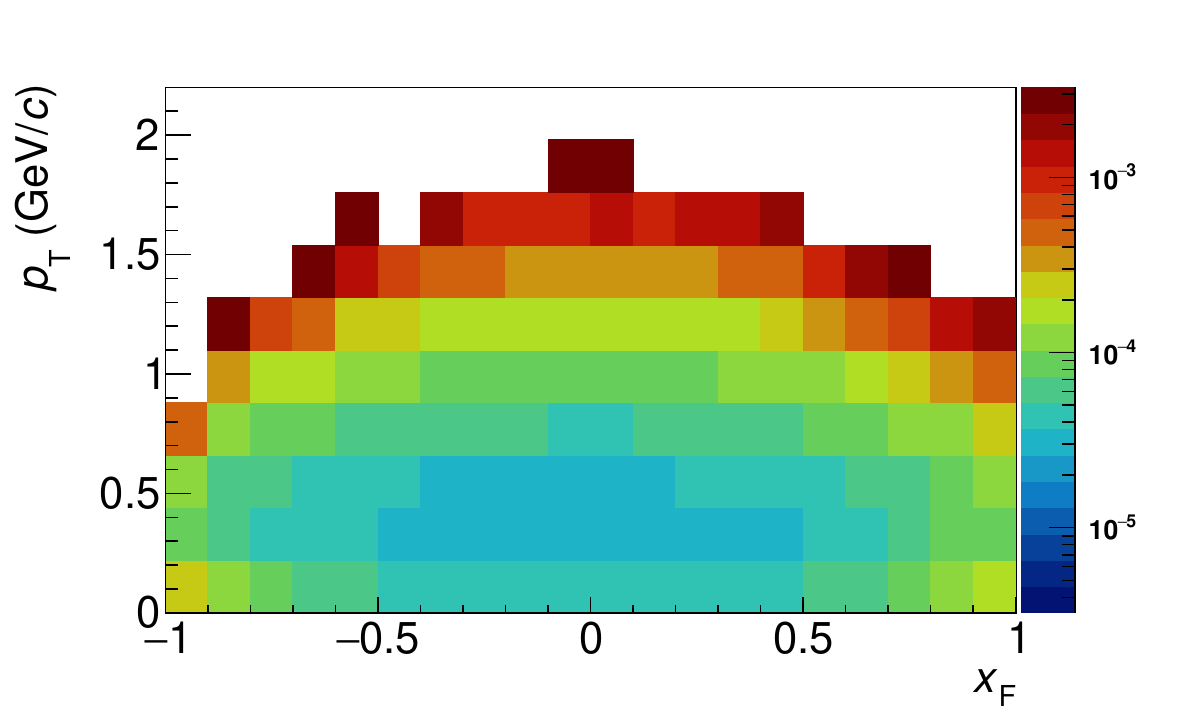}
\caption{The 2D uncertainty of $\Lambda$ polarization of $pp\to X\Lambda$ channel using $10^{13}$ $pp$ events (data from H-NS will take a few months). Left: beam kinetic energy of 3.5~GeV. Right: beam kinetic energy of 9.3~GeV.}
\label{fig:pp2XLambda_pol_xf_pt}
\end{figure}

\begin{figure}[htb]
\centering
\includegraphics[width=0.45\linewidth]{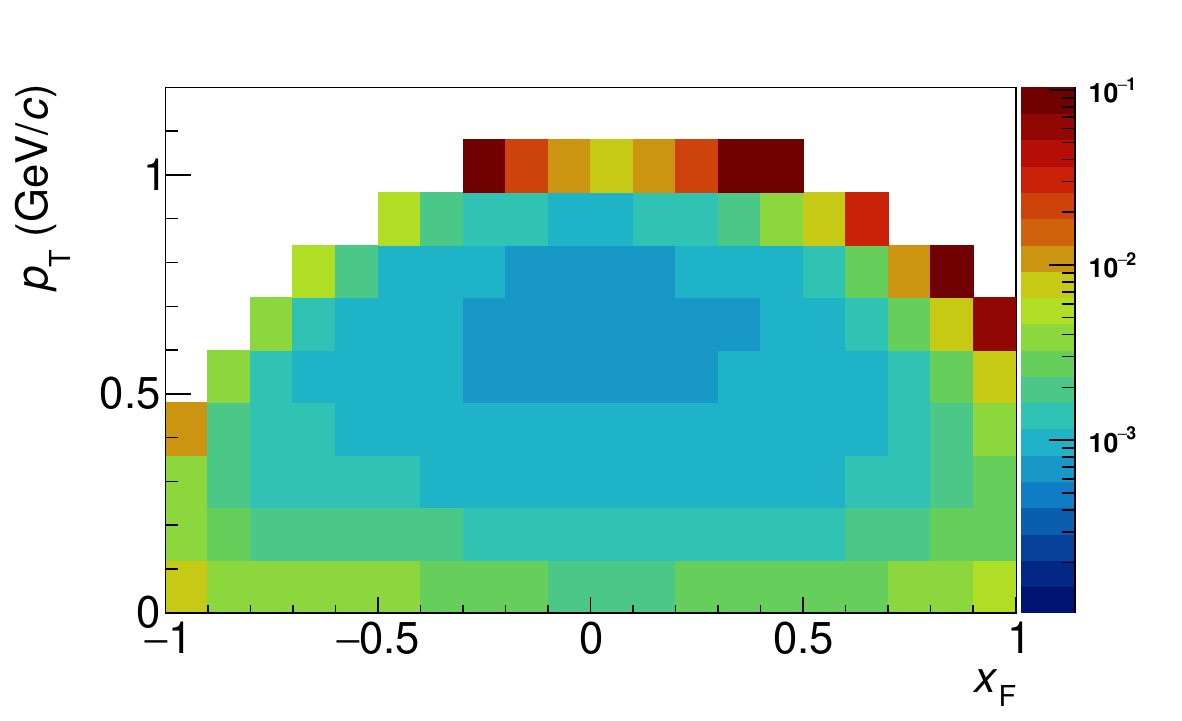}
\includegraphics[width=0.45\linewidth]{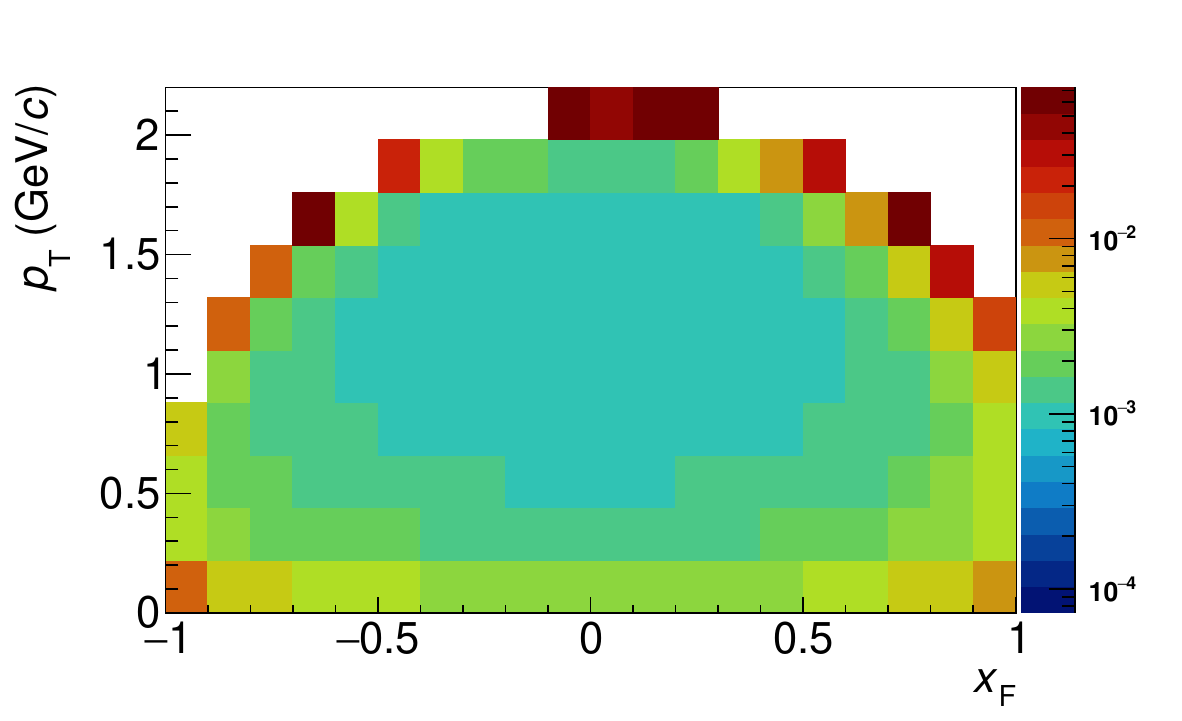}
\caption{The 2D uncertainty of $\Lambda$ polarization using $10^{9}$ simulated $pp\to pK^{+}\Lambda$ events (data from H-NS will take half a month). Left: beam kinetic energy of 3.5~GeV. Right: beam kinetic energy of 9.3~GeV.}
\label{fig:pp2pKLambda_pol_xf_pt}
\end{figure}

%\clearpage
\subsection{Proton polarization in $pp(A)$}
The H-NS detector incorporates a novel integrated polarimeter through the addition of thin carbon scattering targets (1~mm, $<$1\% $X/X_0$) within its tracker system, enabling precision nucleon polarization measurements without compromising conventional detector performance. Extensive simulations validated minimal impact on momentum resolution ($\sim$0.1\% deterioration) and tracking efficiency. Final-state proton polarization is extracted via $p\textrm{C}$ elastic scattering analysis, where a comprehensive event selection strategy combines $\chi^2_{\rm trk} > 100$ cut, $\theta_{\rm sc} > 5^{\circ}$ angular filter, and vertex reconstruction ($\sigma_{R_{xy}} = 0.29$~cm) to achieve 70\% signal efficiency with $4\times10^{-6}$ background survival rate. Critical resolutions include $\Delta p/p$ momentum resolution and milliradian-level angular precision, $\sigma_\theta, \sigma_\phi$, as shown in Fig. \ref{fig:res_mom_recoil_p}, vital for polarization analysis.

\begin{figure}[htb]
\includegraphics[width=0.32\textwidth]{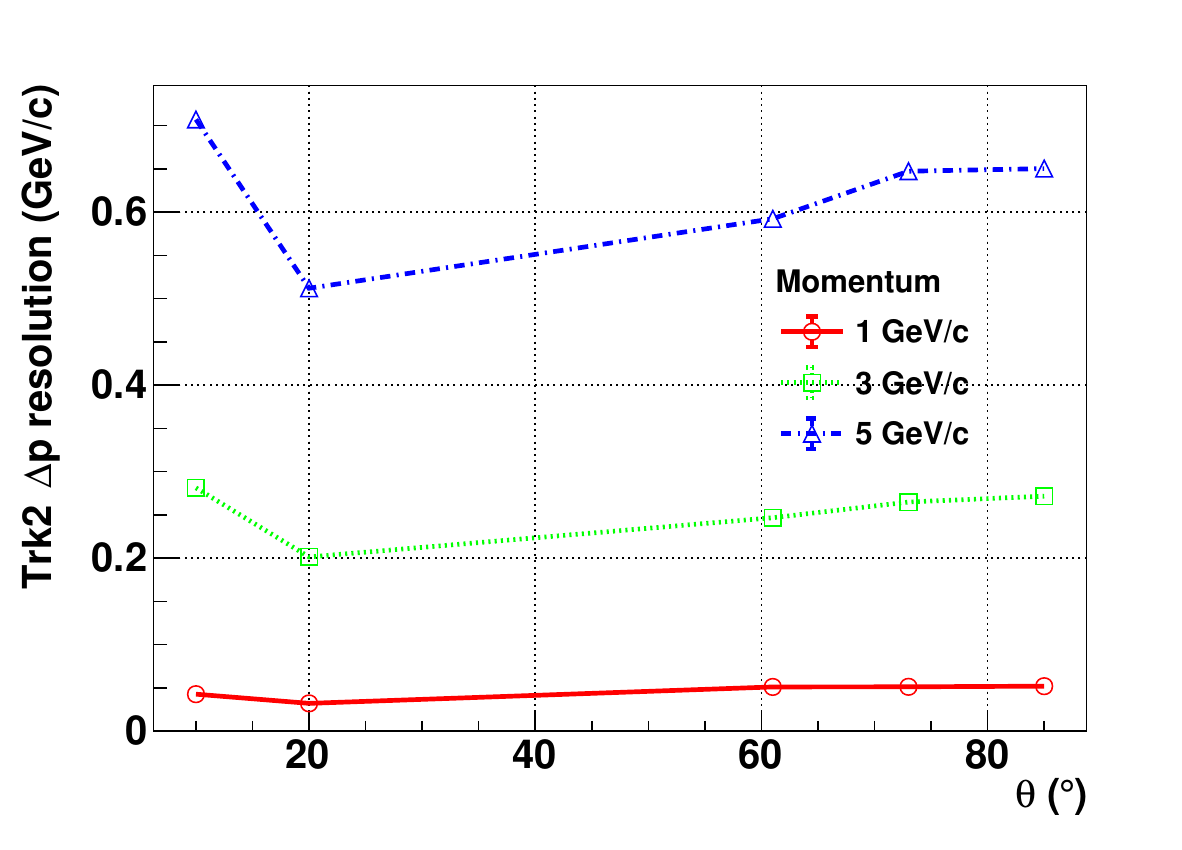}
\includegraphics[width=0.32\textwidth]{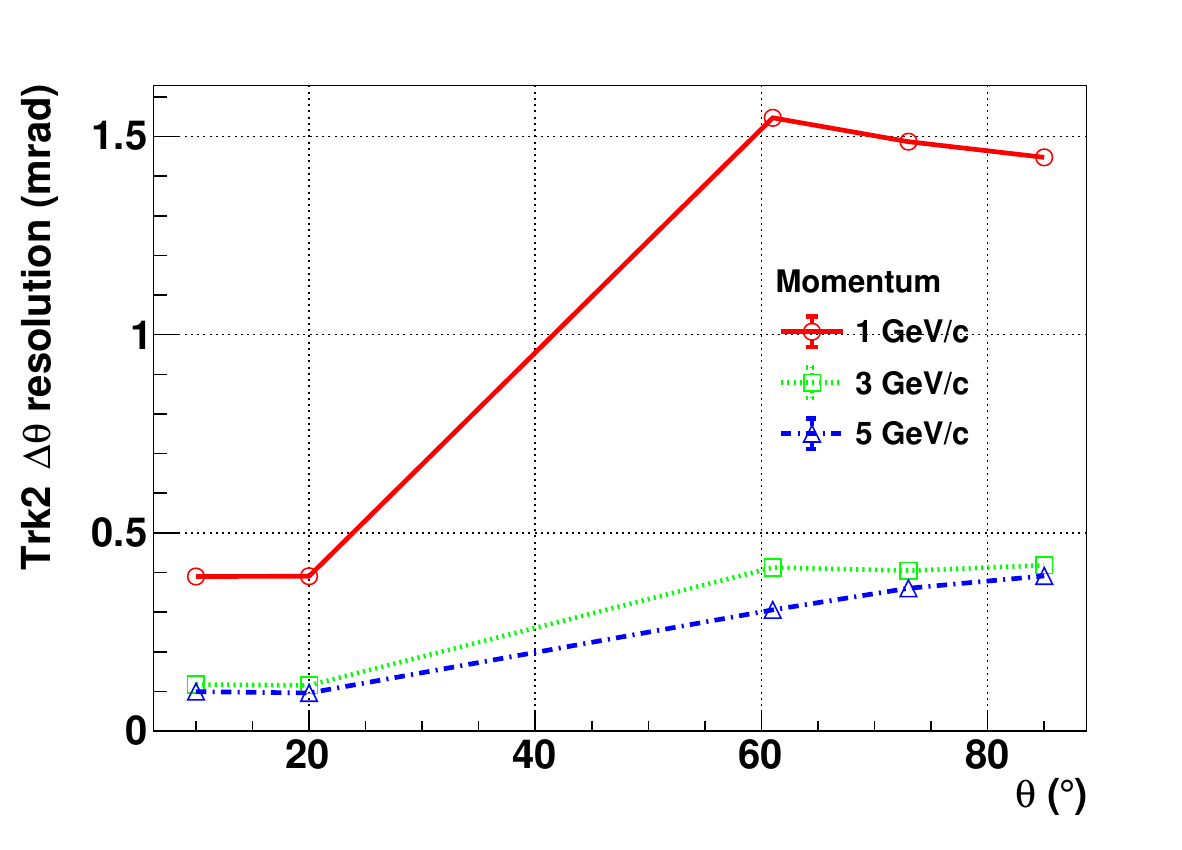}
\includegraphics[width=0.32\textwidth]{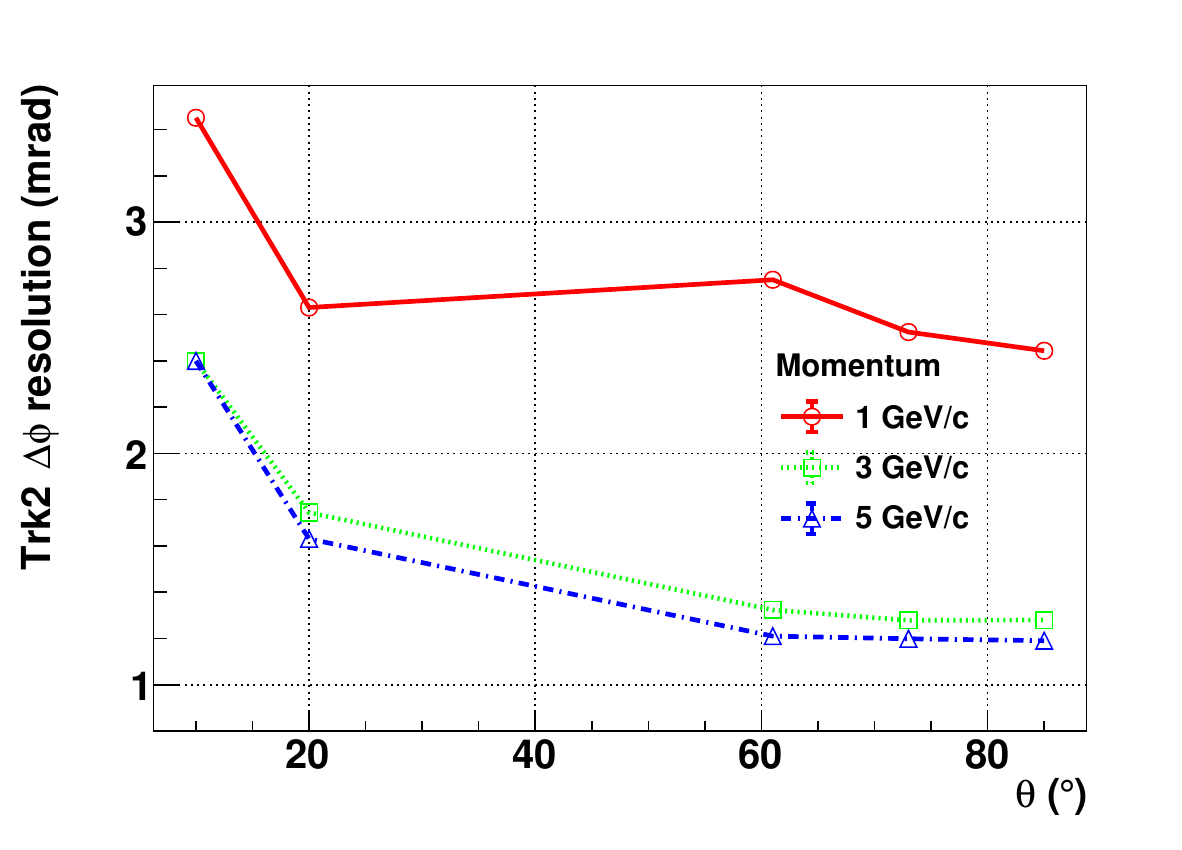}
\caption{\label{fig:res_mom_recoil_p} {Momentum (left), $\theta$ (middle) and $\phi$ (right) resolutions of the scattered proton reconstructed with the outer three tracking layers as a function of polar angle for incident proton with momentum of 1 GeV/{\it c}, 3 GeV/{\it c} and 5 GeV/{\it c}, respectively.
}}
\end{figure}

Polarization extraction leverages $p\textrm{C}$ elastic scattering azimuthal asymmetry ($A_\phi = \mathcal{P}_y A_N^{\rm ave}$), with $\mathcal{P}_y$ determined from fits to $N(\phi) = \frac{\mathcal{N}_{\rm obs}}{2\pi} (1 + \mathcal{P}_y A_N^{\rm ave}\cos\phi)$. For H-NS's 1~MHz luminosity, statistical uncertainties are negligible ($\Delta\mathcal{P}_y/\mathcal{P}_y \sim \sqrt{2/\mathcal{N}_{\rm tot}}/(\mathcal{P}_y A_N^{\rm ave})$), making systematic control of $A_N^{\rm ave}$ the dominant precision factor. 
Figure~\ref{fig:pp2Xp_pol_xf_pt} shows, for beam kinetic energies of 3.5~GeV and 9.3~GeV, the uncertainty of proton polarization as a function of the two‑dimensional $p_{\rm T}$ versus $x_{\rm F}$ distribution after scaling the simulated 20 million $pp$ events up to a total of $10^{13}$ events (i.e., $N_{\rm obs}$ is enlarged proportionally). It can be seen that, when H‑NS detector collects on the order of $10^{13}$ $pp$ events, the estimated uncertainty of the proton polarization measurement can reach the level of $0.02\%$. 
This adaptable design allows target optimization (thickness/material) and establishes a paradigm for future polarization studies at facilities like EIC, CEPC, and EicC, enabling unprecedented probes of strange quark contributions to hyperon polarization through comparative $\Lambda$-proton spin measurements. Further, if the spin of proton can be measured via the polarimeter, one can select the spin state of $\Xi^{-}+p$ and $\Lambda+p+\pi^{-}$ system, which allows us to search for the so-called the $H$-dibaryon~\cite{Jaffe:1976yi} ($J=0$) state via $H \rightarrow \Xi^{-}+p$ or $H \rightarrow \Lambda+p+\pi^{-}$ channel by reconstructing the invariant mass and/or femtoscopy measurements.

\begin{figure}[htb]
\centering
\includegraphics[width=0.45\linewidth]{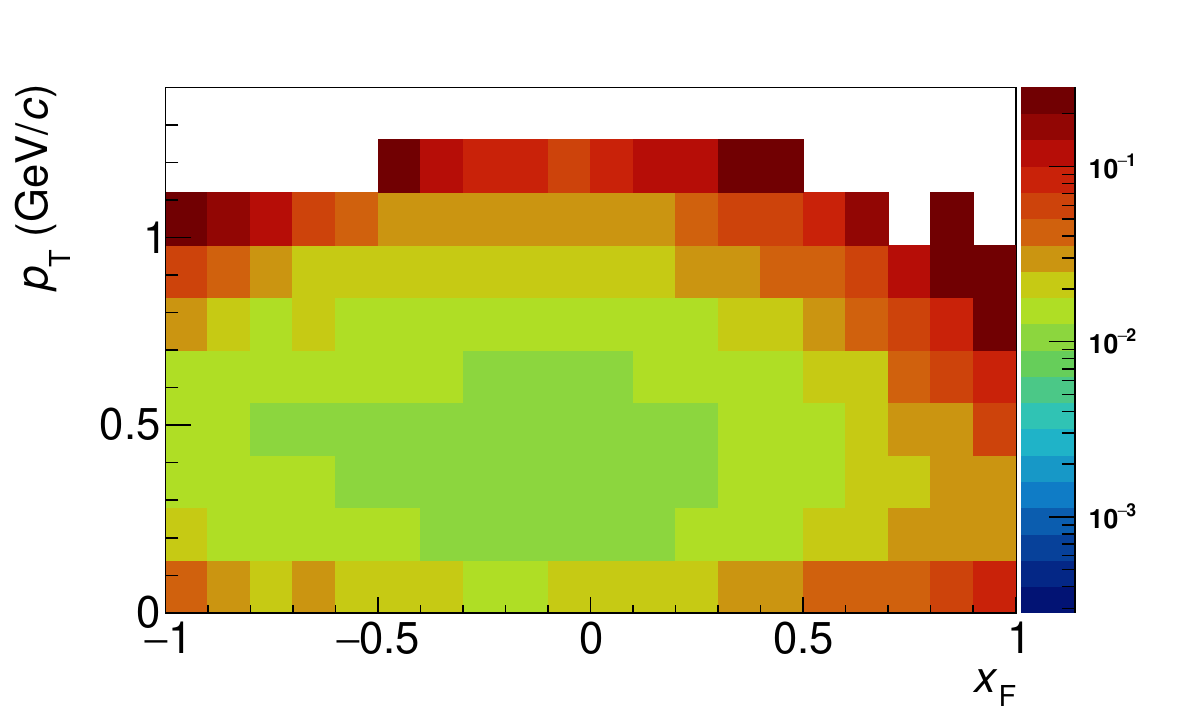}
\includegraphics[width=0.45\linewidth]{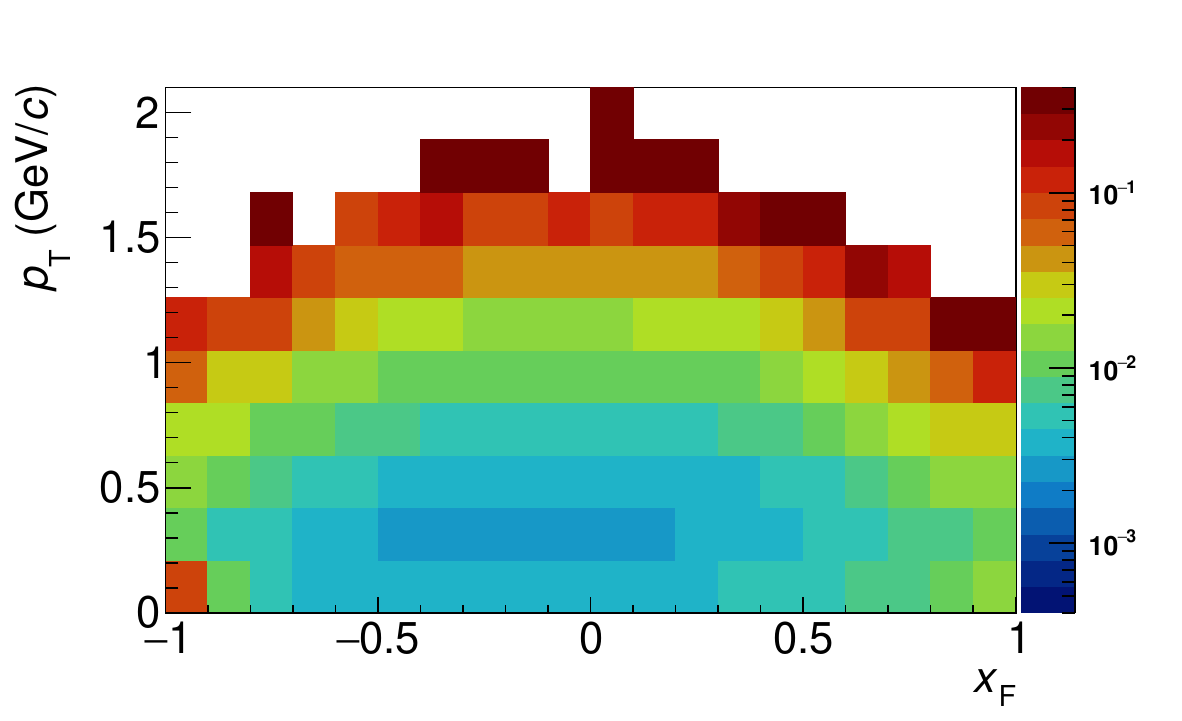}
\caption{The 2D uncertainty of proton polarization from $pp\to X p$ channel using $10^{13}$ $pp$ events (data from H-NS will take a few months). Left: beam kinetic energy of 3.5~GeV. Right: beam kinetic energy of 9.3~GeV.}
\label{fig:pp2Xp_pol_xf_pt}
\end{figure}

\subsection{$\Lambda$ global polarization in $AA$}

The global polarization of $\Lambda$ hyperon can be determined from the angular distribution of hyperon decay products in $\Lambda$’s rest frame with respect to the system orbital angular momenta (OAM)~\cite{STAR:2017ckg}:
\begin{equation}
    P_{H} = \frac{8}{\pi \alpha_{\Lambda}}\frac{1}{R_{\rm EP}^{1}}\left< \sin(\Psi_{1}-\phi_{p}^{*})\right>,
\end{equation}
where $\alpha_{\Lambda} = 0.732 \pm 0.014$ is the $\Lambda$ decay parameter; $\phi^{*}_{p}$ is the azimuthal angle of the daughter proton momentum in the $\Lambda$ rest frame; $\Psi_{1}$ is the first-order event plane angle and the $R_{\rm EP}^{1}$ is the first-order event plane resolution. In this chapter, the first-order event plane is reconstructed with forward MAPS tracker, a typical resolution in mid-central Au+Au collision is $\sim75\%$. The $\Lambda$ candidates, shown in left panel of Fig.~\ref{fig:AuAu_GlobLdPhMethod}, are reconstructed using barrel MAPS tracker to avoid self-correlation.
To extract $\Lambda$ polarization signal, the invariant-mass method is used~\cite{STAR:2018gyt,STAR:2021beb}, in which the $\left< \sin(\Psi_{1}-\phi_{p}^{*})\right>$ is a function of invariant mass which have contributions from both signal and background:
\begin{equation}
    \left< \sin(\Psi_{1}-\phi_{p}^{*})\right>^{\rm obs} = (1-f^{\rm bkg}(M_{\rm inv}))\left< \sin(\Psi_{1}-\phi_{p}^{*})\right>^{\rm sig} + f^{\rm bkg}(M_{\rm inv})\left< \sin(\Psi_{1}-\phi_{p}^{*})\right>^{\rm bkg}.
\end{equation}
An example of polarization signal extraction is shown in the right panel of Fig.~\ref{fig:AuAu_GlobLdPhMethod}.

\begin{figure}[htb]
\centering
\includegraphics[width=0.8\linewidth]{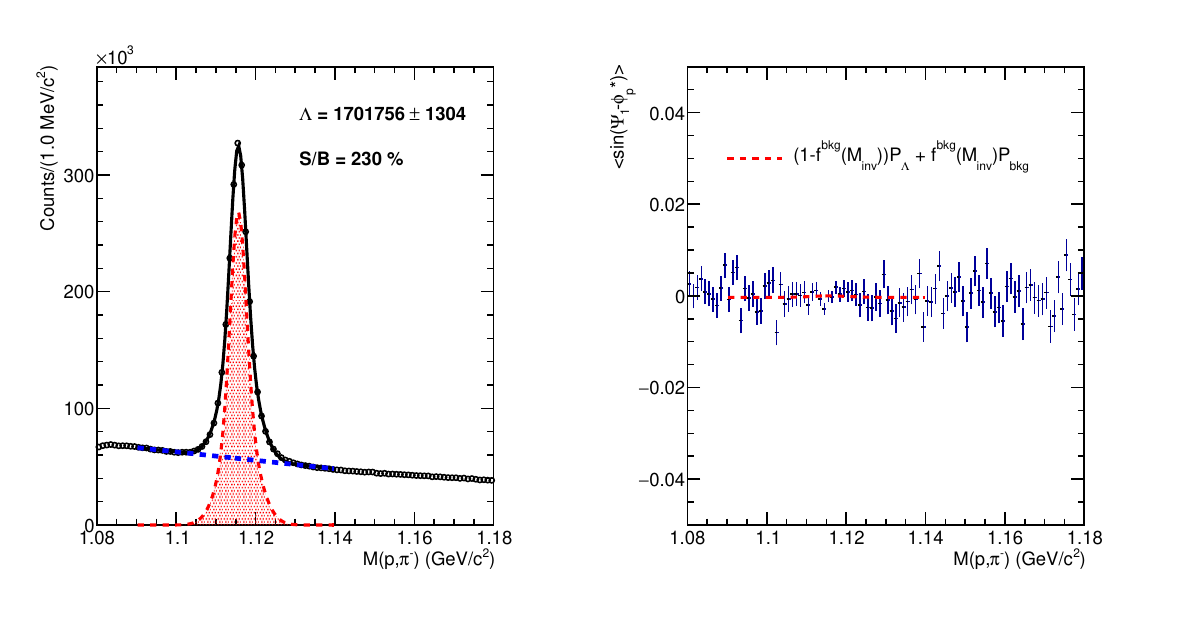}
\caption{The $\Lambda\to p\pi^{-}$ invariant mass distributions from Au+Au events at beam kinetic energy of 2.8 GeV/u after applying all selection criteria. Left: $\Lambda$ signal extraction. Right: example of global polarization signal extraction w.r.t. event plane.}
\label{fig:AuAu_GlobLdPhMethod}
\end{figure}

\begin{figure}[htb]
\centering
\includegraphics[width=0.8\linewidth]{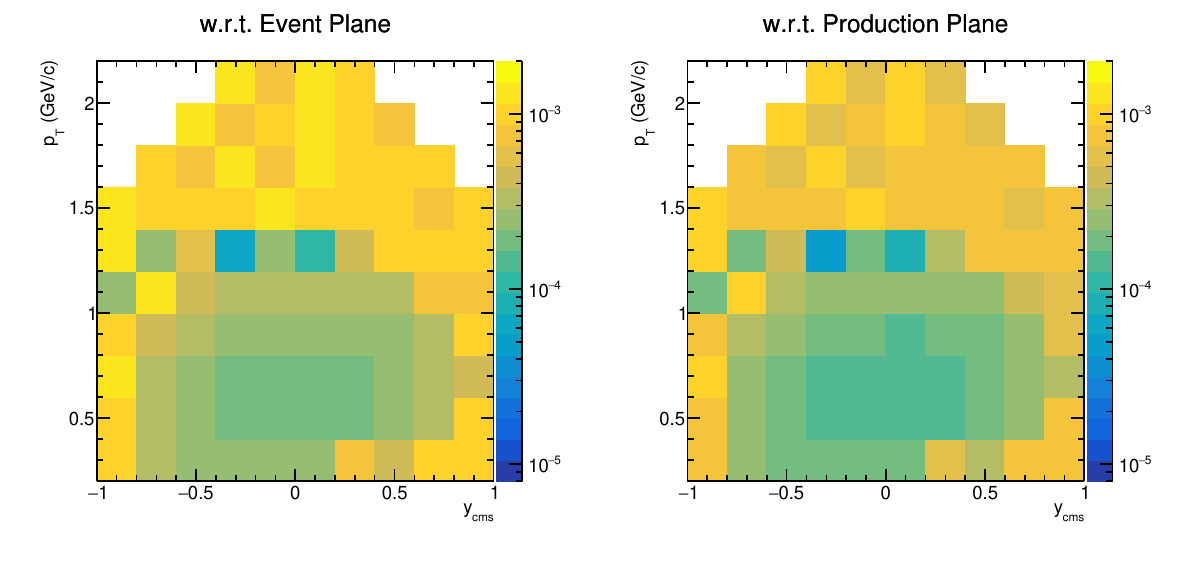}
\caption{The 2D uncertainty of $\Lambda$ polarization using $10^{11}$ Au+Au events at beam kinetic energy of 2.8 GeV/u (data from H-NS will take a few weeks). Left: $\Lambda$ polarization uncertainty w.r.t event plane. Right: $\Lambda$ polarization uncertainty w.r.t production plane.}
\label{fig:AuAu_GlobLdPhErrProj}
\end{figure}

Figure~\ref{fig:AuAu_GlobLdPhErrProj} presents the estimated $\Lambda$ polarization uncertainty, w.r.t. event plane (left) and production plane (right), as a function of the two-dimensional $p_{\rm T}$ versus $y_{\rm cms}$ distribution  at the beam kinetic energy of 2.8 GeV/u. By scaling the simulated 20 million Au+Au events up to a total of $10^{11}$ events, the estimated uncertainty of the $\Lambda$ polarization measurement can reach the level of 0.1$\%$, which takes a few weeks of H-NS running time.
\end{chapter}

\begin{chapter}{Summary}

%\textcolor{red}{This is the summary chapter}

In summary, we propose the construction of the Hyperon-Nucleon Spectrometer at HIAF. The spectrometer is specifically designed to investigate the polarization of  $\Lambda$ hyperons, protons as well as light (hyper-)nuclei in $pp$, $pA$ and $AA$ collisions, with a particular emphasis on the high-$x_{\rm F}$ region. The H-NS will perform systematic and precise multi-dimensional measurements of polarization as a function of $p_{\rm T}$, $x_{\rm F}$, and collision energy across the HIAF coverage, with proton beam energy from 3 GeV up to 32 GeV. The results from the H-NS program will provide unprecedented data to constrain and validate the non-perturbative dynamics of QCD. By systematically mapping hyperon polarization from the hadronic to partonic regime, the experiment will address fundamental questions, directly shedding light on the origin of proton spin and, by extension, the inner structure of the visible matter in the universe.

In addition, The H-NS will be a state-of-the-art spectrometer, leveraging cutting-edge detector technologies. A key feature of its design is the extensive use of advanced silicon detectors. Monolithic Active Pixel Sensors will provide high-resolution vertexing capabilities essential for the precise reconstruction of secondary decay vertices, a prerequisite for analyzing the weak decays of hyperons. Furthermore, the incorporation of Low-Gain Avalanche Diodes will deliver excellent timing resolution, enabling precise event reconstruction and background suppression in the high-rate environment of HIAF. 

Both the physics program and the detector technology are highly compelling. The H-NS collaboration already comprises over 21 institutes from China and several international partners. Most of the detector technologies have completed the R\&D phase, making the project schedule primarily funding-driven. Furthermore, H-NS will ensure a sustainable and smooth transition from HIAF to the upgraded HIAF and the complex of the Electron-ion collider in China (EicC)~\cite{Anderle:2021wcy}. In many ways, H-NS is an important technology verification platform and pre-research project for EicC's key detector technologies, such as high-granularity silicon pixel and precision timing detectors.
The success of the H-NS is strategically vital, extending beyond its core physics program. The expertise developed for H-NS will provide invaluable contributions to the future EicC in both physics program as well as the detector technology. 
\end{chapter}

%\input{EicC_physics_Inclusive}

%\input{EicC_physics_SIDIS}

%\input{EicC_physics_Exclusive}

%\input{EicC_physics_Heavy_Flavor}

%\end{chapter}

%%%%%%%%%%%%%%%%%%%%%%%%%%%%%%%%%%%%%%%%%%%%%%%%%%%%%%%%%%%%%%%%%%%%%%%%%%%%%%%%%%%%%%%

\newpage
\bibliographystyle{unsrt}
\bibliography{reference}

\end{document}